\documentclass[twocolumn,showpacs,aps,floatfix,prd,nofootinbib]{revtex4}
\usepackage{graphicx}
\usepackage{dcolumn}
\usepackage{array, longtable}
\usepackage{epsfig}
\usepackage{amsmath}
\usepackage{colordvi}
\usepackage{hhline}

\newcommand{\BaBarYear}{07}
\newcommand{\BaBarNumber}{021}
\newcommand{\SLACPubNumber}{12435}

 \newcommand{\BaBarType}      {PUB}  

\input pubboard/babarsym

\def\Ecm       {\ensuremath {E_{\rm c.m.}}\xspace}
\def\KKKK      {\ensuremath {\Kp\Km\Kp\Km}\xspace}
\def\KKppch    {\ensuremath {\Kp\Km\pip\pim}\xspace}
\def\KKppnt    {\ensuremath {\Kp\Km\pi^0\pi^0}\xspace}

\def\Kppch     {\ensuremath {KK\pip\pim}\xspace}
\def\Kppnt     {\ensuremath {KK\piz\piz}\xspace}

\def\chifourpi {\ensuremath {\chi^2_{4\pi}}\xspace}
\def\chiKKppch {\ensuremath {\chi^2_{KK\pip\pim}}\xspace}
\def\chiKKppnt {\ensuremath {\chi^2_{KK\pi^0\pi^0}}\xspace}
\def\chifourK  {\ensuremath {\chi^2_{4K}}\xspace}

\long\def\inst#1{\par\nobreak\kern 4pt\nobreak
    {\it #1}\par\vskip 10pt plus 3pt minus 3pt}

\begin{document}

\begin{flushleft}
\babar-\BaBarType-\BaBarYear/\BaBarNumber \\
SLAC-PUB-\SLACPubNumber \\
Phys. Rev. {\bf D76}, 012008 (2007)
\end{flushleft}


\title{\large \bf
\boldmath
The $\epem\to K^+ K^- \pipi$, $K^+ K^-  \ppz$ and $K^+ K^- K^+ K^-$   
Cross Sections 
Measured with Initial-State Radiation 
} 

%
\author{B.~Aubert}
\author{M.~Bona}
\author{D.~Boutigny}
\author{Y.~Karyotakis}
\author{J.~P.~Lees}
\author{V.~Poireau}
\author{X.~Prudent}
\author{V.~Tisserand}
\author{A.~Zghiche}
\affiliation{Laboratoire de Physique des Particules, IN2P3/CNRS et Universit\'e de Savoie, F-74941 Annecy-Le-Vieux, France }
\author{J.~Garra~Tico}
\author{E.~Grauges}
\affiliation{Universitat de Barcelona, Facultat de Fisica, Departament ECM, E-08028 Barcelona, Spain }
% ...
\collaboration{The \babar\ Collaboration}
\noaffiliation

\date{\today}

\begin{abstract}

We study the processes $\epem\to K^+ K^- \pipi\gamma$, 
$K^+K^-\ppz\gamma$ and $K^+ K^- K^+ K^-\gamma$,
where the photon is radiated from the initial state.  
About 34600, 4400 and 2300 fully reconstructed events, respectively, 
are selected from 232~\invfb of \babar\ data. 
The invariant mass of the hadronic final state defines the effective \epem 
center-of-mass energy, 
so that the $K^+ K^- \pipi\gamma$ data can be compared with direct 
measurements of the $\epem\to K^+K^- \pipi$ reaction;
no direct measurements exist for the 
$\epem\to K^+ K^- \ppz$ or $\epem\to K^+ K^-  K^+ K^-$ reactions. 
Studying the structure of these events, we find contributions from a
number of intermediate states, and we extract their cross sections
where possible.
In particular, we isolate the contribution from 
$\epem\to\phi(1020) f_{0}(980)$ and study its structure near threshold.
In the charmonium region,
we observe the $J/\psi$ in all three final states and several
intermediate states, 
as well as the $\psi(2S)$ in some modes,
and measure the corresponding branching fractions.  
We see no signal for the $Y(4260)$ and obtain an
 upper limit of $\BR_{Y(4260)\to\phi\pipi}\cdot\Gamma^{Y}_{ee}<0.4~\ev$ 
at 90\% C.L.

\end{abstract}


\vfill
\maketitle

\setcounter{footnote}{0}

\section{Introduction}
\label{sec:Introduction}

Electron-positron annihilation at fixed center-of-mass (c.m.) energies
has long been a mainstay of research in elementary particle physics.
The idea of utilizing initial-state radiation (ISR) to explore \epem 
reactions below the nominal c.m. energies was outlined in 
Ref.~\cite{baier},
and discussed in the context of high-luminosity $\rm \phi$ and
$B$ factories in Refs.~\cite{arbus, kuehn, ivanch}.
At high energies, \epem annihilation is dominated by quark-level
processes producing two or more hadronic jets.
However, low-multiplicity exclusive processes dominate at energies
below about 2~\gev, and the region near charm threshold, 3.0--4.5~\gev,
features a number of resonances~\cite{PDG}.
These allow us to probe a wealth of physics parameters,
including cross sections,
spectroscopy and form factors.

Of particular current interest are the 
recently observed states in the charmonium region,
such as the $Y(4260)$~\cite{y4260},
and a possible discrepancy between the 
measured value of the anomalous magnetic moment of the muon, $g_\mu-2$, 
and that predicted by the Standard Model~\cite{dehz}.
Charmonium and other states with $J^{PC}=1^{--}$ can be observed as resonances
in the cross section, 
and intermediate states may be present in the hadronic system.
Measurements of the decay modes and their branching fractions are important
in understanding the nature of these states.
For example, the glue-ball model~\cite{shin} predicts a large branching fraction for
$Y(4260)$ into $\phi\pi\pi$.
The prediction for $g_\mu-2$ is based on hadronic-loop corrections measured from
low-energy $\epem\to\,$hadrons data,
and these dominate the uncertainty on the prediction.
Improving this prediction requires not only more precise measurements,
but also measurements over the entire energy range and inclusion of
all the important subprocesses in order to understand possible
acceptance effects.
ISR events at $B$ factories provide independent and contiguous measurements of
hadronic cross sections from the production threshold to
about 5~\gev.

The cross section for the radiation of a photon of energy $E_{\gamma}$ 
followed by the production of a particular hadronic final state $f$ 
is related to the corresponding direct $\epem\to f$ cross section 
$\sigma_f(s)$ by
\begin{equation}
\frac{d\sigma_{\gamma f}(s,x)}{dx} = W(s,x)\cdot \sigma_f(s(1-x))\ ,
\label{eq1}
\end{equation}
where $\sqrt{s}$ is the initial \epem c.m.\@ energy, 
$x\! =\! 2E_{\gamma}/\sqrt{s}$ is the fractional energy of the ISR
photon 
and $\Ecm \!\equiv\! \sqrt{s(1-x)}$ is the effective c.m.\@ energy at
which the final state $f$ is produced. 
The probability density function $W(s,x)$ for ISR photon emission has
been calculated with better than 1\% precision (see e.g.\ Ref.~\cite{ivanch}).
It falls rapidly as $E_{\gamma}$ increases from zero, but has a long
tail, which combines with the increasing $\sigma_f(s(1-x))$ to produce
a sizable cross section at very low \Ecm.
The angular distribution of the ISR photon peaks along the beam directions, 
but 10--15\%~\cite{ivanch} of the photons are within a typical
detector acceptance.

Experimentally, the measured invariant mass of the hadronic final
state defines \Ecm.
An important feature of ISR data is that a wide range of energies is
scanned simultaneously in one experiment, 
so that no structure is missed and
the relative normalization uncertainties in
data from different experiments or 
accelerator parameters are avoided.
Furthermore, for large values of $x$ the hadronic system is collimated, 
reducing acceptance issues and allowing measurements at energies down to 
production threshold.
The mass resolution is not as good as a typical beam energy spread used in 
direct measurements,
but the resolution and absolute energy scale can be monitored
by the width and mass of well known resonances, such as the $J/\psi$
produced in the reaction $\epem \to J/\psi\gamma$. 
Backgrounds from $\epem \!\to\,$hadrons events at the nominal $\sqrt{s}$
and from other ISR processes can be 
suppressed by a combination of particle identification and 
kinematic fitting techniques.
Studies of $\epem\to\mumu\gamma$ and several multi-hadron ISR processes using
\babar\ data have been reported~\cite{Druzhinin1,isr3pi,isr4pi,isr6pi},
demonstrating the viability of such measurements.

The $K^+K^- \pipi$ final state has been measured directly by the
DM1 collaboration~\cite{2k2pidm1} for $\sqrt{s} <\! 2.2~\gev$,
and we have previously published ISR measurements of the
$K^+K^- \pipi$ and $K^+K^-K^+K^-$ final states~\cite{isr4pi} for
$\Ecm \!<\! 4.5~\gev$.
We recently reported~\cite{phif0prd} an updated measurement of the  
$K^+K^- \pipi$ final state with a larger data sample, along with
the first measurement of the $K^+K^- \ppz$ final state, in which we
observed a structure near threshold in the $\phi f_0$ intermediate state.
In this paper we present a more detailed study of these two final states 
along with an updated measurement of the $K^+K^-K^+K^-$ final state.
In all cases we require detection of the ISR photon and perform a set
of kinematic fits.
We are able to suppress backgrounds sufficiently to study these
final states from their respective production thresholds up to 5~\gev.
In addition to measuring the overall cross sections, 
we study the internal structure of the events
and measure cross sections for a number of intermediate states.
We study the charmonium region, 
measure several $J/\psi$ and $\psi(2S)$ branching fractions,
and set limits on other states.

\section{\boldmath The \babar\ detector and dataset}
\label{sec:babar}

The data used in this analysis were collected with the \babar\ detector at
the \pep2\ asymmetric energy \epem\ storage rings. 
The total integrated luminosity used is 232~\invfb, 
which includes 211~\invfb collected at the $\Upsilon(4S)$ peak, 
$\sqrt{s}=10.58~\gev$, 
and 21~\invfb collected below the resonance, at $\sqrt{s}=10.54~\gev$.

The \babar\ detector is described elsewhere~\cite{babar}. 
Here we use charged particles reconstructed in the tracking system,
which comprises the five-layer silicon vertex tracker (SVT) 
and the 40-layer drift chamber (DCH) in a 1.5 T axial magnetic field.
Separation of charged pions, kaons and protons uses a
combination of Cherenkov angles measured in the detector of internally
reflected Cherenkov light (DIRC) and specific ionization measured in
the SVT and DCH. 
For the present study we use a kaon identification algorithm that provides 90--95\% 
efficiency, depending on momentum, and pion and proton rejection
factors in the 20--100 range.
Photon and electron energies are measured in the CsI(Tl)
electromagnetic calorimeter (EMC).
We use muon identification provided by the instrumented flux return (IFR)
to select the $\mumu\gamma$ final state.

To study the detector acceptance and efficiency, 
we use a simulation package developed for radiative processes.
The simulation of hadronic final states, including
$K^+K^- \pipi \gamma$, $K^+K^- \ppz\gamma$ and $K^+K^- K^+K^-\gamma$,
is based on the approach suggested by Czy\.z and K\"uhn\cite{kuehn2}.  
Multiple soft-photon emission from the initial-state charged particles is
implemented with a structure-function technique~\cite{kuraev, strfun},
and photon radiation from the final-state particles is simulated by
the PHOTOS package~\cite{PHOTOS}.  
The accuracy of the radiative corrections is about 1\%.

We simulate the $K^+K^-\pi\pi$ final states both according to phase 
space and with models that include the $\phi(1020)\to K^+K^-$ and/or
$f_{0}(980)\to \pi\pi$ channels,
and the $K^+K^- K^+K^-$ final state both according to phase space and
including the $\phi \to K^+K^-$ channel.
The generated events go
through a detailed detector simulation~\cite{GEANT4}, 
and we reconstruct them with the same software chain as the experimental data. 
Variations in detector and background conditions are taken into account.

We also generate a large number of background processes, 
including the ISR channels 
$\epem \!\!\to\! \pipi\pipi\gamma$ and $\pipi\ppz\gamma$,
which can contribute due to particle misidentification,
and $\phi\eta\gamma$, $\phi\pi^0\gamma$, $\pipi\pi^0\gamma$, 
which have larger cross sections and can contribute via missing or
spurious tracks or photons.
In addition, we study the non-ISR backgrounds
$\epem \!\!\to\! q \qbar$ $(q = u, d, s, c)$ generated by
JETSET~\cite{jetset} and
$\epem \!\!\to\! \tau^+\tau^-$ by KORALB~\cite{koralb}. The contribution from the
\Y4S decays is found to be negligible.
The cross sections for these processes are known with 
about 10\% accuracy or better, which is sufficient for these measurements.

\section{\boldmath Event Selection and Kinematic Fit}
\label{sec:Fits}

In the initial selection of candidate events, we consider
photon candidates in the EMC with energy above 0.03~\gev
and charged tracks reconstructed in the DCH or SVT or both that 
extrapolate within 0.25 cm of the 
beam axis in the transverse plane and within 3 cm of the nominal collision point
along the axis.
These criteria are looser than in our previous analysis~\cite{isr4pi},
and have been chosen to maximize efficiency.
We require a high-energy photon in the event with an energy in the
initial \epem c.m.\ frame of $E_\gamma > 3~\gev$,
and either exactly four charged tracks with zero net charge 
and total momentum roughly opposite to the photon direction,
or exactly two oppositely charged tracks that combine with a set of
other photons to roughly balance the highest-energy photon momentum.
We fit a vertex to the set of charged tracks and use it as the point
of origin to calculate the photon direction.
Most events contain additional soft photons due to 
machine background or interactions in the detector material.

We subject each of these candidate events to a set of constrained 
kinematic fits, and use the fit results,
along with charged-particle identification,
both to select the final states of interest and to measure backgrounds
from other processes.
We assume the photon with the highest $E_\gamma$ in the c.m. frame is the ISR
photon, and the kinematic fits use its direction along with the
four-momenta and covariance matrices of the initial \epem and the
set of selected tracks and photons.
Because of excellent resolution for the momenta in the DCH and good angular
resolution for the photons in the EMC, 
the ISR photon energy is determined with better resolution
through four-momentum conservation than through measurement in the EMC. 
Therefore we do not use its measured energy in the fits,
eliminating the systematic uncertainty due to the EMC calibration for
high energy photons.
The fitted three-momenta for each charged track and photon are used 
in further kinematical calculations.

For the four-track candidates, the fits have three constraints (3C).
We first fit to the $\pipi\pipi$ hypothesis, obtaining a \chifourpi.
If the four tracks include one identified $K^+$ and one $K^-$,
we fit to the $K^+K^-\pipi$ hypothesis and retain the event as a
\KKppch candidate.
For events with one identified kaon, we perform fits with each of
the two oppositely charged tracks given the kaon hypothesis, and the
combination with the lower \chiKKppch is retained if it is lower than
$\chifourpi$.
If the event contains three or four identified $K^\pm$, 
we fit to the $K^+K^-K^+K^-$ hypothesis and retain the event as a 
\KKKK candidate.

For the events with two charged tracks and five or more photon candidates, 
we require both tracks to be identified as kaons
to suppress background from ISR $\pipi\ppz$ and $K^\pm\KS\pi^\mp$ events.
We then pair all non-ISR photon candidates and consider combinations with
invariant mass within $\pm$30~\mevcc of the $\pi^0$ mass as $\pi^0$ candidates.
We perform a six-constraint (6C) fit to each set of two non-overlapping 
$\pi^0$ candidates plus the ISR photon direction, the two tracks and
the beam particles.
Both \piz candidates are constrained to the \piz mass, 
and we retain the combination with the lowest \chiKKppnt.

\section{The {\boldmath $K^+ K^- \pipi$} final state}
\subsection{Final Selection and Backgrounds}
\label{sec:selection1}

\begin{figure}[t]
\includegraphics[width=0.9\linewidth]{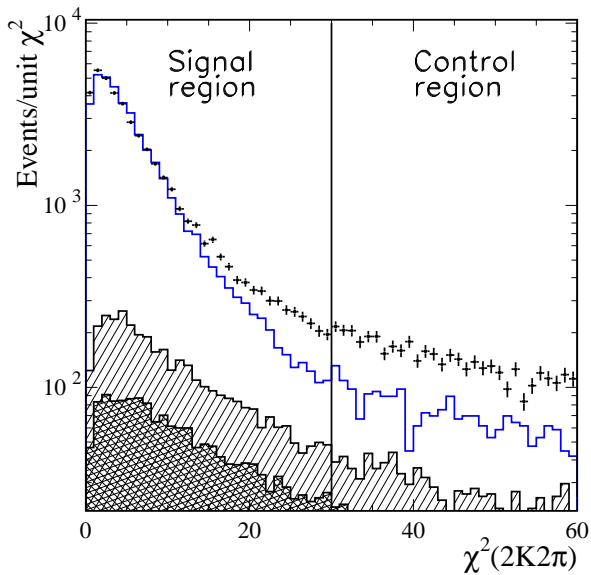}
\vspace{-0.4cm}
\caption{
  Distribution of \chisq from the three-constraint fit for \KKppch candidates
  in the data (points).
  The open histogram is the distribution for simulated signal events, 
  normalized as described in the text.
  The cross-hatched (hatched) histogram represents the background 
  from non-ISR events (plus that from ISR $4\pi$ events), estimated as
  described in the text.
}
\label{2k2pi_chi2_all}
\end{figure}

The experimental \chiKKppch distribution for 
the \KKppch candidates is shown in Fig.~\ref{2k2pi_chi2_all} as points,
and the open histogram is the distribution for the simulated \KKppch events.
The simulated distribution is normalized to the data in the region 
$\chiKKppch\!\! <\! 10$ where the backgrounds and radiative corrections
are insignificant.
The experimental distribution has contributions from background processes,
but the simulated distribution is also 
broader than the expected 3C \chisq distribution. 
This is due to multiple soft-photon emission from the initial state
and radiation from the final-state charged particles, 
which are not taken into account by the fit, 
but are present in both data and simulation. 
The shape of the \chisq distribution at high values was studied in detail
~\cite{isr4pi,isr6pi} using specific ISR processes for which a
very clean sample can be obtained without any limit on the \chisq value.

The cross-hatched histogram in Fig.~\ref{2k2pi_chi2_all} represents
the background from $\epem \!\!\to\! \qqbar$ events, 
which is based on the JETSET simulation.
It is dominated by events with a hard $\pi^0$ producing a fake ISR
photon, 
and the similar kinematics cause it to peak at low values of \chiKKppch.
We evaluate this background in a number of \Ecm ranges by
combining the ISR photon candidate with another photon candidate in 
both data and simulated events,
and comparing the \piz signals in the resulting $\gamma\gamma$
invariant mass distributions.
The simulation gives an \Ecm-dependence consistent with the data,
so we normalize it by an overall factor.
The hatched histogram represents the sum of this background and that from
ISR $\epem \!\!\to\! \pipi\pipi$ events with one or two misidentified 
$\pi^\pm$, 
which also contributes at low \chisq values.
We estimate the contribution as a function of \Ecm from a
simulation using the known cross section~\cite{isr4pi}.

All remaining background sources are either negligible or give a
\chiKKppch distribution that is nearly uniform over the range
shown in Fig.~\ref{2k2pi_chi2_all}.
We therefore define a signal region $\chiKKppch\!\! <\! 30$,
and estimate the sum of the remaining backgrounds from the difference
between the number of data and simulated entries in a control region,
$30\! <\!\chiKKppch\!\! <\!60$.
This difference is normalized to the corresponding difference in the
signal region, as described in detail in Refs.~\cite{isr4pi,isr6pi}.
The signal region contains 34635 data and 14077 simulated events, and 
the control region contains 4634 data and 723 simulated events.

\begin{figure}[tbh]
\includegraphics[width=0.9\linewidth]{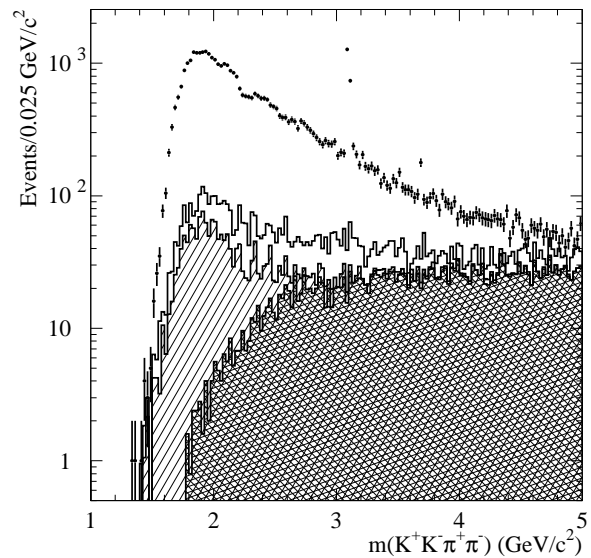}
\vspace{-0.4cm}
\caption{
  The invariant mass distribution for \KKppch candidates in the data (points):
  the cross-hatched, hatched and open histograms represent, cumulatively,
  the non-ISR background, 
  the contribution from ISR $\pipi\pipi$ events,
  and the ISR background from the control region of Fig.~\ref{2k2pi_chi2_all}.
  }
\label{2k2pi_babar}
\end{figure}

Figure~\ref{2k2pi_babar} shows the \KKppch invariant mass distribution 
from threshold up to 5.0~\gevcc for events in the signal region.
Narrow peaks are apparent at the $J/\psi$ and $\psi(2S)$ masses.
The cross-hatched histogram represents the \qqbar background,
which is negligible at low mass but becomes large at higher masses.
The hatched region represents the ISR $\pipi\pipi$ contribution,
which we estimate to be 2.4\% of the selected events on average.
The open histogram represents the sum of all backgrounds,
including those estimated from the control region.
They total 6--8\% at low mass but account for 20-25\% of the
observed data near 4 \gevcc and become the largest contribution
near 5~\gevcc.

We subtract the sum of backgrounds in each mass bin to obtain a number
of signal events.
Considering uncertainties in the cross sections for the background processes, 
the normalization of events in the control region and 
the simulation statistics,
we estimate a systematic uncertainty on the signal yield that is
less than 3\% in the 1.6--3~\gevcc mass region, but 
increases to 3--5\% in the region above 3~\gevcc.

\subsection{Selection Efficiency}
\label{sec:eff1}

\begin{figure}[t]
\includegraphics[width=0.9\linewidth]{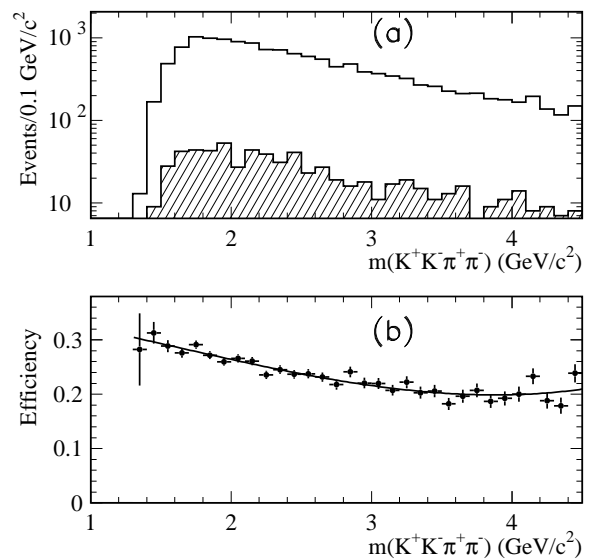}
\vspace{-0.4cm}
\caption{
  (a) The invariant mass distributions for simulated \KKppch events in the phase
  space model, reconstructed in the signal (open) and control 
  (hatched) regions of Fig.~\ref{2k2pi_chi2_all};
  (b) net reconstruction and selection efficiency as a function of
  mass obtained from this simulation 
  (the curve represents a $3^{\rm rd}$ order polynomial fit).
}
\label{mc_acc1}
\end{figure}

The selection procedures applied to the data are also applied to the         
simulated signal samples.
The resulting \KKppch invariant-mass distributions in the signal and
control regions are shown in Fig.~\ref{mc_acc1}(a) for the phase space 
simulation.
The broad, smooth mass distribution is chosen to facilitate the estimation 
of the efficiency as a function of mass,
and this model reproduces the observed distributions of kaon and pion
momenta and polar angles.
We divide the number of reconstructed simulated events in each   
mass interval by the number generated in that interval to obtain the
efficiency shown as the points in Fig.~\ref{mc_acc1}(b).
The 3$^{\rm rd}$ order polynomial fit to the points
is used for further calculations.
We simulate events with the ISR photon confined to the
angular range 20--160$^\circ$ with respect to the electron beam in the
\epem c.m.\ frame, which is about 30\% wider than the EMC acceptance.
This efficiency is for this fiducial region, 
but includes the acceptance for the final-state hadrons, 
the inefficiencies of the detector subsystems, 
and event loss due to additional soft-photon emission.

The simulations including the $\phi(1020)\pipi$ and/or $\Kp\Km f_0(980)$
channels have very different mass and angular distributions in the  
\KKppch rest frame.
However, the angular acceptance is quite uniform for ISR events,
and the efficiencies are consistent with those from the phase space
simulation within 3\%.
To study possible mis-modeling of the acceptance, we repeat the analysis
with the tighter requirements that all charged tracks be
within the DIRC acceptance, $0.45 \!<\! \theta_{\rm ch} \!<\! 2.4$ radians,
and the ISR photon be well away from the edges of the EMC,
$0.35 \!<\! \theta_{\rm ISR} \!<\! 2.4$ radians.
The fraction of selected data events satisfying the tighter
requirements differs from the simulated ratio by 3.7\%.
We conservatively take the sum in quadrature of this variation and the
3\% model variation (5\% total) 
as a systematic uncertainty due to acceptance and model dependence.

\begin{table*}
\caption{Measurements of
  the $\epem \!\to \KKppch$ cross section 
  (errors are statistical only).}
\label{2k2pi_tab}
\begin{ruledtabular}
\begin{tabular}{
  c @{\hspace{-0.07cm}} r@{$\pm$}l @{\hspace{0.35cm}} 
  c @{\hspace{-0.07cm}} r@{$\pm$}l @{\hspace{0.35cm}} 
  c @{\hspace{-0.07cm}} r@{$\pm$}l @{\hspace{0.35cm}}
  c @{\hspace{-0.07cm}} r@{$\pm$}l @{\hspace{0.35cm}} 
  c @{\hspace{-0.07cm}} r@{$\pm$}l }
  \Ecm (GeV) & \multicolumn{2}{c}{$\sigma$ (nb)}
& \Ecm (GeV) & \multicolumn{2}{c}{$\sigma$ (nb)}
& \Ecm (GeV) & \multicolumn{2}{c}{$\sigma$ (nb)}
& \Ecm (GeV) & \multicolumn{2}{c}{$\sigma$ (nb)}
& \Ecm (GeV) & \multicolumn{2}{c}{$\sigma$ (nb)} \\[0.05cm]
\hline
  &  \multicolumn{2}{c}{  } &  &  \multicolumn{2}{c}{  } &  
  &  \multicolumn{2}{c}{  } &  &  \multicolumn{2}{c}{  } &  
  &  \multicolumn{2}{c}{  }   \\[-0.25cm]
1.4125  &  0.00 & 0.02 &  2.1375  &  2.83 & 0.13 &  2.8625  &  0.50 & 0.05 & 
3.5875  &  0.12 & 0.03 &  4.3125  &  0.04 & 0.02 \\
1.4375  &  0.01 & 0.02 &  2.1625  &  2.71 & 0.12 &  2.8875  &  0.51 & 0.05 & 
3.6125  &  0.13 & 0.03 &  4.3375  &  0.04 & 0.02 \\
1.4625  &  0.00 & 0.02 &  2.1875  &  2.46 & 0.12 &  2.9125  &  0.54 & 0.05 & 
3.6375  &  0.12 & 0.03 &  4.3625  &  0.03 & 0.02 \\
1.4875  &  0.04 & 0.02 &  2.2125  &  1.84 & 0.10 &  2.9375  &  0.46 & 0.05 & 
3.6625  &  0.11 & 0.03 &  4.3875  &  0.06 & 0.02 \\
1.5125  &  0.03 & 0.02 &  2.2375  &  1.66 & 0.10 &  2.9625  &  0.45 & 0.05 & 
3.6875  &  0.25 & 0.03 &  4.4125  &  0.01 & 0.02 \\
1.5375  &  0.11 & 0.03 &  2.2625  &  1.59 & 0.09 &  2.9875  &  0.46 & 0.05 & 
3.7125  &  0.07 & 0.03 &  4.4375  &  0.03 & 0.02 \\
1.5625  &  0.15 & 0.04 &  2.2875  &  1.66 & 0.09 &  3.0125  &  0.36 & 0.04 & 
3.7375  &  0.08 & 0.02 &  4.4625  &  0.06 & 0.02 \\
1.5875  &  0.32 & 0.05 &  2.3125  &  1.50 & 0.09 &  3.0375  &  0.39 & 0.04 & 
3.7625  &  0.11 & 0.03 &  4.4875  &  0.03 & 0.02 \\
1.6125  &  0.48 & 0.06 &  2.3375  &  1.65 & 0.09 &  3.0625  &  0.31 & 0.04 & 
3.7875  &  0.11 & 0.03 &  4.5125  &  0.04 & 0.02 \\
1.6375  &  0.85 & 0.08 &  2.3625  &  1.56 & 0.09 &  3.0875  &  2.95 & 0.10 & 
3.8125  &  0.10 & 0.03 &  4.5375  &  0.01 & 0.02 \\
1.6625  &  1.42 & 0.10 &  2.3875  &  1.49 & 0.09 &  3.1125  &  1.51 & 0.08 & 
3.8375  &  0.08 & 0.02 &  4.5625  &  0.02 & 0.02 \\
1.6875  &  1.86 & 0.11 &  2.4125  &  1.46 & 0.09 &  3.1375  &  0.37 & 0.04 & 
3.8625  &  0.12 & 0.03 &  4.5875  &  0.05 & 0.02 \\
1.7125  &  2.36 & 0.13 &  2.4375  &  1.48 & 0.09 &  3.1625  &  0.35 & 0.04 & 
3.8875  &  0.09 & 0.02 &  4.6125  &  0.02 & 0.02 \\
1.7375  &  2.67 & 0.13 &  2.4625  &  1.17 & 0.08 &  3.1875  &  0.28 & 0.04 & 
3.9125  &  0.09 & 0.02 &  4.6375  &  0.01 & 0.02 \\
1.7625  &  3.51 & 0.15 &  2.4875  &  1.16 & 0.08 &  3.2125  &  0.35 & 0.04 & 
3.9375  &  0.08 & 0.02 &  4.6625  &  0.04 & 0.02 \\
1.7875  &  3.98 & 0.16 &  2.5125  &  1.21 & 0.08 &  3.2375  &  0.31 & 0.04 & 
3.9625  &  0.10 & 0.02 &  4.6875  &  0.02 & 0.02 \\
1.8125  &  4.10 & 0.16 &  2.5375  &  0.94 & 0.07 &  3.2625  &  0.30 & 0.04 & 
3.9875  &  0.04 & 0.02 &  4.7125  &  0.03 & 0.02 \\
1.8375  &  4.68 & 0.17 &  2.5625  &  0.95 & 0.07 &  3.2875  &  0.24 & 0.04 & 
4.0125  &  0.06 & 0.02 &  4.7375  &  0.01 & 0.02 \\
1.8625  &  4.49 & 0.17 &  2.5875  &  0.84 & 0.07 &  3.3125  &  0.22 & 0.04 & 
4.0375  &  0.07 & 0.02 &  4.7625  &  0.02 & 0.02 \\
1.8875  &  4.26 & 0.17 &  2.6125  &  0.85 & 0.07 &  3.3375  &  0.25 & 0.04 & 
4.0625  &  0.05 & 0.02 &  4.7875  &  0.01 & 0.02 \\
1.9125  &  4.30 & 0.16 &  2.6375  &  0.90 & 0.07 &  3.3625  &  0.16 & 0.03 & 
4.0875  &  0.06 & 0.02 &  4.8125  &  0.00 & 0.02 \\
1.9375  &  4.20 & 0.16 &  2.6625  &  0.82 & 0.06 &  3.3875  &  0.17 & 0.03 & 
4.1125  &  0.06 & 0.02 &  4.8375  &  0.02 & 0.02 \\
1.9625  &  4.13 & 0.16 &  2.6875  &  0.70 & 0.06 &  3.4125  &  0.18 & 0.03 & 
4.1375  &  0.05 & 0.02 &  4.8625  &  0.00 & 0.02 \\
1.9875  &  3.74 & 0.15 &  2.7125  &  0.86 & 0.06 &  3.4375  &  0.12 & 0.03 & 
4.1625  &  0.06 & 0.02 &  4.8875  &  0.04 & 0.02 \\
2.0125  &  3.45 & 0.15 &  2.7375  &  0.81 & 0.06 &  3.4625  &  0.17 & 0.03 & 
4.1875  &  0.05 & 0.02 &  4.9125  &  0.05 & 0.02 \\
2.0375  &  3.38 & 0.14 &  2.7625  &  0.76 & 0.06 &  3.4875  &  0.17 & 0.03 & 
4.2125  &  0.05 & 0.02 &  4.9375  &  0.02 & 0.02 \\
2.0625  &  3.17 & 0.14 &  2.7875  &  0.73 & 0.06 &  3.5125  &  0.21 & 0.03 & 
4.2375  &  0.08 & 0.02 &  4.9625  &  0.00 & 0.02 \\
2.0875  &  3.23 & 0.14 &  2.8125  &  0.64 & 0.05 &  3.5375  &  0.14 & 0.03 & 
4.2625  &  0.04 & 0.02 &  4.9875  &  0.04 & 0.02 \\
2.1125  &  3.15 & 0.14 &  2.8375  &  0.56 & 0.05 &  3.5625  &  0.16 & 0.03 & 
4.2875  &  0.08 & 0.02 &                         \\
\end{tabular}
\end{ruledtabular}
\end{table*}

\begin{figure}[tbh]
\includegraphics[width=0.9\linewidth]{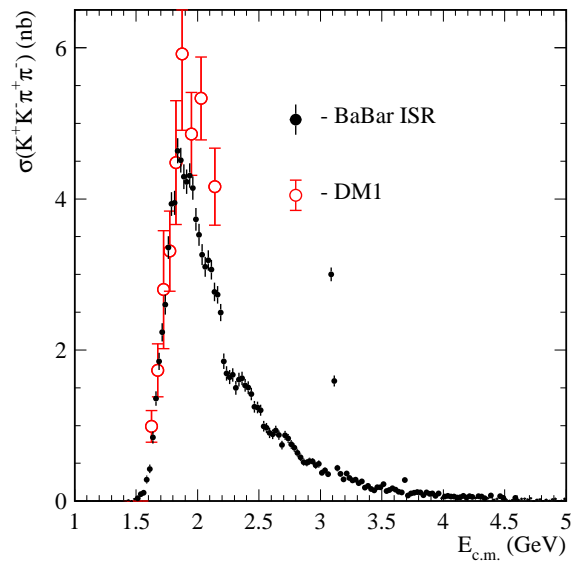}
\vspace{-0.4cm}
\caption{
   The $\epem \!\!\to\! \KKppch$ cross section as a function of the effective 
   \epem c.m.\@ energy measured with ISR data at \babar\ (dots).
   The direct measurements from DM1~\cite{2k2pidm1} are shown as the
   open circles. 
   Only statistical errors are shown.
}
\label{2k2pi_ee_babar}
\end{figure} 

We correct for mis-modeling of the shape of the \chiKKppch distribution
by ($3.0\pm2.0$)\%
and the track finding efficiency following the procedures described in
detail in Ref.~\cite{isr4pi}.
We use a comparison of data and simulated \chifourpi distributions in the 
much larger samples of ISR $\pipi\pipi$ events.
We consider data and simulated events that contain a high-energy photon plus 
exactly three charged tracks and satisfy a set of kinematical criteria, 
including a good $\chi^2$ from a kinematic fit under the hypothesis
that there is exactly one missing track in the event.
We find that the simulated track-finding efficiency is overestimated by
$(0.8\pm0.5)\%$ per track, so we apply a correction of $+(3\pm2)\%$ 
to the signal yield.

We correct the simulated kaon identification efficiency using 
$\epem \!\!\to\! \phi(1020)\gamma \!\to\! \Kp\Km\gamma$ events.
Events with a hard ISR photon and two charged tracks,
one of which is identified as a kaon,
with a $\Kp\Km$ invariant mass near the $\phi$ mass provide a very
clean sample, 
and we compare the fractions of data and simulated events with the
other track also identified as a kaon, as a function of momentum. 
The data-simulation efficiency ratio averages $0.990\pm0.001$ in the 
1--5~\gevc momentum range with variations at the 0.01 level.
We conservatively apply a correction of $+(1.0\pm1.0)$\% per kaon,
or $+(2.0\pm2.0)$\% to the signal yield.

\begin{table}[b]
\caption{
Summary of corrections and systematic uncertainties 
on the $\epem \!\!\to\! \KKppch$  cross section.
The total correction is the linear sum of the components and the
total uncertainty is the sum in quadrature.
  }
\label{error_tab}
\begin{ruledtabular}
\begin{tabular}{l c l} 
     Source             & Correction & Uncertainty        \\
\hline
                        &            &                    \\[-0.2cm]
Rad. Corrections        &  --        & $1\%$              \\
Backgrounds             &  --        & $3\%$, $m_{\Kppch}\! <3~\gevcc$ \\
                        &            & $5\%$, $m_{\Kppch}\! >3~\gevcc$ \\
Model Dependence        &  --        & $5\%$              \\
\chiKKppch Distn.       & $+3\%$     & $2\%$              \\ 
Tracking Efficiency     & $+3\%$     & $2\%$              \\
Kaon ID Efficiency      & $+2\%$     & $2\%$              \\
ISR Luminosity          &  --        & $3\%$              \\[0.1cm]
\hline
                        &            &                    \\[-0.2cm]
Total                   & $+8\%$     & $7\%$, $m_{\Kppch}\! <3~\gevcc$ \\
                        &            & $9\%$, $m_{\Kppch}\! >3~\gevcc$ \\
\end{tabular}
\end{ruledtabular}
\end{table}

\subsection{\boldmath Cross Section for $\epem \!\to \KKppch$}
\label{sec:xs2k2pi}

We calculate the $\epem \!\!\to\! \KKppch$  cross section as a
function of the effective c.m.\ energy from
\begin{equation}
    \sigma_{\Kppch}(\Ecm)
  = \frac{dN_{\Kppch\gamma}(\Ecm)}
         {d{\cal L}(\Ecm) \cdot \epsilon_{\Kppch}(\Ecm)}\ ,
  \label{xseqn}
\end{equation}
where 
$\Ecm \equiv m_{\Kppch}c^2$, 
$m_{\Kppch}$ is the measured invariant mass of the \KKppch system,
$dN_{\Kppch\gamma}$ is the number of selected events after background 
subtraction in the interval $d\Ecm$, 
and
$\epsilon_{\Kppch}(\Ecm)$ is the corrected detection efficiency.
We calculate the differential luminosity, $d{\cal L}(\Ecm)$, 
in each interval $d\Ecm$ from ISR $\mumu\gamma$ events with
the photon in the same  fiducial range used for the
simulation;
the procedure is described in Refs.~\cite{isr4pi, isr6pi}.
From data-simulation comparison we conservatively estimate 
a systematic uncertainty on $d{\cal L}$ of 3\%. 
This $d{\cal L}$ has been corrected for vacuum polarization 
and final-state soft-photon emission;
the former should be excluded when using these data in calculations
of $g_\mu\! -\! 2$.

For the cross section measurement we use the tighter angular criteria
on the charged tracks and the ISR photon,
discussed in Sec.~\ref{sec:eff1}, to exclude possible errors from
incorrect simulation of the EMC and DCH 
edge effects.
We show the cross section as a function of $\Ecm$ in
Fig.~\ref{2k2pi_ee_babar}, with statistical errors only,
and provide a list of our results in Table~\ref{2k2pi_tab}.  
The result is consistent with the direct measurement by DM1~\cite{2k2pidm1},
and with our previous measurement of this channel~\cite{isr4pi}
but has much better statistical precision.
The systematic uncertainties, summarized in Table~\ref{error_tab},
affect the normalization, but have little effect on the energy dependence.

The cross section rises from threshold to a peak value of about 4.7~nb near 
1.85~\gev, then generally decreases with increasing energy.
In addition to narrow peaks at the $J/\psi$ and $\psi(2S)$ masses,
there are several possible wider structures in the 1.8--2.8~\gev
region.
Such structures might be due to thresholds for
intermediate resonant states, such as $\phi f_0(980)$ near 2~\gev.
Gaussian fits to the simulated line shapes give a resolution on the 
measured \KKppch mass that varies between 4.2~\mevcc in the
1.5--2.5~\gevcc region and 5.5~\mevcc in the 2.5--3.5~\gevcc region. 
The resolution function is not purely Gaussian due to soft-photon
radiation, 
but less than 10\% of the signal is outside the 25~\mevcc mass bin.
Since the cross section has no sharp structure other than the $J/\psi$
and $\psi(2S)$ peaks discussed in Sec.~\ref{sec:charmonium} below, 
we apply no correction for resolution.

\subsection{\boldmath Substructure in the \KKppch Final State}
\label{sec:kaons}

\begin{figure}[tbh]
\includegraphics[width=0.9 \linewidth]{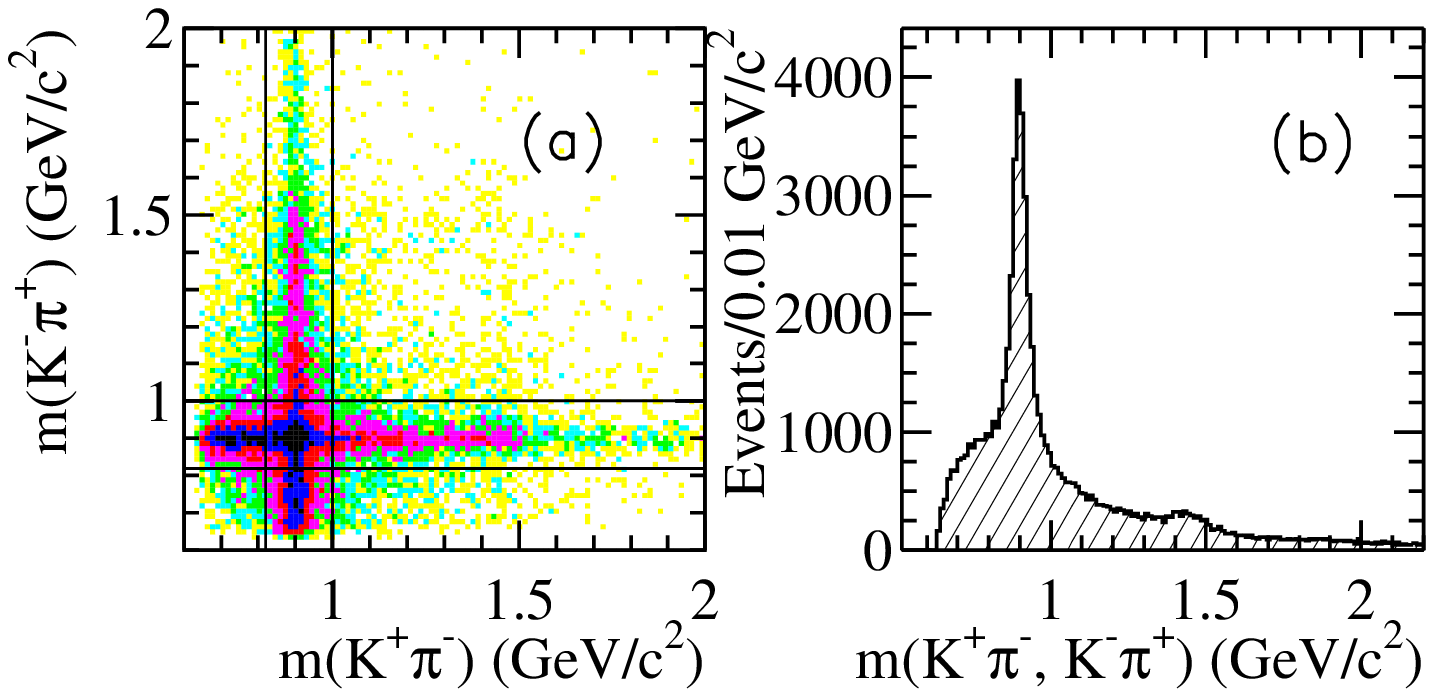}
\includegraphics[width=0.45\linewidth]{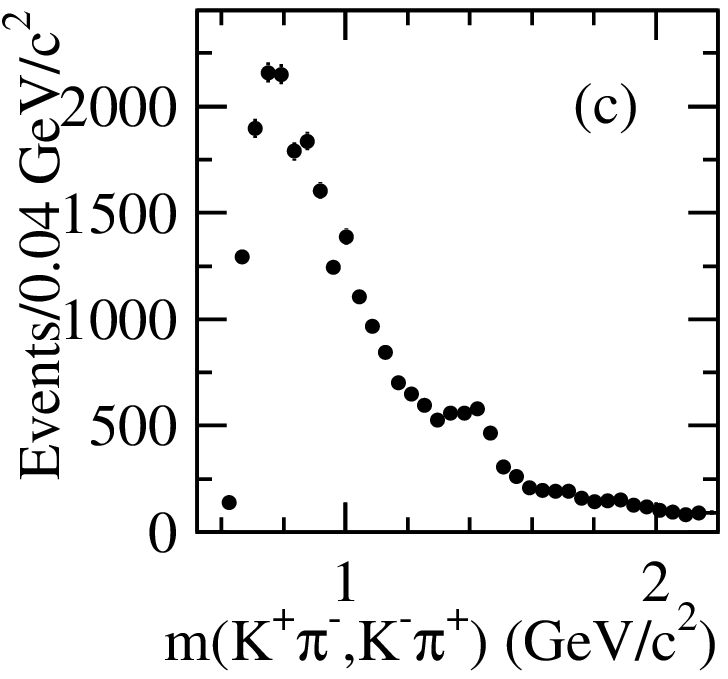}
\vspace{-0.2cm}
\caption{
  (a) Invariant mass of the $\Km\pip$ pair versus that of the $\Kp\pim$ pair;
  (b) sum of projections of (a);
  (c) sum of projections of the $K^{*0}$ bands of (a), with events in
  the overlap region taken only once. The $\phi\pipi$ and $KK\rho$ are vetoed.
  }
\label{kkstar}
\end{figure}

Our previous study~\cite{isr4pi} showed many intermediate resonances 
in the \KKppch final state. 
With the larger data sample used here, they can be seen more clearly
and, in some cases, studied in detail.
Figure~\ref{kkstar}(a) shows a scatter plot of the invariant mass of
the $\Km\pip$ pair versus that of the $\Kp\pim$ pair, and
Fig.~\ref{kkstar}(b) shows the sum of the two projections.
Here we have suppressed the contributions from $\phi\pipi$ and
$\Kp\Km\rho(770)$ by requiring 
$|m(\Kp\Km )\! -\! m(\phi)|\! >\!  10$~\mevcc and
$|m(\pipi)  \! -\! m(\rho)|\! >\! 100$~\mevcc, where $m(\phi)$ and
$m(\rho)$ values are taken from the Particle Data Group (PDG) tables~\cite{PDG}.
Bands and peaks corresponding to the $K^{*0}(892)$ and $K_2^{*0}(1430)$ 
are visible.
In Fig.~\ref{kkstar}(c) we show the sum of projections of the
$K^{*0}(892)$ bands, 
defined by lines in Fig.~\ref{kkstar}(a), 
with events in the overlap region plotted only once.
No $K^{*0}(892)$ signal is seen,
confirming that the $\epem \!\!\to\! K^{*0}(892)\Kbar^{*0}(892)$ 
cross section is small.
We observe associated $K^{*0}(892) \Kbar_2^{*0}(1430)$ production, 
but it is mostly from $J/\psi$ decays (see Sec.~\ref{sec:charmonium}).

\begin{figure}[tbh]
\includegraphics[width=0.9\linewidth]{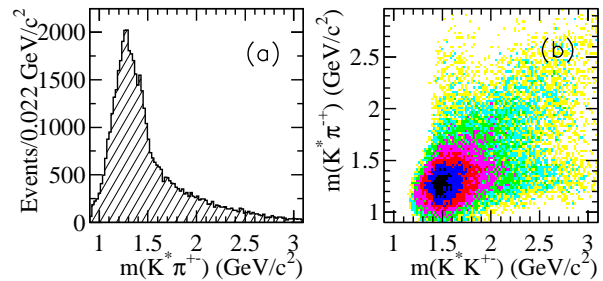}
\vspace{-0.2cm}
\caption{
   (a) The $K^{*0}\pi$ invariant mass distribution;
   (b) the $K^{*0}\pi$ mass versus $K^{*0} K$ mass. 
   }
\label{kstark}
\end{figure}

We combine $K^{*0}/\Kbar^{*0}$ candidates within the lines in 
Fig.~\ref{kkstar}(a) with the remaining pion and kaon to obtain the 
$K^{*0}\pi^{+-}$ invariant mass distribution shown in Fig.~\ref{kstark}(a), 
and the $K^{*0}\pi^{+-}$ vs.\ $K^{*0} K^{-+}$ mass scatter plot in 
Fig.~\ref{kstark}(b).
The bulk of Fig.~\ref{kstark}(b) shows a strong positive correlation,
characteristic of $K^{*0}K\pi$ final states with no higher resonances.
The horizontal band in Fig.~\ref{kstark}(b) corresponds to the peak
region in Fig.~\ref{kstark}(a), and is consistent with contributions
from the $K_1(1270)$ and $K_1(1400)$ resonances.
There is also an indication of a vertical band in Fig.~\ref{kstark}(b),
perhaps corresponding to a $K^{*0} K$ resonance at $\sim$1.5~\gevcc.

\begin{figure}[tbh]
\includegraphics[width=0.9\linewidth]{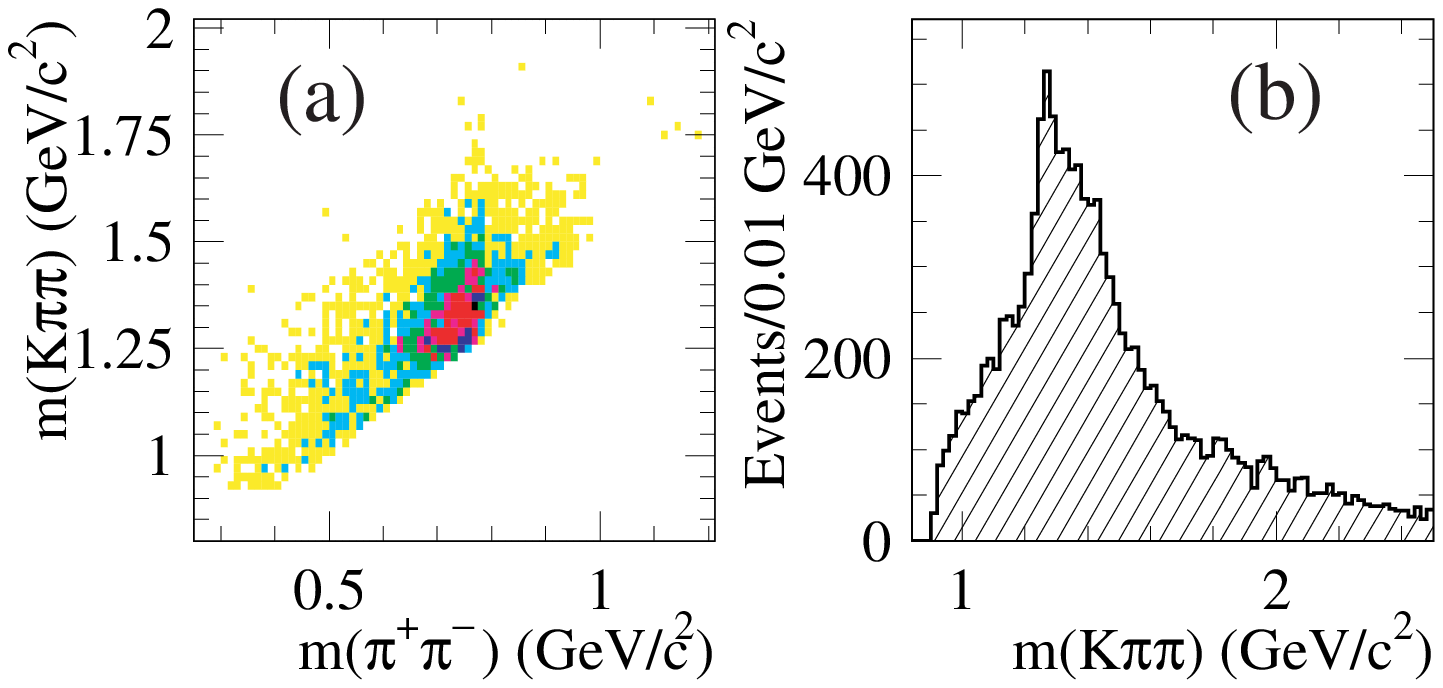}
\includegraphics[width=0.45\linewidth]{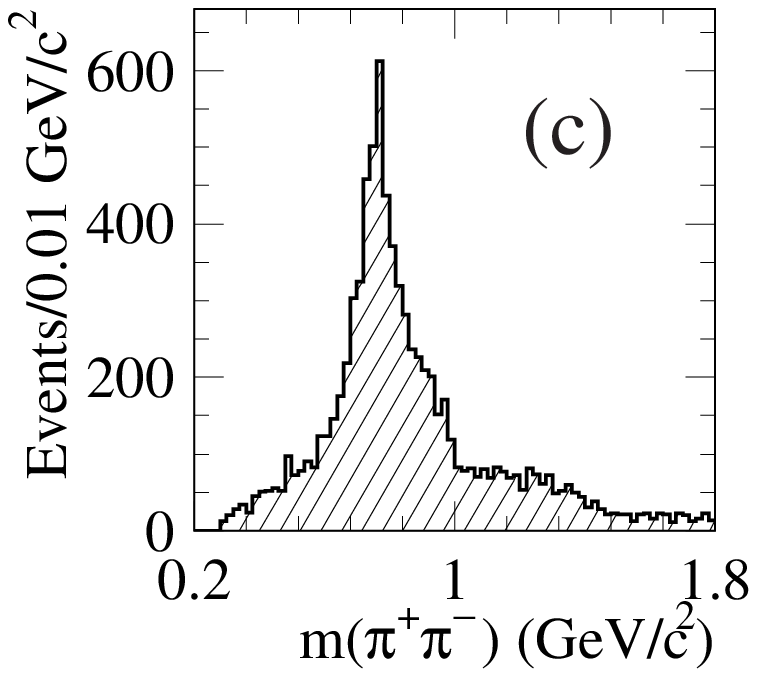}
\vspace{-0.2cm}
\caption{ 
  (a) Invariant mass of the $K^\pm\pipi$ combinations versus that of
  the $\pipi$ pair;
  (b) the $K^\pm\pipi$ and (c) $\pipi$ mass projections of (a).
  }
\label{kpipi}
\end{figure}

We now suppress $K^{*0} K\pi$ by considering only events 
outside the lines in Fig.~\ref{kkstar}(a).
In Fig.~\ref{kpipi} we show the $K^\pm\pipi$ invariant mass (two
entries per event) vs.\ that of the $\pipi$ pair,
along with its two projections.
There is a strong $\rho(770) \!\to\! \pipi$ signal, 
and the $K^\pm\pipi$ mass projection shows further indications of the 
$K_1(1270)$ and $K_1(1400)$ resonances, both of which decay into $K\rho(770)$. 
There are suggestions of additional structure in the \pipi mass distribution, 
including possible $f_0(980)$ shoulder and a
possible enhancement near the $f_2(1270)$, however the current
statistics do not allow us to make definitive statements.

The separation of all these, and any other, intermediate states
involving relatively wide resonances requires a partial wave analysis.
This is beyond the scope of this paper.
Here we present the cross section for the sum of all states including
a $K^{*0}(892)$,
and study intermediate states that include a narrow $\phi$ or $f_0$ resonance.

\subsection{\boldmath  The $\epem \!\!\to\! K^{*0} K \pi$ Cross Section}
\label{kstarxs}

\begin{figure}[t]
\includegraphics[width=0.9\linewidth]{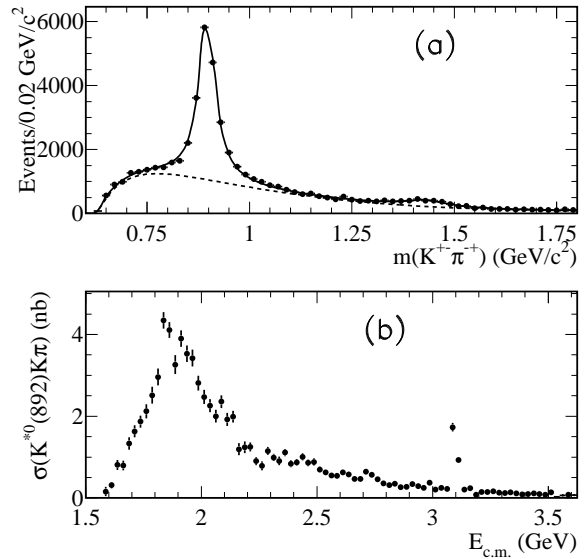}
\vspace{-0.2cm}
\caption{ 
  (a) The $K^\pm\pi^\mp$ mass distribution (two entries per event) for all
  selected \KKppch events:
  the solid line represents a fit including two resonances and a
  polynomial function (see text),
  shown separately as the dashed line;
  (b) the $\epem \!\!\to\! K^{*0} K \pi$  cross section 
  obtained from the $K^{*0}(892)$ signal by a similar fit in
  each 25~\mevcc mass bin.
  }
\label{kstar_sel}
\end{figure}

Signals for the $K^{*0}(892)$ and $K_2^{*0}(1430)$ are clearly visible 
in the $K^\pm\pi^\mp$ mass distributions in Fig.~\ref{kkstar}(b) and,
with a different bin size, in Fig.~\ref{kstar_sel}(a).
We perform a fit to this distribution using P-wave Breit-Wigner (BW) functions 
for the $K^{*0}$ and $K_2^{*0}$ signals and a third-order polynomial
function for the remainder of the distribution taking into account the
$K\pi$ threshold.
The result is shown in Fig.~\ref{kstar_sel}(a).
The fit yields a $K^{*0}$ signal of $19738\pm266$ events with 
$m(K\pi) = 896.2\pm0.3$~\mevcc and $\Gamma(K\pi) = 50.6\pm0.9$~\mev, 
and a $K_2^{*0}$ signal of $1786\pm127$ events with  
$m(K\pi) = 1428.5\pm3.9$~\mevcc and $\Gamma(K\pi) = 113.7\pm9.2$~\mev.
These values are consistent with current world averages~\cite{PDG},
and the fit describes the data well, indicating that contributions
from any other resonances decaying into $K^\pm\pi^\mp$ are small.

\begin{table*}
\caption{Measurements of the $\ep\en\to K^{*0}(892) K \pi$ 
cross section (errors are statistical only).}
\label{kstar_tab}
\begin{ruledtabular}
\begin{tabular}{ c c c c c c c c }
$E_{\rm c.m.}$ (GeV) & $\sigma$ (nb)  
& $E_{\rm c.m.}$ (GeV) & $\sigma$ (nb) 
& $E_{\rm c.m.}$ (GeV) & $\sigma$ (nb) 
& $E_{\rm c.m.}$ (GeV) & $\sigma$ (nb)  
\\
\hline

1.5875& 0.16 $\pm$ 0.11 & 2.0875& 2.36 $\pm$ 0.16 & 2.5875& 0.54 $\pm$ 0.07 
                                                  & 3.0875& 1.73 $\pm$ 0.10 \\
1.6125& 0.31 $\pm$ 0.08 & 2.1125& 1.92 $\pm$ 0.16 & 2.6125& 0.63 $\pm$ 0.06 
                                                  & 3.1125& 0.92 $\pm$ 0.07 \\
1.6375& 0.81 $\pm$ 0.13 & 2.1375& 1.99 $\pm$ 0.14 & 2.6375& 0.57 $\pm$ 0.06 
                                                  & 3.1375& 0.21 $\pm$ 0.04 \\
1.6625& 0.79 $\pm$ 0.12 & 2.1625& 1.19 $\pm$ 0.15 & 2.6625& 0.46 $\pm$ 0.06 
                                                  & 3.1625& 0.24 $\pm$ 0.04 \\
1.6875& 1.33 $\pm$ 0.15 & 2.1875& 1.24 $\pm$ 0.14 & 2.6875& 0.46 $\pm$ 0.06 
                                                  & 3.1875& 0.08 $\pm$ 0.03 \\
1.7125& 1.63 $\pm$ 0.15 & 2.2125& 1.25 $\pm$ 0.11 & 2.7125& 0.64 $\pm$ 0.06 
                                                  & 3.2125& 0.15 $\pm$ 0.03 \\
1.7375& 1.87 $\pm$ 0.14 & 2.2375& 0.90 $\pm$ 0.10 & 2.7375& 0.56 $\pm$ 0.06 
                                                  & 3.2375& 0.14 $\pm$ 0.04 \\
1.7625& 2.12 $\pm$ 0.17 & 2.2625& 0.79 $\pm$ 0.11 & 2.7625& 0.46 $\pm$ 0.06 
                                                  & 3.2625& 0.16 $\pm$ 0.03 \\
1.7875& 2.51 $\pm$ 0.20 & 2.2875& 1.15 $\pm$ 0.10 & 2.7875& 0.36 $\pm$ 0.06 
                                                  & 3.2875& 0.13 $\pm$ 0.03 \\
1.8125& 2.96 $\pm$ 0.21 & 2.3125& 0.99 $\pm$ 0.09 & 2.8125& 0.31 $\pm$ 0.05 
                                                  & 3.3125& 0.12 $\pm$ 0.03 \\
1.8375& 4.35 $\pm$ 0.20 & 2.3375& 0.91 $\pm$ 0.11 & 2.8375& 0.35 $\pm$ 0.05 
                                                  & 3.3375& 0.14 $\pm$ 0.03 \\
1.8625& 4.11 $\pm$ 0.20 & 2.3625& 1.11 $\pm$ 0.09 & 2.8625& 0.27 $\pm$ 0.04 
                                                  & 3.3625& 0.12 $\pm$ 0.06 \\
1.8875& 3.26 $\pm$ 0.23 & 2.3875& 0.83 $\pm$ 0.09 & 2.8875& 0.27 $\pm$ 0.05 
                                                  & 3.3875& 0.09 $\pm$ 0.03 \\
1.9125& 3.90 $\pm$ 0.20 & 2.4125& 0.87 $\pm$ 0.09 & 2.9125& 0.34 $\pm$ 0.05 
                                                  & 3.4125& 0.10 $\pm$ 0.03 \\
1.9375& 3.53 $\pm$ 0.20 & 2.4375& 1.00 $\pm$ 0.09 & 2.9375& 0.29 $\pm$ 0.04 
                                                  & 3.4375& 0.11 $\pm$ 0.03 \\
1.9625& 3.42 $\pm$ 0.21 & 2.4625& 0.86 $\pm$ 0.08 & 2.9625& 0.25 $\pm$ 0.04 
                                                  & 3.4625& 0.10 $\pm$ 0.05 \\
1.9875& 2.81 $\pm$ 0.18 & 2.4875& 0.88 $\pm$ 0.09 & 2.9875& 0.38 $\pm$ 0.05 
                                                  & 3.4875& 0.08 $\pm$ 0.03 \\
2.0125& 2.47 $\pm$ 0.17 & 2.5125& 0.69 $\pm$ 0.07 & 3.0125& 0.21 $\pm$ 0.04 \\ 
2.0375& 2.26 $\pm$ 0.16 & 2.5375& 0.62 $\pm$ 0.07 & 3.0375& 0.24 $\pm$ 0.04 \\
2.0625& 2.00 $\pm$ 0.16 & 2.5625& 0.55 $\pm$ 0.07 & 3.0625& 0.22 $\pm$ 0.04 \\

\end{tabular}
\end{ruledtabular}
\end{table*}

We perform a similar fit to the data in bins of the \KKppch invariant
mass, with the resonance masses and widths fixed to the values
obtained by the overall fit.
Since there is at most one $K^{*0}$ per event, we convert the
resulting $K^{*0}$ yield in each bin into an ``inclusive'' 
$\epem \!\!\to\! K^{*0} K \pi$ cross section,
following the procedure described in Sec.~\ref{sec:xs2k2pi}. 
This cross section is shown in Fig.~\ref{kstar_sel}(b) and listed in
Table~\ref{kstar_tab} for the effective c.m. energies from threshold up to 3.5~\gev.
At higher energies the signals are small and contain an unknown, but
possibly large, contribution from $\epem \!\!\to\! \qqbar$ events.
There is a rapid rise from threshold to a peak value of about 4~nb at 
1.84~\gev, followed by a very rapid decrease with increasing energy.
There are suggestions of narrow structure in the peak region, 
but the only statistically significant structure is the $J/\psi$ peak, 
which is discussed below.

The $\epem \!\!\to\! K^{*0} K \pi$ contribution is a large fraction of
the total \KKppch cross 
section at all energies above its threshold,
and dominates in the 1.8--2.0~\gev region.
We are unable to extract a meaningful measurement of the $K_2^{*0}K\pi$ 
cross section from this data sample because it is more than ten times smaller.
The $\Kp\Km\rho^0(770)$ intermediate state makes up the majority of the
remainder of the cross section and it can be estimated as a difference
of the values in Table~\ref{2k2pi_tab} and Table~\ref{kstar_tab} for
the \KKppch and $K^{*0} K \pi $ final states.

\subsection{\boldmath The $\phi(1020)\pipi$ Intermediate State}
\label{sec:phipipi}
 
\begin{figure}[tbh]
\includegraphics[width=0.9\linewidth]{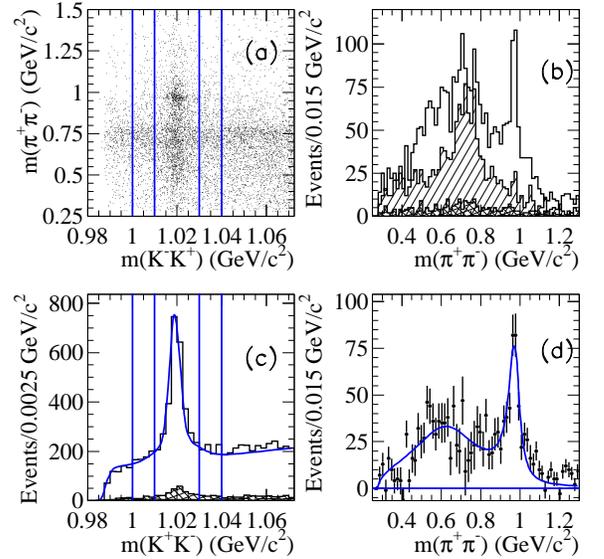}
\vspace{-0.4cm}
\caption{
  (a) The $\pipi$ vs.\ the $\Kp\Km$ invariant
  masses for all selected \KKppch events;
  (b) the $\pipi$ invariant mass projections for events in the $\phi$ peak
  (open histogram), sidebands (hatched) and background control region
  (cross-hatched);
  (c) the $\Kp\Km$ mass projections for all events (open) and control
  region (cross-hatched);
  (d) the difference between the open and the sum of the other histograms in (b). 
  }
\label{phif0_sel}
\end{figure}

Intermediate states containing relatively narrow resonances can be
studied more easily.
Figure~\ref{phif0_sel}(a) shows a scatter plot of the 
invariant mass of the $\pipi$ pair versus that of the $\Kp\Km$ pair.
Horizontal and vertical bands corresponding to the $\rho^0(770)$ and
$\phi$, respectively, are visible,
and there is a concentration of entries on the $\phi$ band corresponding
to the correlated production of $\phi$ and $f_{0}(980)$. 
The $\phi$ signal is also visible in the $\Kp\Km$ mass projection,
Fig.~\ref{phif0_sel}(c).
The large contribution from the $\rho(770)$, 
coming from the $K_{1}$ decay, 
is nearly uniform in the $\Kp\Km$ mass,
and the cross-hatched histogram shows the non-\KKppch background
estimated from the control region in \chiKKppch.
The cross-hatched histogram 
also shows a $\phi$ peak, but this is a small fraction of the events.
Subtracting this background and fitting the remaining data gives 
1706$\pm$56 events produced via the $\phi\pipi$ intermediate state.

To study the $\phi\pipi$ channel, we select candidate events with a
$\Kp\Km$ invariant mass    
within 10~\mevcc of the $\phi$ mass, 
indicated by the inner vertical lines in Figs.~\ref{phif0_sel}(a,c),
and estimate the non-$\phi$ contribution from the mass sidebands between
the inner and outer vertical lines.
In Fig.~\ref{phif0_sel}(b) we show the \pipi invariant mass distributions
for $\phi$ candidate events, sideband events and \chisq control region events
as the open, hatched and cross-hatched histograms, respectively,
and in Fig.~\ref{phif0_sel}(d) we show the numbers of entries from the
candidate events minus those from the sideband and control region. 
There is a clear $f_0$ peak over a broad mass distribution, with no
indication of associated $\rho^0$ production.

A coherent sum of two Breit-Wigner functions is sufficient to describe the
invariant mass distribution of the \pipi pair recoiling against a
$\phi$ in Fig.~\ref{phif0_sel}(d).
We fit with the function:
\begin{eqnarray}
  F(m) & = & \sqrt{1-4m_{\pi}^2/m^2} \cdot |A_1(m) +e^{i\psi}A_2(m)|^2 ~,
\label{f0fit}\\
A_i(m) & = & m_i \Gamma_i \sqrt{N_i}/(m_i^2 - m^2 +im_i\Gamma_i)~,\nonumber
\end{eqnarray}
where $m$ is the \pipi invariant mass, 
$m_i$ and $\Gamma_i$ are the parameters of the $i^{th}$ resonance, 
$\psi$ is their relative phase 
and $N_i$ are normalization parameters, corresponding to the number of
events under each BW.
One BW corresponds to the $f_0(980)$, but a wide range of values of
the other parameters can describe the data.
Fixing the relative phase to $\psi = \pi$ 
and the parameters of the first BW to
$m_1=0.6$~\gevcc and $\Gamma_1 = 0.45$~\gev 
(which could be interpreted as the $f_0(600)$~\cite{PDG}),
we obtain the fit shown in Fig.~\ref{phif0_sel}(d).
It describes the data well and gives an $f_{0}(980)$ signal of
$262 \pm 30$ events,
with $m_2 = 0.973 \pm 0.003~\gevcc$ and $\Gamma_2 = 0.065 \pm 0.013~\gev$,
consistent with the PDG values~\cite{PDG}.
There is a suggestion of an $f_2(1270)$ peak in the data, but it is
much smaller than the $f_0$ peak and we do not consider it further.

We obtain the number of $\epem \!\!\to\! \phi\pipi$ events in bins of
$\phi\pipi$ invariant mass by fitting the $K^+ K^-$ invariant mass
projection in that bin after subtracting non-\KKppch background.
Each projection is a subset of Fig.~\ref{phif0_sel}(c), 
where the curve represent a fit to the full sample.
In each mass bin, all parameters are fixed to the values obtained from
the overall fit except the numbers of events in the $\phi$ peak and the
non-$\phi$ component.

\begin{figure}[tbh]
\includegraphics[width=0.9\linewidth]{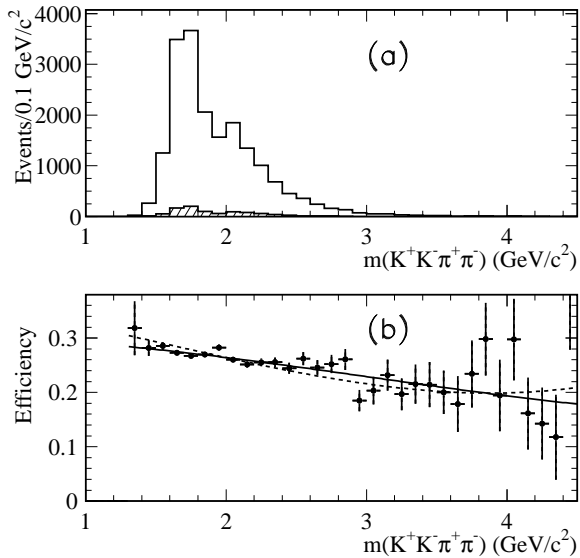}
\vspace{-0.4cm}
\caption{
  (a) The \KKppch invariant mass distributions from the $\phi\pipi$  
  simulation described in the text,
  reconstructed in the signal (open) and control (hatched) regions;
  (b) net reconstruction and selection efficiency as a function of mass:
  the solid line represents a cubic fit, and the dashed line the
  corresponding fit for the space phase model shown in Fig.~\ref{mc_acc1}.
}
\label{mc_acc2}
\end{figure}

The efficiency may depend on the details of the production mechanism.
Using the two-pion mass distribution in Fig.~\ref{phif0_sel}(d) 
as input, 
we simulate the \pipi system as an S-wave comprising two scalar
resonances,
with parameters set to the values given above.
To describe the $\phi\pipi$ mass distribution we use a simple model
with one resonance, the $\phi(1680)$, of mass 1.68~\gevcc and
width 0.2~\gev, decaying to $\phi f_0$. 
The simulated reconstructed spectrum is shown in Fig.~\ref{mc_acc2}(a).
There is a sharp increase at about 2~\gevcc due to the 
$\phi f_0(980)$ threshold.
All other structure is determined by phase space and a $m^{-2}$
falloff with increasing mass.

\begin{figure}[tbh]
\includegraphics[width=0.9\linewidth]{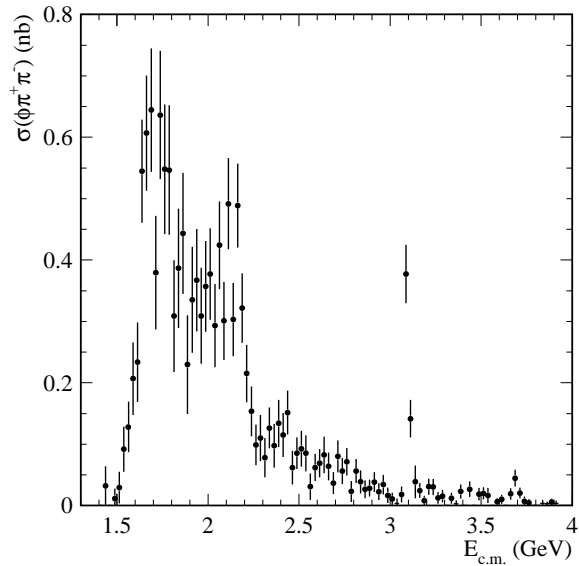}
\vspace{-0.4cm}
\caption{
  The $\epem \!\!\to\! \phi\pipi$  cross section 
as a  function of the effective \epem c.m.\ energy.
  }
\label{phipipixs}
\end{figure}

Dividing the number of reconstructed events in each bin by the number 
of generated ones, 
we obtain the efficiency as a function of $\phi\pipi$ mass shown in
Fig.~\ref{mc_acc2}(b). 
The solid line represents a fit to a third order polynomial,
and the dashed line the corresponding fit to the phase space model
from Fig.~\ref{mc_acc1}. 
The model dependence is weak, giving confidence in the efficiency calculation.
We calculate the $\epem \!\!\to\! \phi\pipi$ cross section as described
in Sec.~\ref{sec:xs2k2pi} but using the efficiency from the fit to
Fig.~\ref{mc_acc2}(b) and dividing by the $\phi \!\to\! K^+ K^-$
branching fraction of 0.491~\cite{PDG}.
We show our results as a function of energy in Fig.~\ref{phipipixs} and list them
in Table~\ref{phi2pi_tab}.
The cross section has a peak value of about 0.6~nb at about 1.7~\gev, 
then decreases with increasing energy until $\phi(1020) f_0(980)$ 
threshold, around 2.0~\gev.
From this point it rises, falls sharply at about 2.2~\gev,
and then decreases slowly.
Except in the charmonium region, the results
at energies above 2.9~\gev are not meaningful due to 
small signals and potentially large backgrounds, and are omitted from
Table~\ref{phi2pi_tab}.
Figure~\ref{phipipixs}
displays the cross-section up to 4.5~\gev to show the signals from
the $J/\psi$ and $\psi(2S)$ decays.
They are discussed in Sec.~\ref{sec:charmonium}.
There are no previous measurements of this cross section,
and our results are consistent 
with the upper limits given in Ref.~\cite{2k2pidm1}.

\begin{table*}
\caption{Measurements of the $\ep\en\to\phi(1020)\pipi$ 
cross section (errors are statistical only).}
\label{phi2pi_tab}
\begin{ruledtabular}
\begin{tabular}{ c c c c c c c c }
$E_{\rm c.m.}$ (GeV) & $\sigma$ (nb)  
& $E_{\rm c.m.}$ (GeV) & $\sigma$ (nb) 
& $E_{\rm c.m.}$ (GeV) & $\sigma$ (nb) 
& $E_{\rm c.m.}$ (GeV) & $\sigma$ (nb)  
\\
\hline
 1.4875 &  0.01 $\pm$  0.02 & 1.8375 &  0.39 $\pm$  0.10 & 2.1875 &  0.32 $\pm$  0.06 & 2.5375 &  0.09 $\pm$  0.03 \\
 1.5125 &  0.03 $\pm$  0.03 & 1.8625 &  0.44 $\pm$  0.10 & 2.2125 &  0.22 $\pm$  0.05 & 2.5625 &  0.03 $\pm$  0.02 \\
 1.5375 &  0.09 $\pm$  0.04 & 1.8875 &  0.23 $\pm$  0.08 & 2.2375 &  0.15 $\pm$  0.04 & 2.5875 &  0.06 $\pm$  0.02 \\
 1.5625 &  0.13 $\pm$  0.04 & 1.9125 &  0.34 $\pm$  0.09 & 2.2625 &  0.10 $\pm$  0.03 & 2.6125 &  0.07 $\pm$  0.02 \\
 1.5875 &  0.21 $\pm$  0.06 & 1.9375 &  0.37 $\pm$  0.08 & 2.2875 &  0.11 $\pm$  0.04 & 2.6375 &  0.08 $\pm$  0.03 \\
 1.6125 &  0.23 $\pm$  0.06 & 1.9625 &  0.31 $\pm$  0.08 & 2.3125 &  0.08 $\pm$  0.03 & 2.6625 &  0.06 $\pm$  0.02 \\
 1.6375 &  0.54 $\pm$  0.08 & 1.9875 &  0.36 $\pm$  0.07 & 2.3375 &  0.13 $\pm$  0.03 & 2.6875 &  0.04 $\pm$  0.02 \\
 1.6625 &  0.61 $\pm$  0.09 & 2.0125 &  0.38 $\pm$  0.07 & 2.3625 &  0.10 $\pm$  0.04 & 2.7125 &  0.08 $\pm$  0.03 \\
 1.6875 &  0.64 $\pm$  0.10 & 2.0375 &  0.29 $\pm$  0.07 & 2.3875 &  0.13 $\pm$  0.04 & 2.7375 &  0.06 $\pm$  0.02 \\
 1.7125 &  0.38 $\pm$  0.09 & 2.0625 &  0.42 $\pm$  0.07 & 2.4125 &  0.12 $\pm$  0.04 & 2.7625 &  0.07 $\pm$  0.02 \\
 1.7375 &  0.64 $\pm$  0.10 & 2.0875 &  0.30 $\pm$  0.06 & 2.4375 &  0.15 $\pm$  0.04 & 2.7875 &  0.02 $\pm$  0.02 \\
 1.7625 &  0.55 $\pm$  0.11 & 2.1125 &  0.49 $\pm$  0.07 & 2.4625 &  0.06 $\pm$  0.03 & 2.8125 &  0.06 $\pm$  0.02 \\
 1.7875 &  0.55 $\pm$  0.11 & 2.1375 &  0.30 $\pm$  0.06 & 2.4875 &  0.09 $\pm$  0.03 & 2.8375 &  0.04 $\pm$  0.02 \\
 1.8125 &  0.31 $\pm$  0.09 & 2.1625 &  0.49 $\pm$  0.07 & 2.5125 &  0.09 $\pm$  0.03 & 2.8625 &  0.03 $\pm$  0.01 \\
\end{tabular}
\end{ruledtabular}
\end{table*}

\begin{figure*}[tbh]
\includegraphics[width=0.32\linewidth]{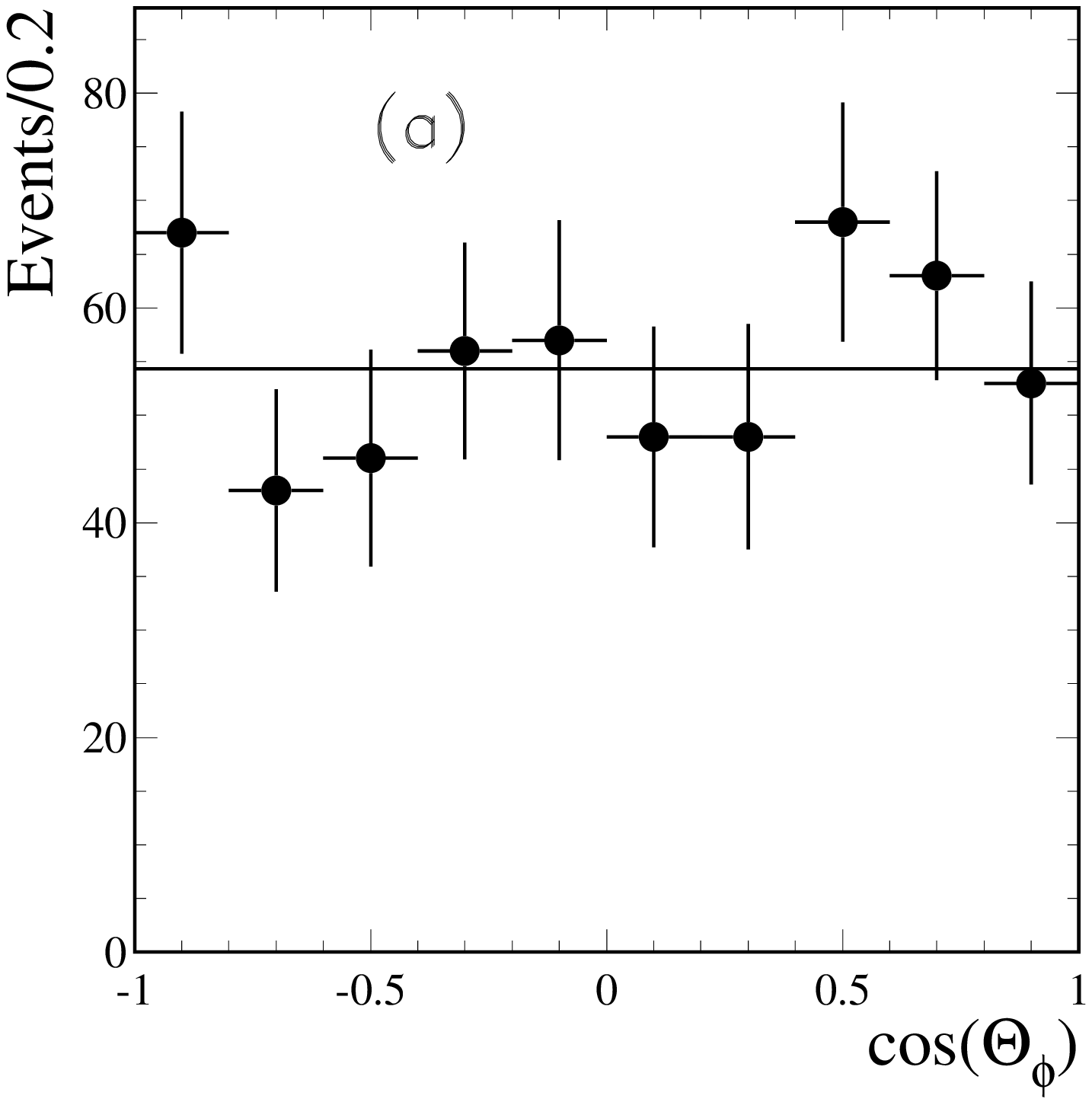}
\includegraphics[width=0.32\linewidth]{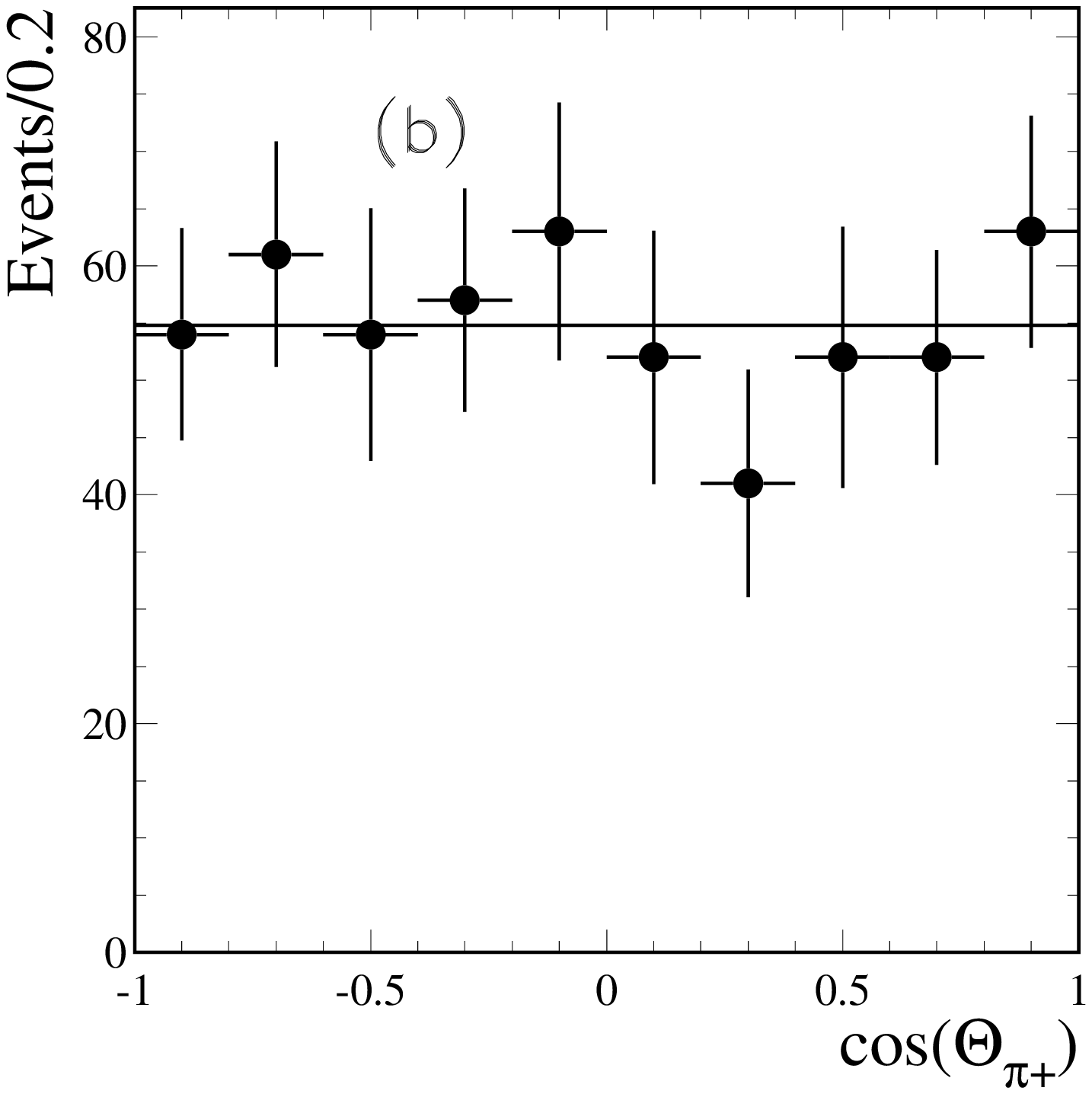}
\includegraphics[width=0.32\linewidth]{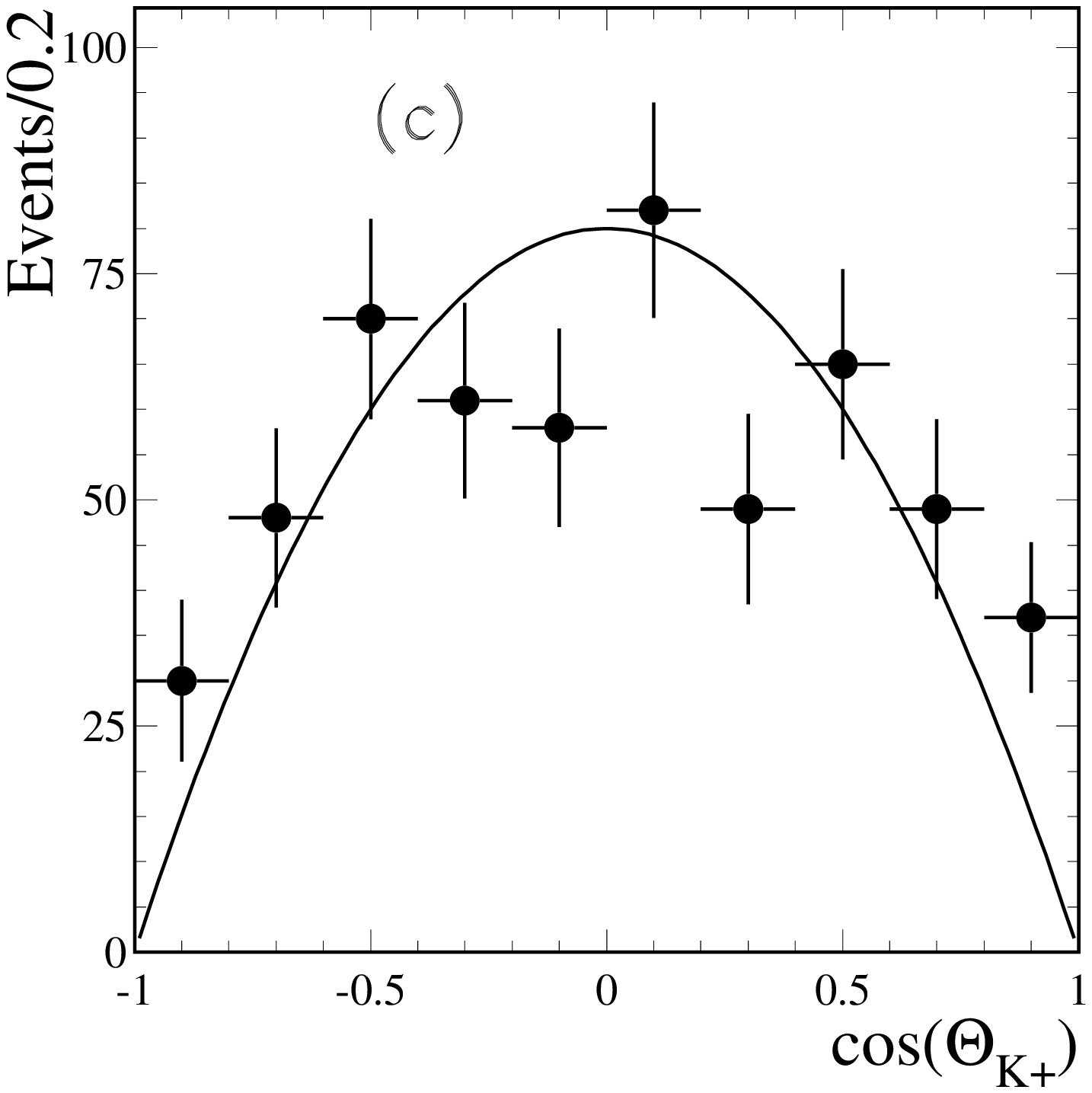}
\vspace{-0.2cm}
\caption{
  Distributions of the cosines of
  (a) the $\phi$ production angle,
  (b) the pion helicity angle, and
  (c) the kaon helicity angle (see text) for $\epem \!\!\to\! \phi\pipi$
  events:
  the lines represent the distributions expected if the \pipi system
  recoiling against a vector $\phi$ meson is produced in an S-wave, 
  normalized to the number of events in the data.
  }
\label{phi_angle}
\end{figure*}
\begin{figure}[tbh]
\includegraphics[width=0.9\linewidth]{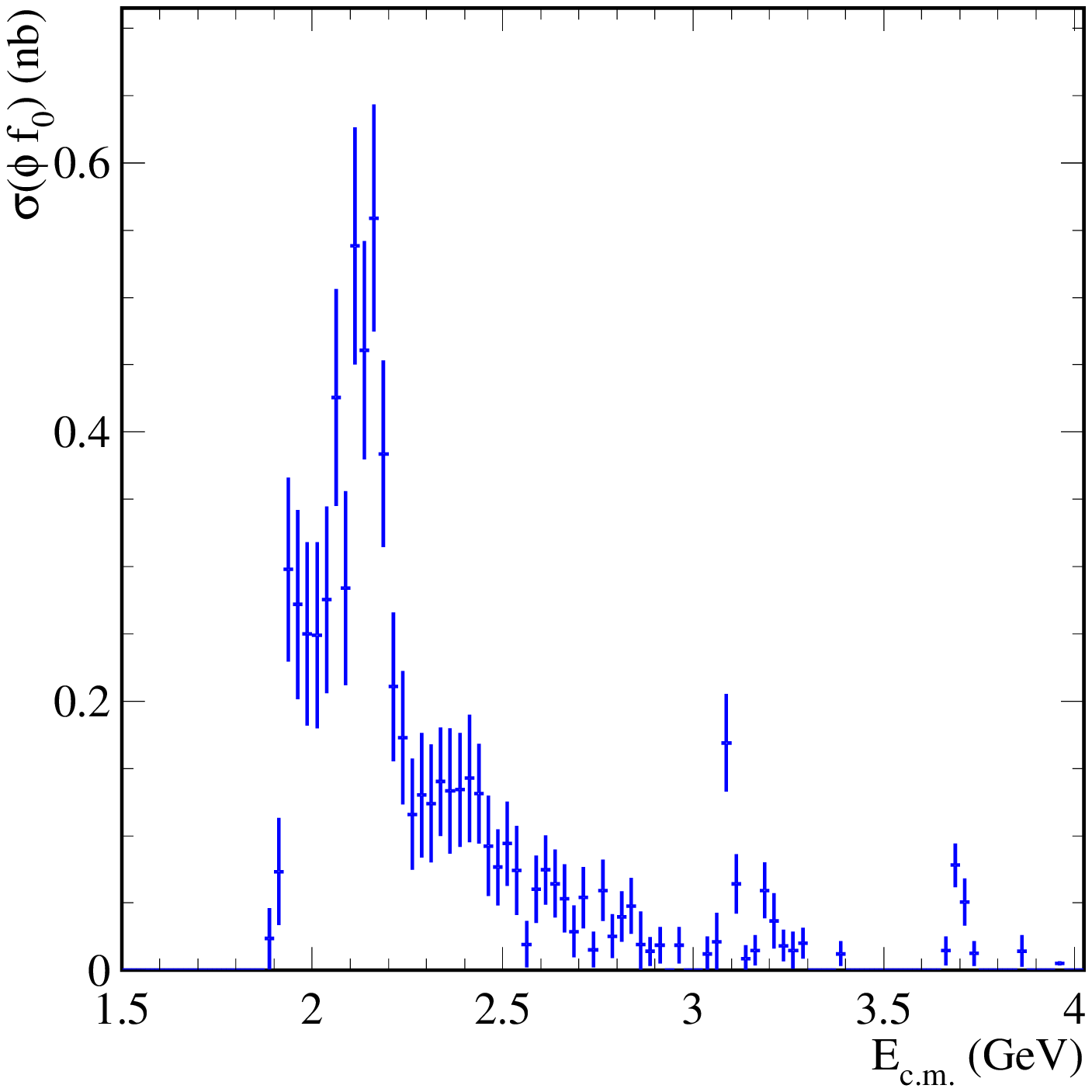}
\vspace{-0.4cm}
\caption{
The $\epem \!\!\to\! \phi(1020) f_{0}(980)$  cross section 
  as a function of the effective \epem c.m.\ energy obtained from the
\KKppch final state.
  }
\label{phif0xs}
\end{figure}

We perform a study of the angular distributions in the $\phi(1020)\pipi$ 
final state by considering all \KKppch candidate events with mass below 3~\gevcc,
binning them in terms of the cosine of the angles defined below,
and fitting the background-subtracted $\Kp\Km$ mass projections.
The efficiency is nearly uniform in these angles,
so we study the number of events in each bin.
We define the $\phi$ production angle, $\Theta_\phi$ as the angle
between the $\phi$ momentum and the $e^-$ beam direction in the rest  
frame of the $\phi\pipi$ system. 
The distribution of $\cos\Theta_\phi$, shown in Fig.~\ref{phi_angle}(a),
is consistent with the uniform distribution expected if S-wave
two-body channels $\phi X$, $X \!\to\! \pipi$ dominate the $\phi\pipi$ system.
We define the pion and kaon helicity angles, $\Theta_{\pip}$ and
$\Theta_{\Kp}$ as those between the \pip and the \pipi-system momenta in
the \pipi rest frame and between the \Kp and ISR photon momenta in the
$\phi$ rest frame, respectively.
The distributions of $\cos\Theta_{\pip}$ and $\cos\Theta_{\Kp}$, 
shown in Figs.~\ref{phi_angle}(b) and~\ref{phi_angle}(c), respectively,
are consistent with those expected from scalar and vector meson decays.

\subsection{\boldmath The $\phi(1020) f_{0}(980)$ Intermediate State}
\label{sec:phif01}

The narrow $f_0(980)$ peak seen in Fig.~\ref{phif0_sel}(d) allows the
selection of a fairly clean sample of $\phi f_0$ events.
We repeat the analysis just described with the additional requirement
that the \pipi invariant mass be in the range 0.85--1.10~\gevcc.
The fit to the full sample yields about 700 events;
all of these contain a true $\phi$, but about 10\% are from 
$\epem \!\to \phi\pipi$ events where the pion pair is not produced through
the $f_0(980)$.

We convert the numbers of fitted events in each mass bin 
into a measurement of the $\epem \!\!\to\! \phi(1020) f_{0}(980)$
cross section as described above and dividing by the 
$f_0 \!\!\to\! \pipi$ branching fraction of two-thirds.
The cross section is shown in Fig.~\ref{phif0xs} as a function of
the effective c.m. energy
and is listed in Table~\ref{phif0_tab}.
Its behavior near threshold does not appear to be smooth, 
but is more consistent with a steep rise to a value of about 0.3~nb
at 1.95~\gev followed by a slow decrease that is interrupted
by a structure around 2.175~\gev.
Possible interpretations of this structure are discussed 
in Sec.~\ref{phif0bump}.
Again, the values are not meaningful for the effective c.m.\ above about 2.9~\gevcc, 
except for the $J/\psi$ and $\psi(2S)$ signals,
discussed in Sec.~\ref{sec:charmonium}.

\begin{table*}
\caption{Summary of the $\ep\en\to\phi(1020) f_{0}(980)$ 
cross section measurement. Errors are statistical only.}
\label{phif0_tab}
\begin{ruledtabular}
\begin{tabular}{ c c c c c c c c }
$E_{\rm c.m.}$ (GeV) & $\sigma$ (nb)  
& $E_{\rm c.m.}$ (GeV) & $\sigma$ (nb) 
& $E_{\rm c.m.}$ (GeV) & $\sigma$ (nb) 
& $E_{\rm c.m.}$ (GeV) & $\sigma$ (nb)  
\\
\hline

 1.8875 &  0.02 $\pm$  0.02 & 2.1625 &  0.56 $\pm$  0.08 & 2.4375 &  0.13 $\pm$  0.04 & 2.7125 &  0.05 $\pm$  0.02 \\
 1.9125 &  0.07 $\pm$  0.04 & 2.1875 &  0.38 $\pm$  0.07 & 2.4625 &  0.09 $\pm$  0.04 & 2.7375 &  0.02 $\pm$  0.01 \\
 1.9375 &  0.30 $\pm$  0.07 & 2.2125 &  0.21 $\pm$  0.06 & 2.4875 &  0.08 $\pm$  0.03 & 2.7625 &  0.06 $\pm$  0.02 \\
 1.9625 &  0.27 $\pm$  0.07 & 2.2375 &  0.17 $\pm$  0.05 & 2.5125 &  0.09 $\pm$  0.03 & 2.7875 &  0.03 $\pm$  0.02 \\
 1.9875 &  0.25 $\pm$  0.07 & 2.2625 &  0.12 $\pm$  0.04 & 2.5375 &  0.07 $\pm$  0.03 & 2.8125 &  0.04 $\pm$  0.02 \\
 2.0125 &  0.25 $\pm$  0.07 & 2.2875 &  0.13 $\pm$  0.05 & 2.5625 &  0.02 $\pm$  0.02 & 2.8375 &  0.05 $\pm$  0.02 \\
 2.0375 &  0.28 $\pm$  0.07 & 2.3125 &  0.12 $\pm$  0.04 & 2.5875 &  0.06 $\pm$  0.03 & 2.8625 &  0.02 $\pm$  0.02 \\
 2.0625 &  0.43 $\pm$  0.08 & 2.3375 &  0.14 $\pm$  0.04 & 2.6125 &  0.07 $\pm$  0.03 & 2.8875 &  0.01 $\pm$  0.01 \\
 2.0875 &  0.28 $\pm$  0.07 & 2.3625 &  0.13 $\pm$  0.05 & 2.6375 &  0.06 $\pm$  0.03 & 2.9125 &  0.02 $\pm$  0.01 \\
 2.1125 &  0.54 $\pm$  0.09 & 2.3875 &  0.13 $\pm$  0.04 & 2.6625 &  0.05 $\pm$  0.03 & 2.9375 &  0.00 $\pm$  0.00 \\
 2.1375 &  0.46 $\pm$  0.08 & 2.4125 &  0.14 $\pm$  0.05 & 2.6875 &  0.03 $\pm$  0.02 & 2.9625 &  0.02 $\pm$  0.01 \\

\end{tabular}
\end{ruledtabular}
\end{table*}

\section{The {\boldmath $K^+ K^-\ppz$} Final State}
\subsection{Final Selection and Backgrounds}

The \KKppnt sample contains background from the ISR processes
$\epem \!\!\to\! \Kp\Km\piz\gamma$ and $\Kp\Km\eta\gamma$,
in which two soft photon candidates from machine- or detector-related
background combine with the relatively energetic photons from
the \piz or $\eta$ to form two fake \piz candidates.
We reduce this background using the helicity angle between each 
reconstructed \piz direction and the direction of its higher-energy photon
daughter calculated in its rest frame.
If the cosines of both helicity angles are higher than 0.85,
we remove the event.

\begin{figure}[tbh]
\includegraphics[width=0.9\linewidth]{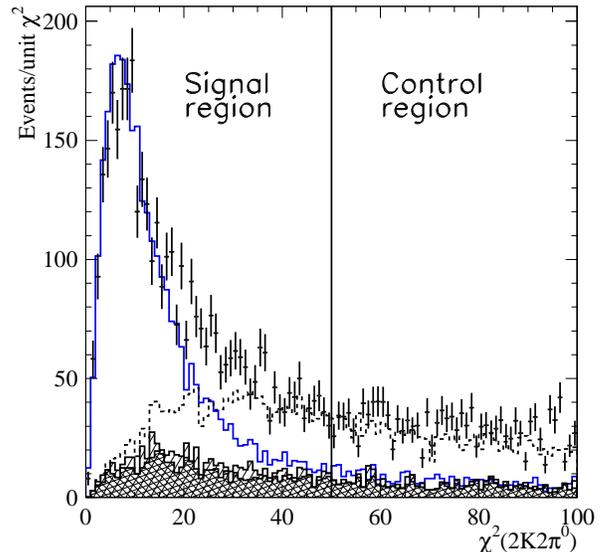}
\vspace{-0.4cm}
\caption{
  Distribution of \chisq from the six-constraint fit for \KKppnt candidates
  in the data (points).
  The open histogram is the distribution for simulated signal events, 
  normalized as described in the text.
  The cross-hatched, hatched and dashed histograms represent, cumulatively, 
  the backgrounds from non-ISR events, ISR $\pipi\piz\piz$ events,
  and ISR $\Kp\Km\piz$, $\Kp\Km\eta$ and $\KpKm\ppz\piz$ events.
}
\label{2k2pi0_chi2_all}
\end{figure}

Figure~\ref{2k2pi0_chi2_all} shows the distribution of \chiKKppnt for
the remaining candidates together with the
simulated \KKppnt events.
Again, the distributions are broader than those for a typical 6C \chisq
due to higher order ISR, 
and we normalize the histogram to the data in the region 
$\chiKKppnt\! <\! 10$.
The cross-hatched histogram in Fig.~\ref{2k2pi0_chi2_all} represents 
background from $\epem \!\!\to\! \qqbar$ events, 
evaluated in the same way as for the \KKppch final state.
The hatched histogram represents the sum of this background and
that from ISR $\pipi\ppz$ events with both charged pions misidentified
as kaons, evaluated using the simulation.

The dominant background in this case is from residual ISR $\Kp\Km\piz$
and $\Kp\Km\eta$ events, 
as well as ISR $\KpKm\ppz\piz$ events.
Their simulated contribution, 
shown as the dashed histogram in Fig.~\ref{2k2pi0_chi2_all}, 
is consistent with the data in the high \chiKKppnt region.
All other backgrounds are either negligible or distributed uniformly 
in \chiKKppnt.
We define a signal region, $\chiKKppnt\! <\! 50$, 
containing 4425 data and 6948 simulated events, 
and a control region, $50\! <\! \chiKKppnt\! <\! 100$,
containing 1751 data and 848 simulated events.

\begin{figure}[tbh]
\includegraphics[width=0.9\linewidth]{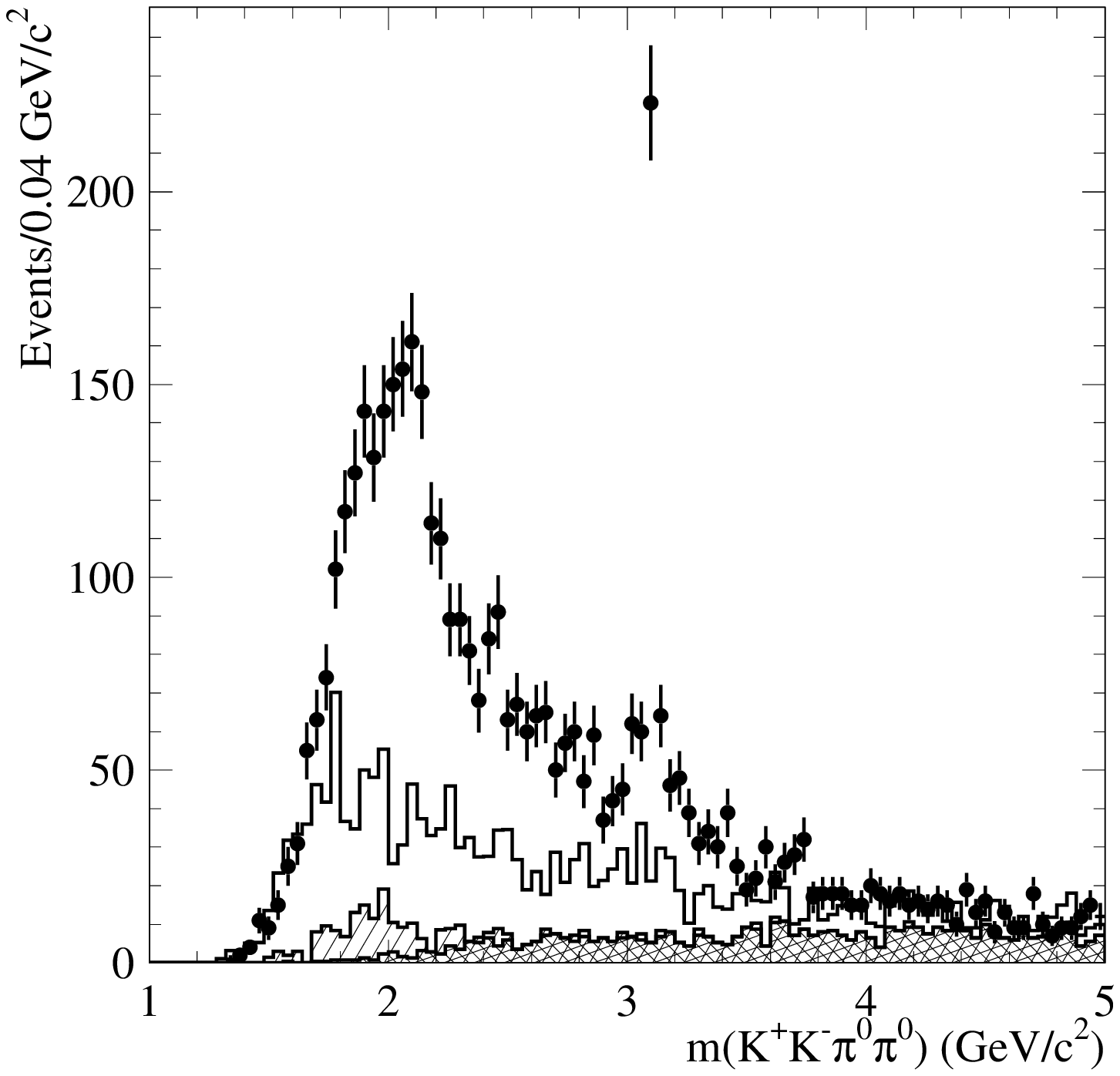}
\vspace{-0.4cm}
\caption{
  Invariant mass distribution for \KKppnt candidates in the data (points).
  The cross-hatched, hatched and open histograms represent, cumulatively, 
  the non-ISR background, the contribution from ISR $\pipi\ppz$ events,
  and the ISR background from the control region of Fig.~\ref{2k2pi0_chi2_all}.
  }
\label{2k2pi0_babar}
\end{figure}

Figure~\ref{2k2pi0_babar} shows the \KKppnt invariant mass
distribution from threshold up to 5~\gevcc for events in the signal region.
The \qqbar background (cross-hatched histogram) is negligible at low 
masses but forms a large fraction of the selected events above about 4~\gevcc.
The ISR $\pipi\ppz$ contribution (hatched region) is negligible except
in the 1.5--2.5~\gevcc region.
The sum of all other backgrounds, estimated from the control region,
is the dominant contribution below 1.6~\gevcc and non negligible everywhere.
The total background in the 1.6--2.5~\gevcc region
is 15--20\% (open histogram).

We subtract the sum of backgrounds from the number of selected events 
in each mass bin to obtain a number of signal events.
Considering uncertainties in the cross sections for the background processes,
the normalization of events in the control region and the simulation
statistics, we estimate a systematic uncertainty on the signal yield
after background subtraction as
less than 5\% in the 1.6--3.0~\gevcc region, 
but increases to 10\% in the region above 3~\gevcc.

\subsection{Selection Efficiency}

\begin{figure}[tbh]
\includegraphics[width=0.9\linewidth]{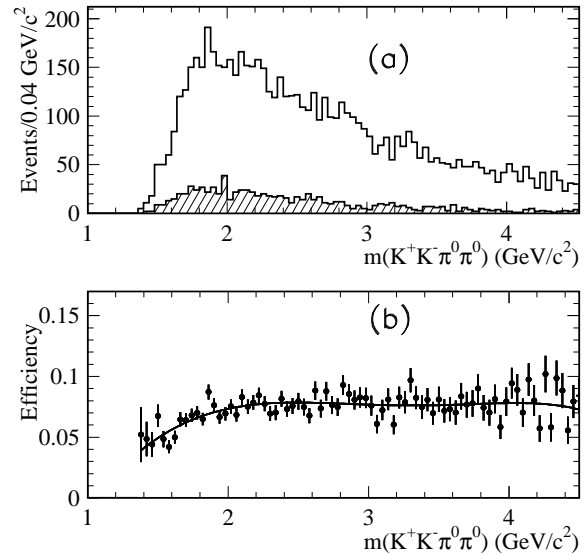}
\vspace{-0.4cm}
\caption{
  (a) Invariant mass distribution for simulated \KKppnt events in the
  signal (open) and control (hatched) regions (see Fig.~\ref{2k2pi0_chi2_all});
  (b) net reconstruction and selection efficiency as a function of mass
  obtained from this simulation
  (the curve represents a third order polynomial fit).
  }
\label{mc_acc3}
\end{figure}

The detection efficiency is determined in the same manner as in 
Sec.~\ref{sec:eff1}.
Figure~\ref{mc_acc3}(a) shows the simulated \KKppnt invariant mass
distributions in the signal and control regions from the phase space
model.
We divide the number of reconstructed events in each
40~\mevcc mass interval by the number generated ones in that interval to
obtain the efficiency shown as the points in Fig.~\ref{mc_acc3}(b);
a third order polynomial fit to the efficiency
is used to calculate the cross section.
Again, the simulation of the ISR photon
covers a limited angular range, 
about 30\% wider than EMC acceptance,
and shown efficiency is factor 0.7 lower than for the hadronic system
alone.
Simulations assuming dominance of the $\phi \!\to\! \Kp\Km$ and/or
$f_0 \!\!\to\! \ppz$ channels give consistent results, and we apply
the same 5\% systematic uncertainty for possible model dependence 
as in Sec.~\ref{sec:eff1}.

We correct for mis-modeling of the track finding and kaon identification 
efficiencies as in Sec.~\ref{sec:eff1},
and for the shape of the \chiKKppnt distribution analogously,
using the result in Ref.~\cite{isr6pi}, $(0\pm 6)$\%. 
We correct the \piz-finding efficiency using the procedure described in
detail in Ref.~\cite{isr6pi}.
From ISR $\epem \!\!\to\! \omega\piz\gamma \!\!\to\! \pipi\ppz\gamma$ events
selected with and without the \piz from the $\omega$ decay,
we find that the simulated efficiency for one \piz is too high by (2.8$\pm$1.4)\%.
Conservatively we apply a correction of $+(5.6\pm 2.8)$\% for two \piz
in the event.

\begin{table*}
\caption{Measurements of the $\ep\en\to K^+ K^- \ppz$ 
cross section (errors are statistical only).}
\label{2k2pi0_tab}
\begin{ruledtabular}
\begin{tabular}{ c c c c c c c c }
$E_{\rm c.m.}$ (GeV) & $\sigma$ (nb)  
& $E_{\rm c.m.}$ (GeV) & $\sigma$ (nb) 
& $E_{\rm c.m.}$ (GeV) & $\sigma$ (nb) 
& $E_{\rm c.m.}$ (GeV) & $\sigma$ (nb)  
\\
\hline
 1.4200 &  0.00 $\pm$  0.05 & 2.3400 &  0.35 $\pm$  0.06 & 3.2600 &  0.13 $\pm$  0.03 & 4.1800 &  0.02 $\pm$  0.01 \\
 1.4600 &  0.12 $\pm$  0.07 & 2.3800 &  0.29 $\pm$  0.06 & 3.3000 &  0.09 $\pm$  0.03 & 4.2200 &  0.03 $\pm$  0.01 \\
 1.5000 &  0.00 $\pm$  0.07 & 2.4200 &  0.38 $\pm$  0.06 & 3.3400 &  0.09 $\pm$  0.03 & 4.2600 &  0.03 $\pm$  0.01 \\
 1.5400 &  0.01 $\pm$  0.08 & 2.4600 &  0.38 $\pm$  0.06 & 3.3800 &  0.08 $\pm$  0.02 & 4.3000 &  0.03 $\pm$  0.01 \\
 1.5800 &  0.03 $\pm$  0.09 & 2.5000 &  0.22 $\pm$  0.05 & 3.4200 &  0.11 $\pm$  0.03 & 4.3400 &  0.03 $\pm$  0.01 \\
 1.6200 &  0.09 $\pm$  0.09 & 2.5400 &  0.25 $\pm$  0.05 & 3.4600 &  0.06 $\pm$  0.02 & 4.3800 &  0.01 $\pm$  0.01 \\
 1.6600 &  0.31 $\pm$  0.11 & 2.5800 &  0.25 $\pm$  0.05 & 3.5000 &  0.04 $\pm$  0.02 & 4.4200 &  0.05 $\pm$  0.01 \\
 1.7000 &  0.35 $\pm$  0.11 & 2.6200 &  0.25 $\pm$  0.05 & 3.5400 &  0.06 $\pm$  0.02 & 4.4600 &  0.03 $\pm$  0.01 \\
 1.7400 &  0.49 $\pm$  0.11 & 2.6600 &  0.28 $\pm$  0.05 & 3.5800 &  0.07 $\pm$  0.02 & 4.5000 &  0.04 $\pm$  0.01 \\
 1.7800 &  0.51 $\pm$  0.12 & 2.7000 &  0.16 $\pm$  0.04 & 3.6200 &  0.04 $\pm$  0.02 & 4.5400 &  0.00 $\pm$  0.01 \\
 1.8200 &  0.84 $\pm$  0.12 & 2.7400 &  0.22 $\pm$  0.04 & 3.6600 &  0.06 $\pm$  0.02 & 4.5800 &  0.02 $\pm$  0.01 \\
 1.8600 &  0.94 $\pm$  0.11 & 2.7800 &  0.21 $\pm$  0.04 & 3.7000 &  0.08 $\pm$  0.02 & 4.6200 &  0.02 $\pm$  0.01 \\
 1.9000 &  0.95 $\pm$  0.12 & 2.8200 &  0.13 $\pm$  0.04 & 3.7400 &  0.09 $\pm$  0.02 & 4.6600 &  0.02 $\pm$  0.01 \\
 1.9400 &  0.80 $\pm$  0.11 & 2.8600 &  0.21 $\pm$  0.04 & 3.7800 &  0.02 $\pm$  0.02 & 4.7000 &  0.04 $\pm$  0.01 \\
 1.9800 &  0.87 $\pm$  0.11 & 2.9000 &  0.11 $\pm$  0.03 & 3.8200 &  0.05 $\pm$  0.01 & 4.7400 &  0.02 $\pm$  0.01 \\
 2.0200 &  1.00 $\pm$  0.10 & 2.9400 &  0.12 $\pm$  0.04 & 3.8600 &  0.04 $\pm$  0.01 & 4.7800 &  0.01 $\pm$  0.01 \\
 2.0600 &  0.96 $\pm$  0.10 & 2.9800 &  0.12 $\pm$  0.04 & 3.9000 &  0.03 $\pm$  0.02 & 4.8200 &  0.01 $\pm$  0.01 \\
 2.1000 &  0.90 $\pm$  0.10 & 3.0200 &  0.21 $\pm$  0.04 & 3.9400 &  0.02 $\pm$  0.01 & 4.8600 &  0.01 $\pm$  0.01 \\
 2.1400 &  0.82 $\pm$  0.10 & 3.0600 &  0.16 $\pm$  0.04 & 3.9800 &  0.03 $\pm$  0.01 & 4.9000 &  0.03 $\pm$  0.01 \\
 2.1800 &  0.58 $\pm$  0.08 & 3.1000 &  0.92 $\pm$  0.07 & 4.0200 &  0.05 $\pm$  0.01 & 4.9400 &  0.04 $\pm$  0.02 \\
 2.2200 &  0.56 $\pm$  0.08 & 3.1400 &  0.19 $\pm$  0.04 & 4.0600 &  0.04 $\pm$  0.01 & 4.9800 &  0.04 $\pm$  0.02 \\
 2.2600 &  0.37 $\pm$  0.07 & 3.1800 &  0.12 $\pm$  0.03 & 4.1000 &  0.03 $\pm$  0.01 &  &    \\
 2.3000 &  0.43 $\pm$  0.07 & 3.2200 &  0.14 $\pm$  0.03 & 4.1400 &  0.03 $\pm$  0.01 &  &    \\

\end{tabular}
\end{ruledtabular}
\end{table*}

\begin{table}[tbh]
\caption{
Summary of corrections and systematic uncertainties 
on the $\epem \!\!\to\! \KKppnt$  cross  section.
The total correction is the linear sum of the components and the
total uncertainty is the sum in quadrature.
  }
\label{error2_tab}
\begin{ruledtabular}
\begin{tabular}{l c r@{}l} 
     Source             & Correction & \multicolumn{2}{c}{Uncertainty}   \\
\hline
                        &            &       &              \\[-0.2cm]
Rad. Corrections        &  --        &  $1\%$&              \\
Backgrounds             &  --        &  $5\%$&, $m_{\Kppnt}\! <3~\gevcc$ \\
                        &            & $10\%$&, $m_{\Kppnt}\! >3~\gevcc$ \\
Model Dependence        &  --        &  $5\%$&              \\
\chiKKppnt Distn.       &   0\%      &  $6\%$&              \\ 
Tracking Efficiency     & $+1.6\%$   &  $0.8\%$&              \\
Kaon ID Efficiency      & $+2\%$     &  $2\%$&              \\
\piz Efficiency         & $+5.6\%$   &  $2.8\%$&              \\
ISR Luminosity          &  --        &  $3\%$&              \\[0.1cm]
\hline
                        &            &       &             \\[-0.2cm]
Total                   & $+9.2\%$   & $10\%$&, $m_{\Kppnt}\! <3~\gevcc$ \\
                        &            & $14\%$&, $m_{\Kppnt}\! >3~\gevcc$ \\
\end{tabular}
\end{ruledtabular}
\end{table}

\subsection{\boldmath Cross Section for $\epem \to \KKppnt$}
\label{sec:2k2pi0xs}

\begin{figure}[tbh]
\includegraphics[width=0.9\linewidth]{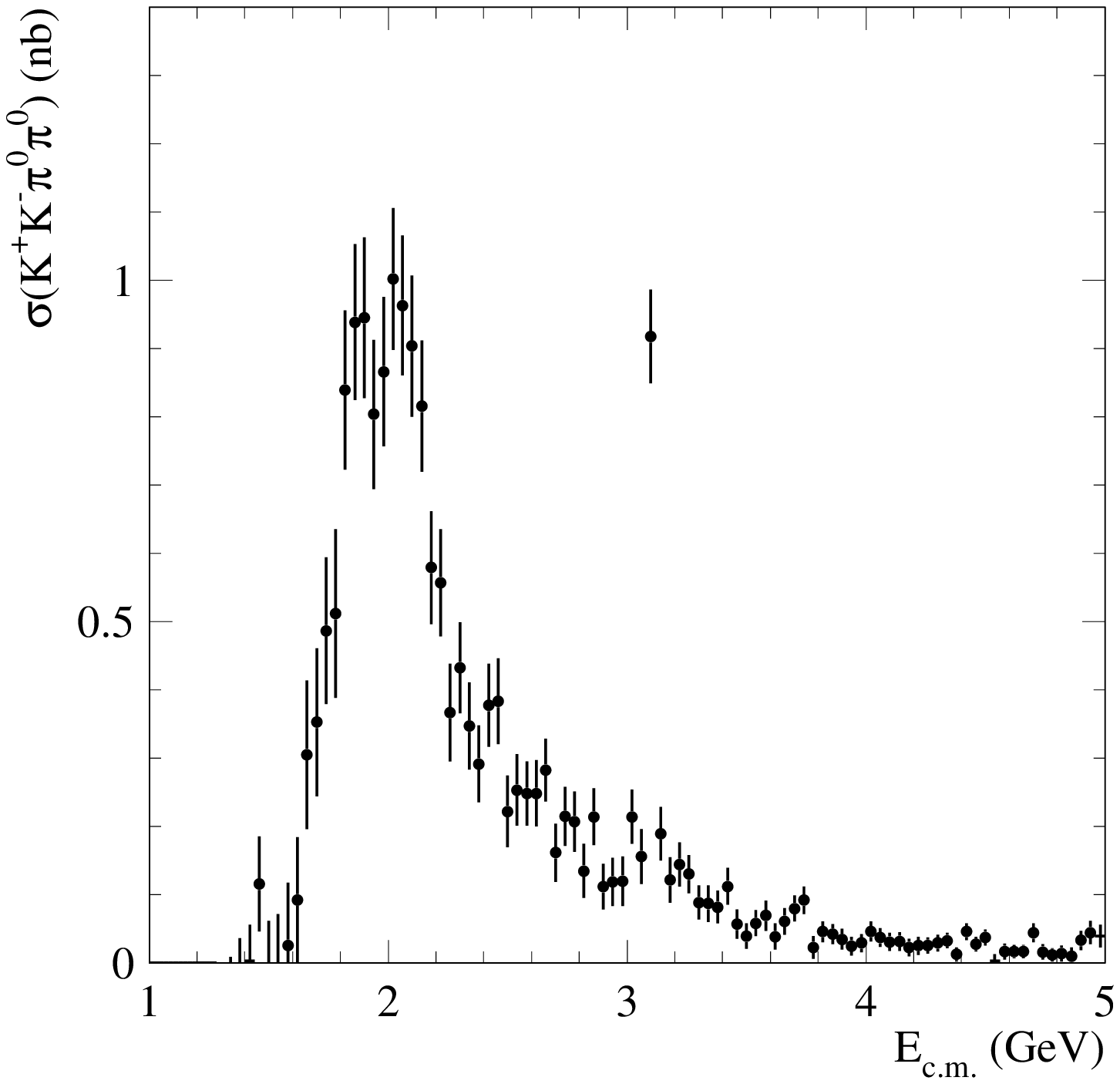}
\vspace{-0.5cm}
\caption{
  The $\epem \!\!\to\! \KKppnt$ cross section as a function of the
  effective \epem
  c.m.\ energy measured with ISR data at \babar.
  The errors are statistical only.
  }
\label{2k2pi0_ee_babar}
\end{figure}

We calculate the cross section for $\epem \to \KKppnt$ in 40~\mev \Ecm
intervals from the analog of Eq.~\ref{xseqn},
using the invariant mass of the \KKppnt system to determine the
effective c.m.\ energy.
We show the first measurement of this cross section
in Fig.~\ref{2k2pi0_ee_babar} and list the results obtained in Table~\ref{2k2pi0_tab}.
The cross section rises to a peak value near 1~nb at 2~\gev, falls sharply at
2.2~\gev, then decreases slowly.
The only statistically significant structure is the $J/\psi$ peak.
The drop at 2.2~\gev is similar to that seen in the \KKppch mode.
Again, $d{\cal L}$ includes corrections for vacuum polarization that
should be omitted from calculations of $g_\mu\! -\! 2$.

The simulated \KKppnt invariant mass resolution is 8.8~\mevcc in the 
1.5--2.5~\gevcc mass range,
and increases with mass to 11.2~\mevcc in the 2.5--3.5~\gevcc range. 
Since less than 20\% of the events in a 40~\mevcc bin are
reconstructed outside that bin
and the cross section has no sharp structure other than the $J/\psi$ peak,
we again make no correction for resolution.
The point-to-point systematic errors are much smaller than statistical
ones, and
the errors on the normalization are summarized in Table~\ref{error2_tab},
along with the corrections that were applied to the measurements.
The total correction is $+9.2$\%, and the total systematic uncertainty
is 10\% at low mass, increasing to 14\% above 3~\gevcc.

\subsection{\boldmath Substructure in the \KKppnt Final State}
\label{sec:kaons2}

\begin{figure}[tbh]
\includegraphics[width=0.9\linewidth]{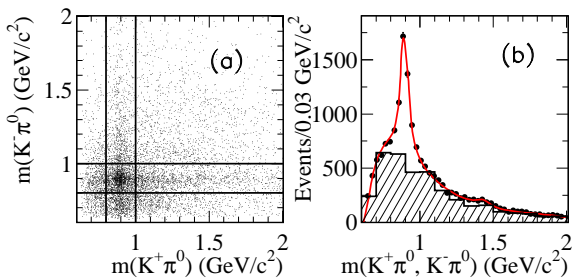}
\vspace{-0.4cm}
\caption{
  (a) Invariant mass of the $\Km\piz$ pair versus that of the
  $\Kp\piz$ pair in selected \KKppnt events (two entries per event);
  (b) sum of projections of (a) (dots, four entries per event).
  The curve represents the result of the fit described in the text.
  The hatched histogram is the $K^\pm\piz$ distribution for events in
  which the other $K^\mp\piz$ combination is within the
  $K^{*\pm}(892)$ bands indicated in (a), with events in the overlap
  region taken only once.
  }
\label{kkstarpi0}
\end{figure}

A scatter plot of the invariant mass 
of the $K^-\pi^0$ versus that of the $K^+\pi^0$ pair is shown in
Fig.~\ref{kkstarpi0}(a)  
with two entries per event selected in the \chisq signal region. 
Horizontal and vertical bands corresponding to the $K^{*+}(892)$ and
$K^{*-}(892)$, respectively, are visible.
Figure~\ref{kkstarpi0}(b) shows as points the sum of the two projections of 
Fig.~\ref{kkstarpi0}(a);
a large $K^{*\pm}(892)$ signal is evident.
Fitting this distribution with the function discussed in
Sec.~\ref{kstarxs} gives a good \chisq and the curve shown on 
Fig.~\ref{kkstarpi0}(b).
The $K^{*\pm}(1430)$:$K^{*\pm}(892)$ ratio is consistent with that
for neutral $K^*$ seen in the \KKppch channel,
and the number of $K^{*\pm}(892)$ in the peak is consistent
with one per selected event.
The hatched histogram in Fig.~\ref{kkstarpi0}(b) represents the $K^\pm\piz$
mass in events with the other $K^\mp\piz$ mass within the lines in
Fig.~\ref{kkstarpi0}(a), 
but with events in the overlap region used only once,
and shows no $K^{*\pm}(892)$ signal.
These results indicate that the $\epem \!\!\to\! K^{*\pm} K^{*\mp}$  
cross section is small and that the $K^{*\pm}(892) K^{\mp}\pi^0$ channels
dominate the overall cross section.

We find no signals for resonances 
in the $\Kp\Km\piz$ or $K^\pm\ppz$ decay modes.
Since the $K^{*\pm}(892) K^{\mp}\pi^0$ channels dominate and the
statistics are low in any mass bin, we do not attempt to extract a
separate $K^{*\pm}(892) K^{\mp}\pi^0$ cross section.
The total \KKppnt cross section is roughly a factor of four lower than
the $K^{*0}(892) K^{\pm}\pi^\mp$ cross section observed in the \KKppch
final state.
This is consistent with what one might expect from isospin and the
charged vs.\ neutral $K^*$ branching fractions into charged kaons.

\begin{table*}
\caption{Measurements of the $\ep\en\to\phi(1020) f_{0}(980)$ 
cross section ($f_{0}\to\ppz$,  errors are statistical only).}
\label{phif0_tab2}
\begin{ruledtabular}
\begin{tabular}{ c c c c c c }
$E_{\rm c.m.}$ (GeV) & $\sigma$ (nb)  
& $E_{\rm c.m.}$ (GeV) & $\sigma$ (nb) 
& $E_{\rm c.m.}$ (GeV) & $\sigma$ (nb) 
\\
\hline

 1.88-1.92 &  $0.078^{+0.082}_{-0.053}$ & 2.12-2.16 &  $0.397^{+0.164}_{-0.137}$ & 2.44-2.52 &  
$0.120^{+0.053}_{-0.042}$ \\
 1.92-1.96 &  $0.220^{+0.114}_{-0.085}$ & 2.16-2.20 &  $0.408^{+0.143}_{-0.118}$ & 2.52-2.68 &  
$0.050^{+0.024}_{-0.019}$ \\
 1.96-2.00 &  $0.136^{+0.104}_{-0.075}$ & 2.20-2.24 &  $0.070^{+0.064}_{-0.042}$ & 2.68-2.84 &  
$0.026^{+0.017}_{-0.012}$ \\
 2.00-2.04 &  $0.446^{+0.160}_{-0.131}$ & 2.24-2.28 &  $0.174^{+0.095}_{-0.071}$ & 2.84-3.00 &  
$0.026^{+0.015}_{-0.011}$ \\
 2.04-2.08 &  $0.315^{+0.142}_{-0.113}$ & 2.28-2.36 &  $0.069^{+0.042}_{-0.030}$ & 3.00-3.16 &  
$0.032^{+0.017}_{-0.013}$ \\
 2.08-2.12 &  $0.519^{+0.169}_{-0.141}$ & 2.36-2.44 &  $0.112^{+0.051}_{-0.040}$ & 3.16-3.32 &  
$0.012^{+0.012}_{-0.008}$ 
\end{tabular}
\end{ruledtabular}
\end{table*}

\subsection{\boldmath The $\phi(1020)\ppz$ Intermediate State}
\label{sec:phipipi2}

\begin{figure}[tbh]
\includegraphics[width=0.9\linewidth]{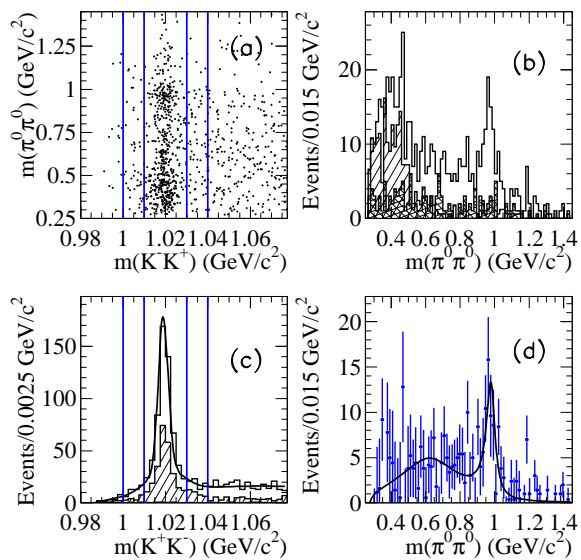}
\vspace{-0.4cm}
\caption{
  (a) Scatter plot of the $\ppz$ invariant mass vs.\ the $\Kp\Km$ invariant
  mass for all selected \KKppnt events;
  (b) the $\ppz$ invariant mass projections for events in the $\phi$ peak
  (open histogram), sidebands (cross-hatched) and control region
  (hatched);
  (c) the $\Kp\Km$ mass projection for events in the signal (open) and
  control (hatched) regions;
  (d) difference between the open and other histograms in (b).
  }
\label{phif0_sel2}
\end{figure}

The selection of events containing a $\phi(1020) \!\to\! \KpKm$ decay
follows that in Section~\ref{sec:phipipi}. 
Figure~\ref{phif0_sel2}(a) shows a
scatter plot of the invariant mass of the $\ppz$ pair versus that of the 
\KpKm pair.
A vertical band corresponding to the $\phi$ is visible,
whose intensity decreases with increasing \ppz mass except for an enhancement
in the $f_{0}(980)$ region.
The $\phi$ signal is also visible in the $\Kp\Km$ invariant mass
projection shown in Fig.~\ref{phif0_sel2}(c).
The relative non-$\phi$ background is smaller than in the \KKppch
mode, 
but there is a large background from ISR $\phi\pi^0$, $\phi\eta$
and/or $\phi\ppz\piz$ events, 
as indicated by the control region histogram (hatched) in 
Fig.~\ref{phif0_sel2}(c).
The contributions from non-ISR and ISR $\pipi\ppz$ events are negligible.
Selecting $\phi$ candidate and side band events as for the \KKppch mode
(vertical lines in Figs.~\ref{phif0_sel2}(a,c)),
we obtain the \ppz mass projections shown as the open and cross-hatched
histograms, respectively, in Fig.~\ref{phif0_sel2}(b).
Control region events (hatched histogram) are concentrated at low masses.
A peak corresponding to the $f_{0}(980)$ is visible over a relatively
low background. 

In Fig.~\ref{phif0_sel2}(d) we show the numbers of entries from the
candidate events minus those from the sideband and control regions.
A sum of two Breit-Wigner functions is again sufficient to describe
the data.
Fitting Eq.~\ref{f0fit} with the parameters of one BW fixed to the  
values given in Sec.~\ref{sec:phipipi}, corresponding to the $f_{0}(600)$,
we obtain a good fit, shown as the curve in Fig.~\ref{phif0_sel2}(d).
This fit yields a $f_{0}(980)$ signal of $54 \pm 9$ events with a
mass $m = 0.970 \pm 0.007~\gevcc$ and width $\Gamma = 0.081 \pm 0.021~\gev$  
consistent with PDG values~\cite{PDG}.
Due to low statistics and high backgrounds, we do not extract an
$\epem\to\phi(1020)\ppz$ cross section.

\subsection{\boldmath The $\phi(1020) f_0(980)$ Intermediate State}
\label{sec:phif02}

Since the background under the $f_0(980)$ peak in Figs.~\ref{phif0_sel2}(b,d)
is relatively low we are able to extract the $\phi(1020) f_0(980)$ 
contribution. 
As in Sec.~\ref{sec:phif01}, we require the dipion mass to be in the
range 0.85--1.10~\gevcc and fit the background-subtracted $\Kp\Km$ 
mass projection in each bin of \KKppnt mass to obtain a number of   
$\phi f_0$ events.
Again, about 10\% of these are $\phi\ppz$ events in which the \ppz
pair is not produced through the $f_0$, but this does not affect the
conclusions.

\begin{figure}[tbh]
\includegraphics[width=0.9\linewidth]{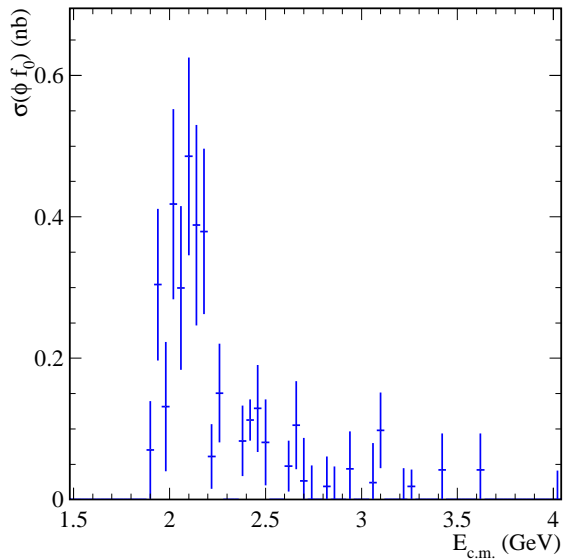}
\vspace{-0.4cm}
\caption{
  Cross section for the reaction $\epem \!\!\to\! \phi(1020) f_{0}(980)$ 
  as a function of effective \epem c.m.\ energy obtained from the
  \KKppnt final state.
}
\label{phif0xs2}
\end{figure}

We convert the number of fitted events in each mass bin into a
measurement of the $\epem \!\!\to\! \phi(1020) f_{0}(980)$ cross
section as described above and dividing by the
$f_{0}(980) \!\!\to\! \ppz$ branching fraction of one-third.
The cross section is shown in Fig.~\ref{phif0xs2} as a function of
\Ecm and is listed in Table~\ref{phif0_tab2}.
Due to smaller number of events, we have used larger bins at higher energies.
The overall shape is consistent with that obtained in the \KKppch mode 
(see Fig.~\ref{phif0xs}),
and there is a sharp drop near 2.2~\gevcc, 
but the statistical errors are large and no conclusion can be drawn
from this mode alone.
Possible interpretations are discussed in Section~\ref{phif0bump}.

\section{\boldmath The \KKKK Final State}
\subsection{Final Selection and Background}

\begin{figure}[tbh]
\includegraphics[width=0.9\linewidth]{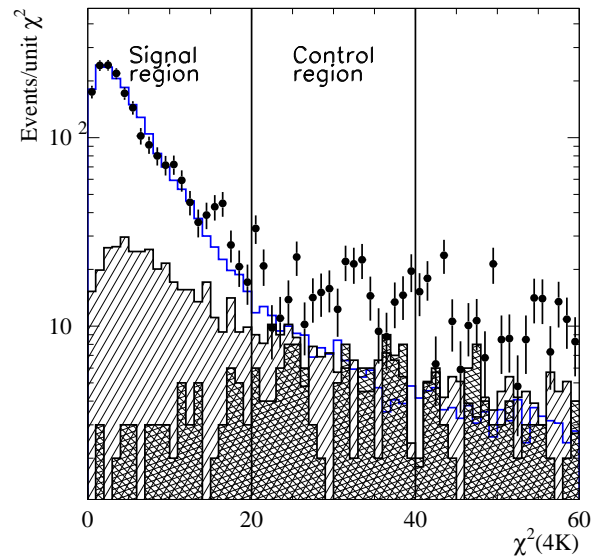}
\vspace{-0.4cm}
\caption{
  Distribution of \chisq from the three-constraint fit for \KKKK
  candidates in the data (points).
  The open histogram is the distribution for simulated signal events,
  normalized as described in the text.
  The hatched histogram represents the background from
  non-ISR events, estimated
  as described in the text. The cross-hatched histograms is for
  simulated ISR $\Kp\Km\pipi$ events. 
  }
\label{4k_chi2_all}
\end{figure}

Figure~\ref{4k_chi2_all} shows the distribution of \chifourK for the
\KKKK candidates as points, and the open histogram is the distribution for
simulated \KKKK events,
normalized to the data in the region $\chifourK \! <\! 5$ where
the backgrounds and radiative corrections are small.
The hatched histogram represents the background from 
$\epem \!\!\to\! \qqbar$ events, evaluated as for the other modes.
The cross-hatched histogram represents the background
from simulated ISR \KKppch events with both charged pions 
misidentified as kaons. 

\begin{figure}[tbh]
\includegraphics[width=0.9\linewidth]{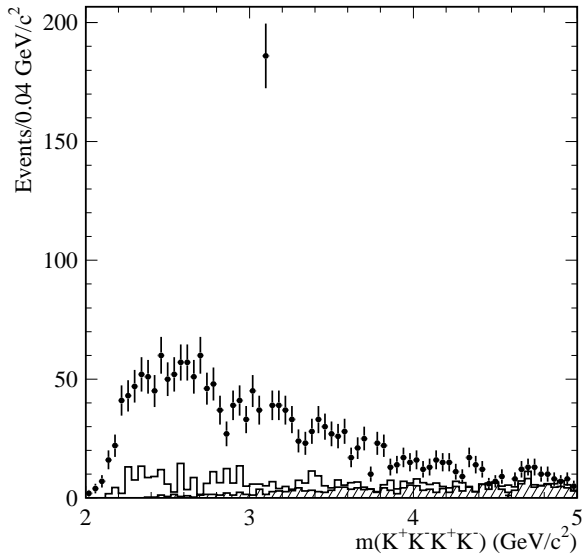}
\vspace{-0.4cm}
\caption{
  Invariant mass distribution for \KKKK candidates in the data (points).
  The hatched and open histograms represent, cumulatively, 
  the non-ISR background and the ISR background from the control region,
  which is dominated by the contribution from ISR \KKppch events.
  }
\label{4k_babar}
\end{figure}

We define signal and control regions of $\chifourK\! <\! 20$ and 
$20\! <\! \chifourK\! <\! 40$, respectively.
The signal region contains 2,305 data and 20,616 simulated events, 
and the control region contains 463 data and 1,601 simulated events.
Figure~\ref{4k_babar} shows the \KKKK invariant mass distribution from
threshold up to 5~\gevcc for events in the signal region as points
with errors.
The \qqbar background (hatched histogram) is small at low masses, but
dominant above about 4.5~\gevcc.
Since the ISR \KKppch background does not peak at low \chifourK values, 
we include it in the background evaluated from the control region,
according to the method explained in Sec.~\ref{sec:selection1}.
It dominates this background, which is 10\% or lower at all masses.
The total background is shown as the open histogram in Fig.~\ref{4k_babar}.

We subtract the sum of backgrounds from the number of selected events
in each mass bin to obtain a number of signal events.
Considering uncertainties in the cross sections for the background processes,
the normalization of events in the control region, 
and the simulation statistics,
we estimate a systematic uncertainty on the signal yield of less than
5\% in the 2--3~\gevcc region, increasing to 10\% in the region above 3~\gevcc.

\begin{figure}[tbh]
\includegraphics[width=0.9\linewidth]{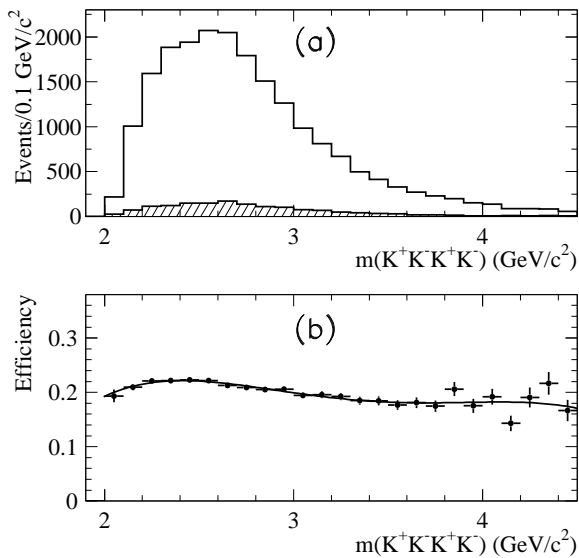}
\vspace{-0.4cm}
\caption{
  (a) Invariant mass distributions for simulated \KKKK events in the
  signal (open) and control (hatched) regions (see Fig.~\ref{4k_chi2_all});
  (b) net reconstruction and selection efficiency as a function of mass
  obtained from this simulation 
  (the curve represents a 3$^{\rm rd}$ order polynomial fit).
  }
\label{mc_acc4}
\end{figure}

\subsection{Selection Efficiency}

The detection efficiency is determined as for the other two final states.
Figure~\ref{mc_acc4}(a) shows the simulated \KKKK invariant-mass
distributions in the signal and control regions from the phase space
model.
We divide the number of reconstructed  events in each
mass interval by the number of generated ones in that interval
to obtain the efficiency shown as the points in Fig.~\ref{mc_acc4}(b).
It is quite uniform, and we fit a third order polynomial, which we use
to extract the cross section.
A factor of 0.7 is again applicable 
for only hadronic system efficiency
due to the limited angular
coverage of the ISR photon simulation.
A simulation assuming dominance of the $\phi\Kp\Km$ channel, 
with the \KpKm pair in an S-wave, gives consistent results,
and we apply the same 5\% systematic  uncertainty as for the other
final states.
We correct for mis-modeling of the track finding and kaon identification 
efficiencies as in Sec.~\ref{sec:eff1},
and for the shape of the \chifourK distribution analogously,
using the result in Ref.~\cite{isr4pi}, $(3.0\pm 2.0)$\%.

\subsection{\boldmath Cross Section for $\epem\to K^+ K^- K^+ K^-$}
\label{sec:4kxs}

\begin{figure}[tbh]
\includegraphics[width=0.9\linewidth]{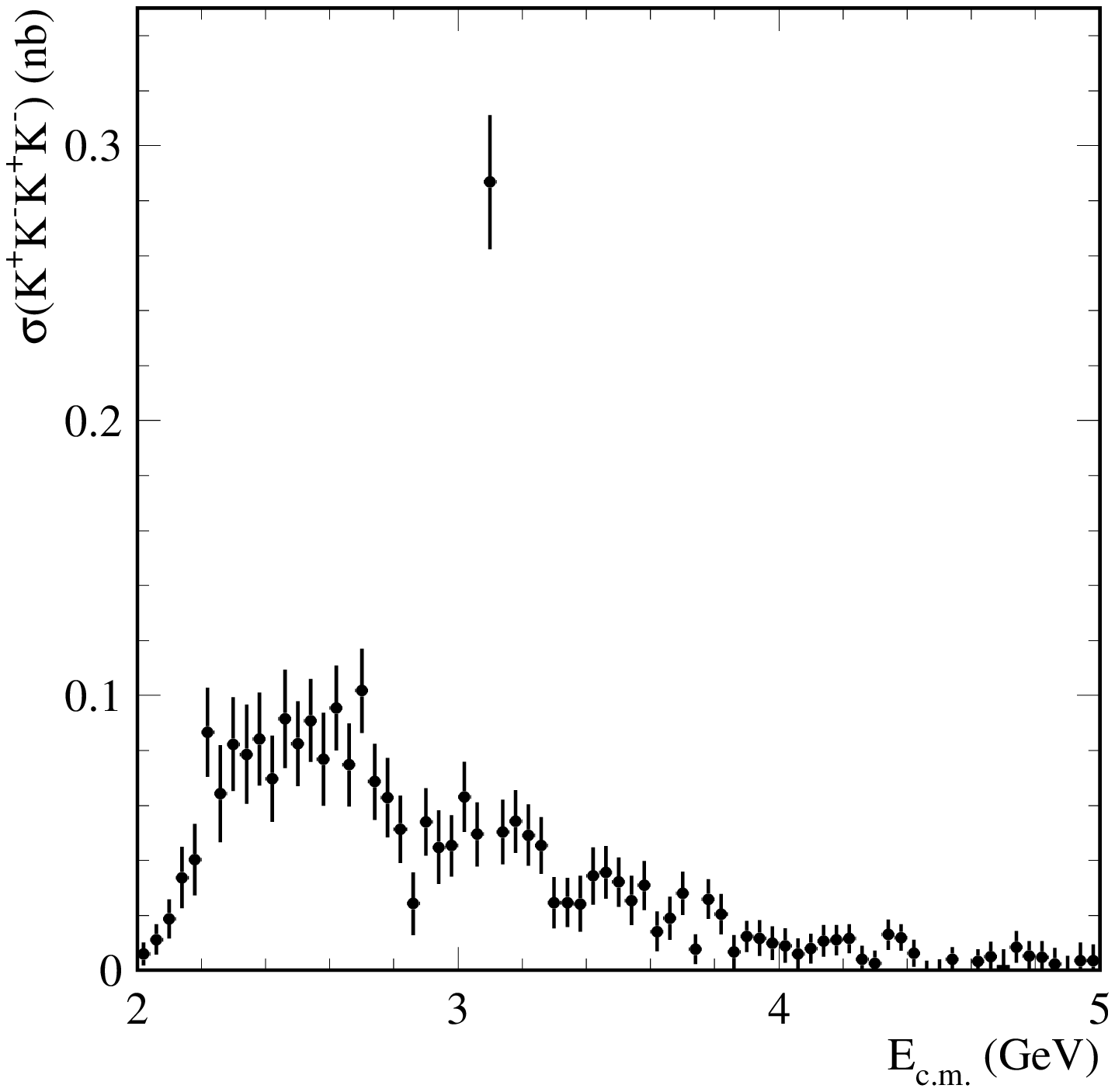}
\vspace{-0.5cm}
\caption{
  The $\epem \!\!\to\! \KKKK$ cross section as a function of the effective \epem
  c.m.\ energy measured with ISR data at \babar.
  The errors are statistical only.
  }
\label{4k_ee_babar}
\end{figure}

We calculate the $\epem \!\!\to\! \KKKK$ cross section in
40~\mev \Ecm intervals from the analog of Eq.~\ref{xseqn},
using the invariant mass of the \KKKK system to determine the
effective c.m.\ energy.
We show this cross section in Fig.~\ref{4k_ee_babar} and list it in
Table~\ref{4k_tab}.  
It rises to a peak value near 0.1~nb in the 2.3--2.7~\gev region, 
then decreases slowly with increasing energy.
The only statistically significant narrow structure is the $J/\psi$ peak.
Again, $d{\cal L}$ includes corrections for vacuum polarization that
should be omitted from calculations of $g_\mu\! -\! 2$.
This supersedes our previous result~\cite{isr4pi}.

\begin{table*}
\caption{Measurements of the $\ep\en\to K^+ K^- K^+ K^-$ 
cross section (errors are statistical only).}
\label{4k_tab}
\begin{ruledtabular}
\begin{tabular}{ c c c c c c c c }
$E_{\rm c.m.}$ (GeV) & $\sigma$ (nb)  
& $E_{\rm c.m.}$ (GeV) & $\sigma$ (nb) 
& $E_{\rm c.m.}$ (GeV) & $\sigma$ (nb) 
& $E_{\rm c.m.}$ (GeV) & $\sigma$ (nb)  
\\
\hline
   2.02 & 0.006 $\pm$ 0.004 &   2.66 & 0.075 $\pm$ 0.015 &   3.30 & 0.025 $\pm$ 0.009 &   3.94 & 0.012 $\pm$ 0.006 \\
   2.06 & 0.011 $\pm$ 0.006 &   2.70 & 0.102 $\pm$ 0.015 &   3.34 & 0.025 $\pm$ 0.009 &   3.98 & 0.012 $\pm$ 0.006 \\
   2.10 & 0.019 $\pm$ 0.007 &   2.74 & 0.069 $\pm$ 0.014 &   3.38 & 0.024 $\pm$ 0.010 &   4.02 & 0.010 $\pm$ 0.006 \\
   2.14 & 0.034 $\pm$ 0.011 &   2.78 & 0.063 $\pm$ 0.014 &   3.42 & 0.034 $\pm$ 0.010 &   4.06 & 0.009 $\pm$ 0.005 \\
   2.18 & 0.040 $\pm$ 0.013 &   2.82 & 0.051 $\pm$ 0.012 &   3.46 & 0.036 $\pm$ 0.009 &   4.10 & 0.006 $\pm$ 0.005 \\
   2.22 & 0.087 $\pm$ 0.016 &   2.86 & 0.024 $\pm$ 0.011 &   3.50 & 0.032 $\pm$ 0.009 &   4.14 & 0.008 $\pm$ 0.006 \\
   2.26 & 0.064 $\pm$ 0.018 &   2.90 & 0.054 $\pm$ 0.012 &   3.54 & 0.025 $\pm$ 0.009 &   4.18 & 0.011 $\pm$ 0.005 \\
   2.30 & 0.082 $\pm$ 0.017 &   2.94 & 0.045 $\pm$ 0.013 &   3.58 & 0.031 $\pm$ 0.009 &   4.22 & 0.011 $\pm$ 0.005 \\
   2.34 & 0.079 $\pm$ 0.018 &   2.98 & 0.045 $\pm$ 0.011 &   3.62 & 0.014 $\pm$ 0.007 &   4.26 & 0.012 $\pm$ 0.005 \\
   2.38 & 0.084 $\pm$ 0.017 &   3.02 & 0.063 $\pm$ 0.013 &   3.66 & 0.019 $\pm$ 0.008 &   4.30 & 0.004 $\pm$ 0.005 \\
   2.42 & 0.070 $\pm$ 0.016 &   3.06 & 0.049 $\pm$ 0.012 &   3.70 & 0.028 $\pm$ 0.008 &   4.34 & 0.003 $\pm$ 0.006 \\
   2.46 & 0.092 $\pm$ 0.018 &   3.10 & 0.287 $\pm$ 0.024 &   3.74 & 0.008 $\pm$ 0.005 &   4.38 & 0.013 $\pm$ 0.005 \\
   2.50 & 0.082 $\pm$ 0.015 &   3.14 & 0.050 $\pm$ 0.012 &   3.78 & 0.026 $\pm$ 0.007 &   4.42 & 0.012 $\pm$ 0.005 \\
   2.54 & 0.091 $\pm$ 0.015 &   3.18 & 0.054 $\pm$ 0.011 &   3.82 & 0.020 $\pm$ 0.007 &   4.46 & 0.006 $\pm$-0.004 \\
   2.58 & 0.077 $\pm$ 0.017 &   3.22 & 0.049 $\pm$ 0.011 &   3.86 & 0.007 $\pm$ 0.006 &   4.50 &-0.001 $\pm$-0.004 \\
   2.62 & 0.095 $\pm$ 0.015 &   3.26 & 0.045 $\pm$ 0.010 &   3.90 & 0.012 $\pm$ 0.006 &   4.54 & 0.000 $\pm$ 0.004 \\
\end{tabular}
\end{ruledtabular}
\end{table*}

The simulated \KKKK invariant mass resolution is 3.0~\mevcc in the 
2.0--2.5~\gevcc range, increasing with mass to 4.7~\mevcc in the 
2.5--3.5~\gevcc range. 
Since the cross section has no sharp structure except for the  $J/\psi$
peak, we again make no correction for resolution.
The errors shown in Fig.~\ref{4k_ee_babar} and Table~\ref{4k_tab} are
statistical only.
The point-to-point systematic errors are much smaller than this, and
the errors on the normalization are summarized in Table~\ref{error3_tab},
along with the corrections applied to the measurement.
The total correction is $+$10\%, 
and the total systematic uncertainty is 9\% at low mass, increasing to
13\% above 3~\gevcc.

\begin{table}[tbh]
\caption{
Summary of corrections and systematic uncertainties 
on the $\epem \!\!\to\! \KKKK$    cross section.
The total correction is the linear sum of the components and the
total uncertainty is the sum in quadrature.
  }
\label{error3_tab}
\begin{ruledtabular}
\begin{tabular}{l c r@{}l} 
     Source             & Correction & \multicolumn{2}{c}{Uncertainty}   \\
\hline
                        &            &       &              \\[-0.2cm]
Rad. Corrections        &  --        &  $1\%$&              \\
Backgrounds             &  --        &  $5\%$& $m_{4K} <3~\gevcc$ \\
                        &            & $10\%$& $m_{4K} >3~\gevcc$ \\
Model Dependence        &  --        &  $5\%$&              \\
\chifourK Distribution  & $+3\%$     &  $2\%$&              \\ 
Tracking Efficiency     & $+3\%$     &  $2\%$&              \\
Kaon ID Efficiency      & $+4\%$     &  $4\%$&              \\
ISR Luminosity          &  --        &  $3\%$&              \\[0.1cm]
\hline
                        &            &       &             \\[-0.2cm]
Total                   & $+10\%$    &  $9\%$& $m_{4K} <3~\gevcc$ \\
                        &            & $13\%$& $m_{4K} >3~\gevcc$ \\
\end{tabular}
\end{ruledtabular}
\end{table}

\begin{figure}[tbh]
\includegraphics[width=0.9\linewidth]{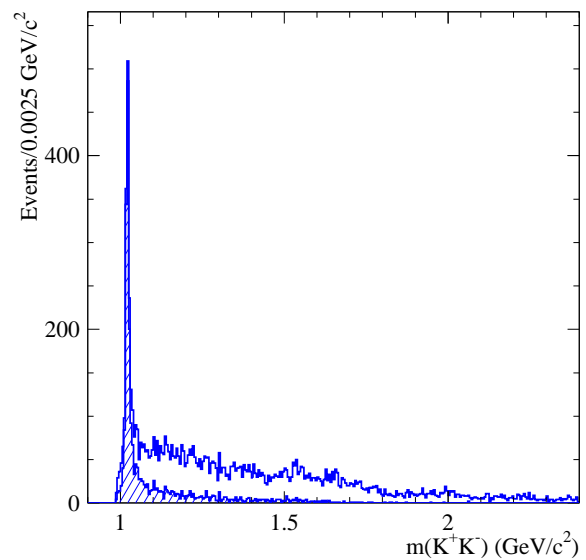}
\vspace{-0.4cm}
\caption{
  Invariant mass distributions for all \KpKm pairs in selected
  $\epem \!\!\to\! \KKKK$ events (open histogram) and for the 
  combination in each event closest to the $\phi$-meson mass (hatched).
  }
\label{mkk_phi}
\end{figure}

\begin{figure}[tbh]
\includegraphics[width=0.9\linewidth]{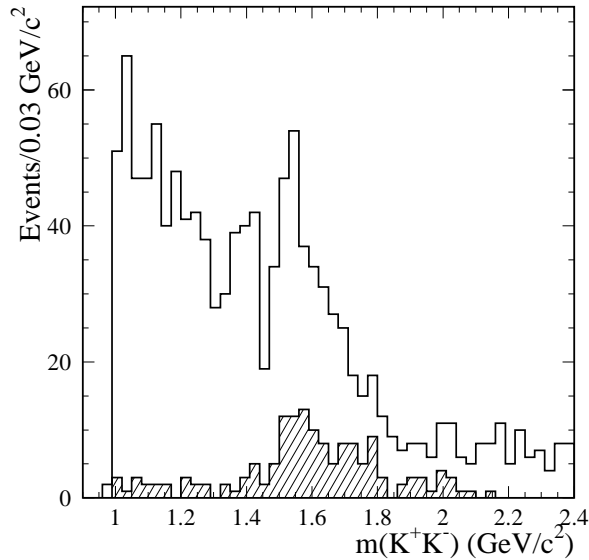}
\vspace{-0.4cm}
\caption{
  Invariant mass distribution for \KpKm pairs in events in which
  the other \KpKm pair has mass closest to and within 10~\mevcc of 
  the nominal $\phi$ mass (open histogram).
  The hatched histogram is for the subset with a \KKKK mass in the
  $J/\psi$ peak.
  }
\label{mkk_notphi}
\end{figure}

\subsection{\boldmath The $\phi(1020) \KpKm$  Intermediate State}
\label{sec:phif03}

Figure~\ref{mkk_phi} shows the invariant mass distribution for all
\KpKm pairs in the selected \KKKK events (4 entries per event) as the
open histogram.
A prominent $\phi$ peak is visible, along with possible peaks at 1.5,
1.7 and 2.0~\gevcc.
The hatched histogram is for the pair in each event with mass closest
to the nominal $\phi$ mass,
and indicates that the $\phi \KpKm$ channel dominates the \KKKK final state. 
Our previous finding of very little $\phi$ signal~\cite{isr4pi} was
incorrect due to an error in the analysis algorithm.

If the pair mass closest to the $\phi$ mass is within 10~\mevcc of the
$\phi$ mass, 
then we include the invariant mass of the other \KpKm combination 
in Fig.~\ref{mkk_notphi}.
The contribution from events in the $J/\psi$ peak is shown as the
hatched histogram which is in agreement with the BES
experiment~\cite{bes4k} which studied
the structures around 1.5, 1.7 and 2.0~\gevcc in detail.
There is no evidence for the $\phi f_0$ channel, 
but there is an enhancement at threshold that can be
interpreted as the tail of the $f_0(980)$.
This is expected in light of the $\phi f_0$ cross sections measured
above in the \KKppch and \KKppnt final states.
However the statistics and uncertainties in the $f_0(980) \!\to\! \KpKm$
lineshape do not allow a meaningful extraction of the cross section in
this final state.

We observe no significant structure in the $K^+ K^- K^{\pm}$ mass distribution.
We use these events to study the possibility that part of our
$\phi\pipi$ signal is due to $\phi\KpKm$ events with the two kaons not from
the $\phi$ taken as pions.
No structure is present in the reconstructed \KKppch invariant mass
distribution from these events.

\section{\boldmath $\epem \!\to \phi f_0$ Near Threshold}
\label{phif0bump}

\begin{figure}[tbh]
\includegraphics[width=0.9\linewidth]{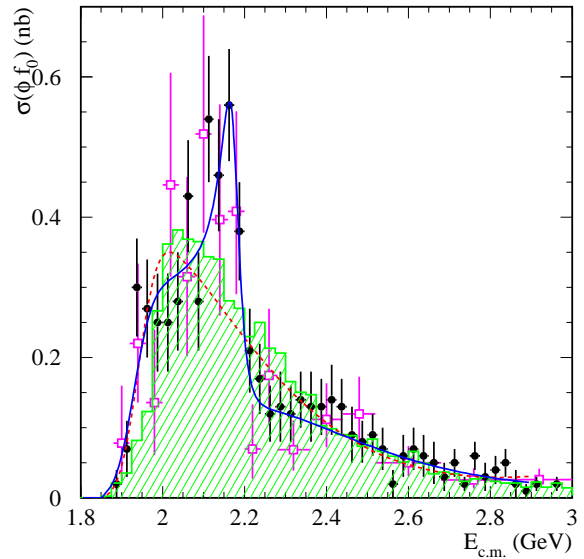}
\vspace{-0.3cm}
\caption{
  The $\epem \!\!\to\! \phi(1020) f_{0}(980)$ cross section measured 
  in the \KKppch (circles) and \KKppnt (squares) final states.
  The hatched histogram shows the simulated cross section, assuming no 
  resonant structure.
  The solid (dashed) line represents the result of the one-resonance
  (no-resonance) fit described in the text. 
  }
\label{phif0xsall}
\end{figure}

The behavior of the $\epem \!\!\to\! \phi f_0$ cross section near 
threshold shows a structure near 2150~\mevcc,
and we have published this result in Ref.~\cite{phif0prd}.
Here we provide a more detailed study of this cross section in the
1.8--3~\gev region.
In Fig.~\ref{phif0xsall} we superimpose the cross sections measured in the \KKppch 
and \KKppnt final states (shown in Figs.~\ref{phif0xs} and~\ref{phif0xs2});
they are consistent with each other.
The \KKKK cross section (Sec.~\ref{sec:phif03}) is also consistent
with the presence of a structure near 2150~\mevcc and shows a
contribution from the $\phi f_0$ channel, 
but since we cannot extract a meaningful $\phi f_0$ cross section, we
do not discuss this final state further.

First, we attempt to reproduce this spectrum with a smooth threshold
function.
In the absence of resonances, the only theoretical constraint on the
cross section well above threshold is that it should decrease smoothly
with increasing \Ecm.
However the form of the cutoff at threshold is determined by
the properties of the intermediate resonances and the final state
particle spins, phase space and detector resolution.
The model discussed in Sec.~\ref{sec:phipipi} takes the $\phi$ and 
$f_0(980)$ lineshapes, the spins of all particles and their phase
space into account, and postulates a simple $\Ecm^{-4}$ dependence of
the cross section.
For the $\epem \!\!\to\! \phi f_0$ reaction, it predicts the cross
section shown as the hatched histogram in Fig.~\ref{phif0xsall},
normalized to the same total area as the data.
It shows a sharp rise from the threshold with a peak near 2070~\mev 
and is inconsistent with the data.

To account for uncertainties in the $f_0$ width and the shape of
the cross section well above threshold, 
we seek a functional form that describes the simulation and whose
parameters can be varied to cover a reasonable range of possibilities.
This can be achieved by the product of a phase space term,
an exponential rise and a second order polynomial:
\begin{eqnarray}
\sigma_{nr}(\mu) &\! =\! & P(\mu) \cdot A_{nr}(\mu) ,        \label{nrfcn} \\
     A_{nr}(\mu) &\! =\! & \sigma_0 \cdot (1-e^{-(\mu / a_1)^4}) 
                           \cdot (1 + a_2\mu + a_3\mu^2) ,   \nonumber     \\
        P(\mu)   &\! =\! & \sqrt{1 - m_0^2/(m_0 +\mu)^2}, ~~
          \mu        =    \Ecm\! - m_0 ,    \nonumber
\end{eqnarray}
where
the $a_i$ are free parameters,
$\sigma_0$ is a normalization factor,
and $P(\mu)$ is a good approximation of the two-body phase space 
for particles with similar masses.
Both the $\phi(1020)$ and $f_0(980)$ have small but finite widths, 
and our selection criterion of $m(\pi\pi)\! >\! 0.85~\gevcc$ defines 
an effective minimum mass, $m_0\! =\! 1.8$~\gevcc.
Fitting Eq.~\ref{nrfcn} to the simulated cross section yields the
$a_i$ values listed in the first column of Table~\ref{fittabc}.
Fitting to the data with all $a_i$ fixed to these values and $\sigma_0$ 
floating yields \chisq/n.d.f.$=\! 86/(56-2)$. 
Floating $a_2$ and $a_3$ in addition, 
we obtain \chisq/n.d.f.$=\! 85/(56-4)$ with a
confidence level (C.L.) of 0.0025. 
If we float all parameters in Eq.~\ref{nrfcn}, the fit yields 
\chisq/n.d.f.$= 80/(56-5)$ with C.L. of 0.0053.
The results of these fits are listed in Table~\ref{fittabc}, 
and the latter is shown as the dashed curve on Fig.~\ref{phif0xsall};
all fits are inconsistent with the data.

\begin{table}[hbt]
\caption{
  Parameter values, \chisq values and confidence levels 
  from the fits of Eq.~\ref{nrfcn} to the data described in the text.
  An asterisk denotes a value that was fixed in that fit.
  }
\label{fittabc}
\begin{ruledtabular}
\begin{tabular}{l c r@{$\pm$}l c r@{$\pm$}l c r@{$\pm$}l} 
   Fit   & \hspace*{0.4cm} & \multicolumn{2}{c}{All $a_i$ fixed}  
         & \hspace*{0.4cm} & \multicolumn{2}{c}{Only $a_1$ fixed}
         & \hspace*{0.4cm} & \multicolumn{2}{c}{All $a_i$ free}       \\
\hline
$\sigma_0$    && 1.19 & 0.03  &&   1.23 & 0.03  &&   1.09 & 0.01  \\
$a_1$         && \multicolumn{2}{c}{0.218*}
                              && \multicolumn{2}{c}{0.218*}
                                                &&   0.174& 0.012 \\
$a_2$         && \multicolumn{2}{c}{$-$1.68*}
                              &&$-$1.51 & 0.15  &&$-$1.49 & 0.12  \\
$a_3$         && \multicolumn{2}{c}{0.81*}
                              &&   0.66 & 0.14  &&   0.63 & 0.11  \\
\chisq /n.d.f.&& \multicolumn{2}{c}{86.4/54}
                              && \multicolumn{2}{c}{85.3/52}
                              && \multicolumn{2}{c}{80.5/51}\\
P(\chisq)     && \multicolumn{2}{c}{0.0035}
                              && \multicolumn{2}{c}{0.0025}
                              && \multicolumn{2}{c}{0.0053} \\
\end{tabular}
\end{ruledtabular}
\end{table}

We now add a resonance and fit the data with the function
\begin{eqnarray}
  \sigma_{1r}(\mu) &  =  &  \frac{P(\mu)}{P(m_1)}  \cdot
  \left| A_{nr}(\mu)e^{i\psi_1} + 
         A_{r1}(\mu) \right|^2 ,                  \label{bwsig} \\
    A_{r1}(\mu)    &  =  &   \frac{\sqrt{\sigma_1} m_1 \Gamma_1}
                     {m_1^2- \Ecm^2\! -i \Ecm \Gamma_1} ,    \nonumber
\end{eqnarray}
where 
$m_1$ and $\Gamma_1$ are the mass and width of the resonance,
$\sigma_1$ is its peak cross section,
and $\psi_1$ is its phase relative to the non-resonant component.
We obtain good fits both assuming no interference between the two
components, $\psi_1\! =\! \pi$, and with $\psi_1$ floating.
The result of the latter fit is shown as the solid curve on 
Fig.~\ref{phif0xsall}.
The data are somewhat above this curve near 2.4~\gevcc
and a fit with two resonances can also describe the data.
Due to the sharp drop near 2.2~\gevcc, the single-resonance fit with
interference gives a resonance mass about 30~\mevcc higher than the
other two fits.
All these fits, with or without resonances, give a peak non-resonant
cross section in the range 0.3--0.4~nb,
which is of independent theoretical interest, because it can be related
to the $\phi\to f_0(980)\gamma$ decay studied at the
$\phi$-factory~\cite{phif0theory}. 

Under the hypothesis of one resonance interfering with the
non-resonant component, the fit gives the resonance parameters
\begin{eqnarray*}
 \sigma_1 & = &  0.13  \pm 0.04~{\rm nb,} \\
      m_1 & = &  2.175 \pm 0.010~\gevcc , \\
 \Gamma_1 & = &  0.058 \pm 0.016~\gev , {\rm and} \\
   \psi_1 & = & -0.57  \pm 0.30~{\rm radians,} 
\end{eqnarray*}
along with \chisq/n.d.f.$=\! 37.6/(56-9)$ (C.L. 0.84). 
We can estimate the product of its electronic width and branching 
fraction to $\phi f_0$ as
$$
  \BR_{\phi f_0} \cdot \Gamma_{ee} = 
  \frac{\Gamma_1 \sigma_1 m_1^2}{12\pi C } = (2.5\pm 0.8\pm 0.4)~\ev\ ,
$$
where we fit the product $\Gamma_1 \sigma_1$ to reduce correlations, 
and the conversion constant $C\! =\! 0.389$~mb(\gevcc)$^2$.
The second error is systematic and corresponds to the normalization
errors on the cross section.

The significance of the structure calculated from the change in \chisq
between the best fit and the null hypothesis is 6.2 standard
deviations.
Since this calculation can be unreliable in the case of low statistics
and functions that vary rapidly on the scale of the bin size,
we perform a set of simulations in which we generate a number of
events according to a Poisson distribution about the number observed
in the data and with a mass distribution given by either the
simulation or fitted function in Fig.~\ref{phif0xsall} without
resonant structure.
On each sample, we perform fits to Eqs.~\ref{nrfcn} and~\ref{bwsig}
and calculate the difference in \chisq.
The fraction of trials giving a \chisq difference larger than that seen
in the data corresponds to a significance of approximately 5
standard deviations. 

\begin{figure}[tbh]
\includegraphics[width=0.9\linewidth]{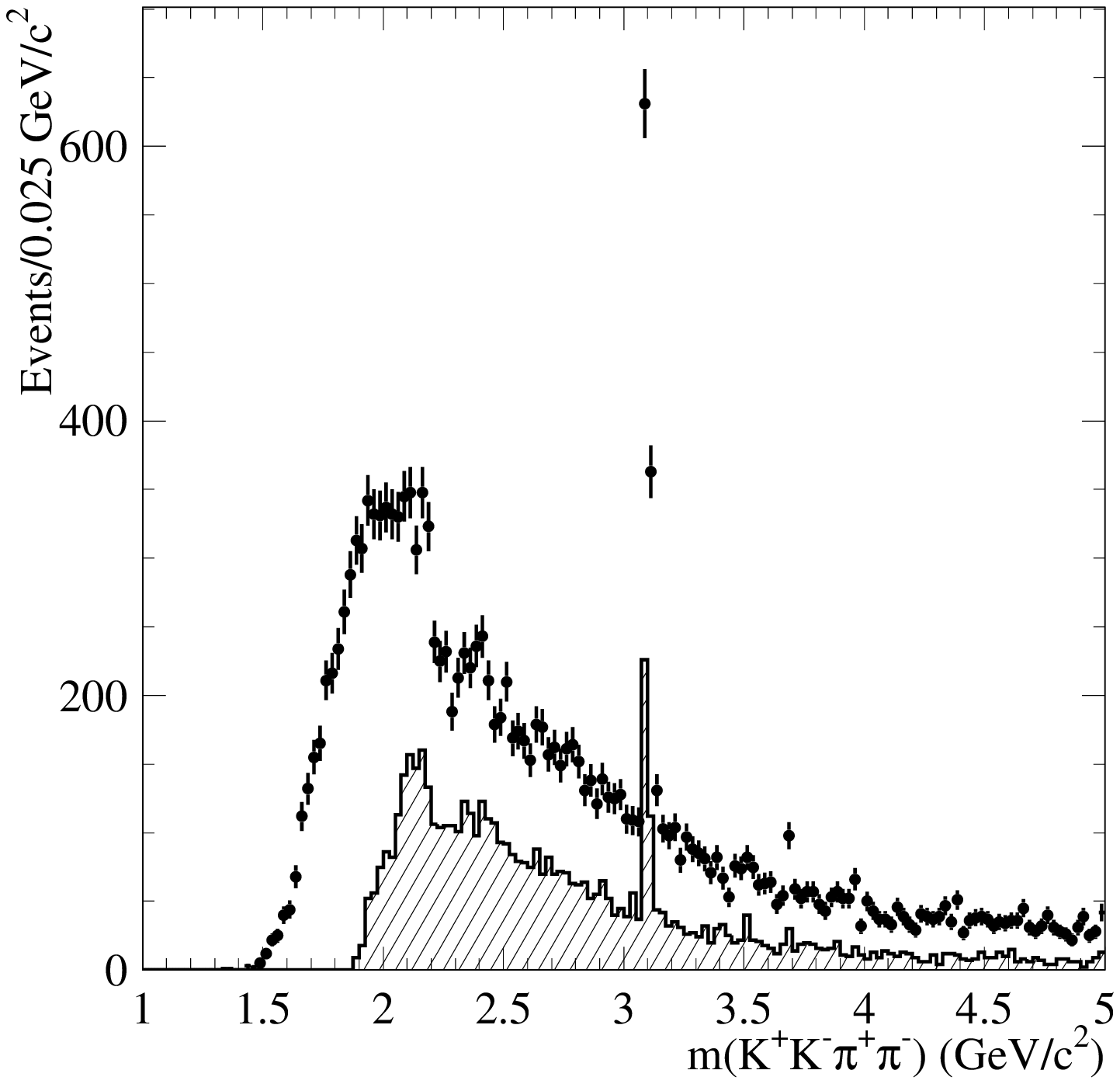}
\vspace{-0.6cm}
\caption{
  Invariant mass distribution for all selected \KKppch events lying
  outside the $K^{*0}(892)$ bands of Fig.~\ref{kkstar} (points),
  and the subset of these events with $0.85<m(\pipi)<1.10$~\gevcc (hatched).
  }
\label{2k2pi_nok*}
\includegraphics[width=0.9\linewidth]{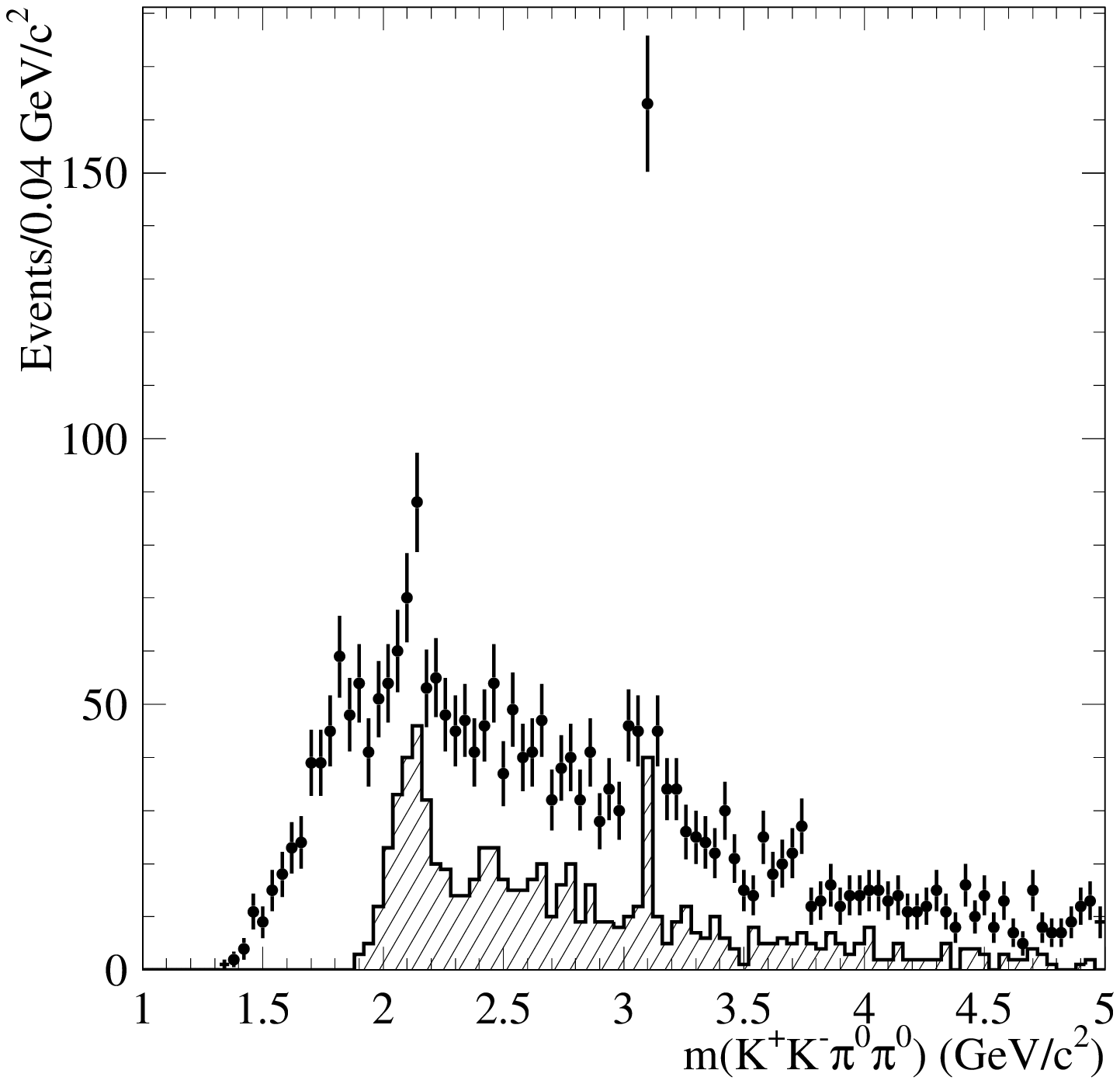}
\vspace{-0.6cm}
\caption{
  Invariant mass distribution for all selected \KKppnt events lying
  outside the $K^{*0}(892)$ bands of Fig.~\ref{kkstarpi0} (points),
  and the subset of these events with $0.85<m(\ppz)<1.10$~\gevcc (hatched).
  }
\label{2k2pi0_nok*}

\end{figure}

We search for this structure in other submodes with different and/or
fewer intermediate resonances.
The total cross sections are dominated by $K^*K\pi$ channels,
and the $K^{*0}\Kp\pim$ cross section is shown
in Fig.~\ref{kstar_sel}.
There is no significant structure in the 2.1--2.5~\gev region, 
but the point-to-point statistical uncertainties are large.
If we remove events within the bands in Figs.~\ref{kkstar}
and~\ref{kkstarpi0}, 
then most of the events containing a $K^*$ are eliminated and we
obtain the raw mass distributions shown as the points with errors in 
Figs.~\ref{2k2pi_nok*} and~\ref{2k2pi0_nok*}, respectively.
Both distribution show evidence of a structure around 2.15~\gevcc
and the \KKppch distribution also shows a structure near 2.4~\gevcc. 
We cannot exclude the presence of these structures in events with a
$K^*$, but we can conclude that they do not dominate those events,
whereas they comprise a substantial fraction of the remaining events 
in that mass region.

Applying the further requirement that the dipion mass be in the range
0.85--1.10~\gevcc,
we remove most of the events without an $f_0$, and obtain the mass
distributions shown as the hatched histograms in 
Figs.~\ref{2k2pi_nok*} and~\ref{2k2pi0_nok*}.
Peaks are visible at both 2.15~\gevcc and 2.4~\gevcc in both
distributions, and they contain enough events to account for the
corresponding structures in the distributions for all non-$K^*$ events.
These peaks contain at least as many events as are present in the $\phi f_0$
samples, but the non-resonant components are higher and there is a
substantial kinematic overlap between $\Kp\Km f_0$ events and
$K^*K\pi$ events in this mass range.

\begin{figure}[tbh]
\includegraphics[width=0.9\linewidth]{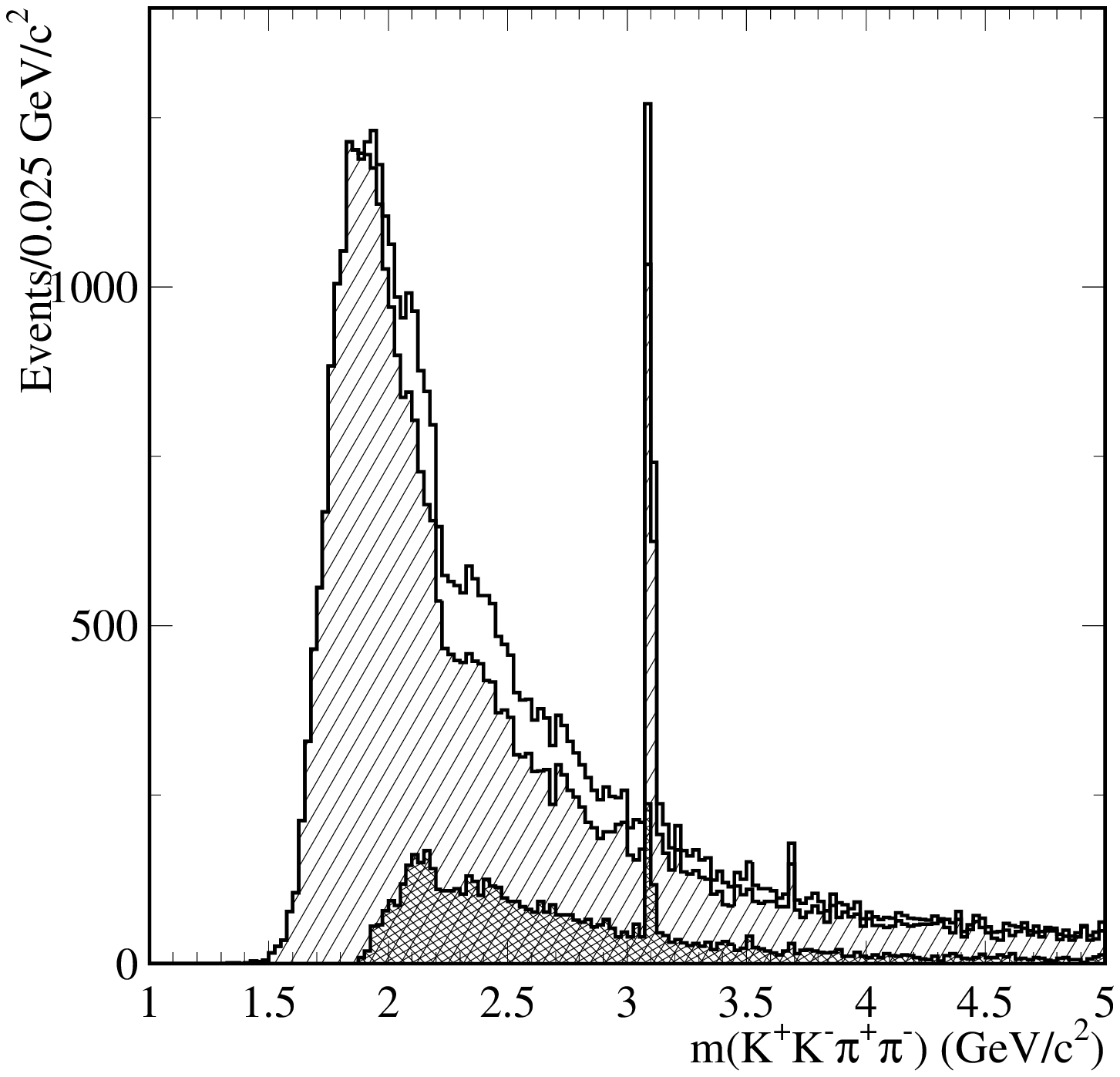}
\vspace{-0.6cm}
\caption{ 
   The $K^+K^-\pipi$ invariant mass distribution for all selected
   events (open histogram), and for those with a \pipi mass
   inside (cross-hatched) or outside (hatched) the $f_0$ band
   as defined in the text.
   }
\label{f0ch}
%
\includegraphics[width=0.9\linewidth]{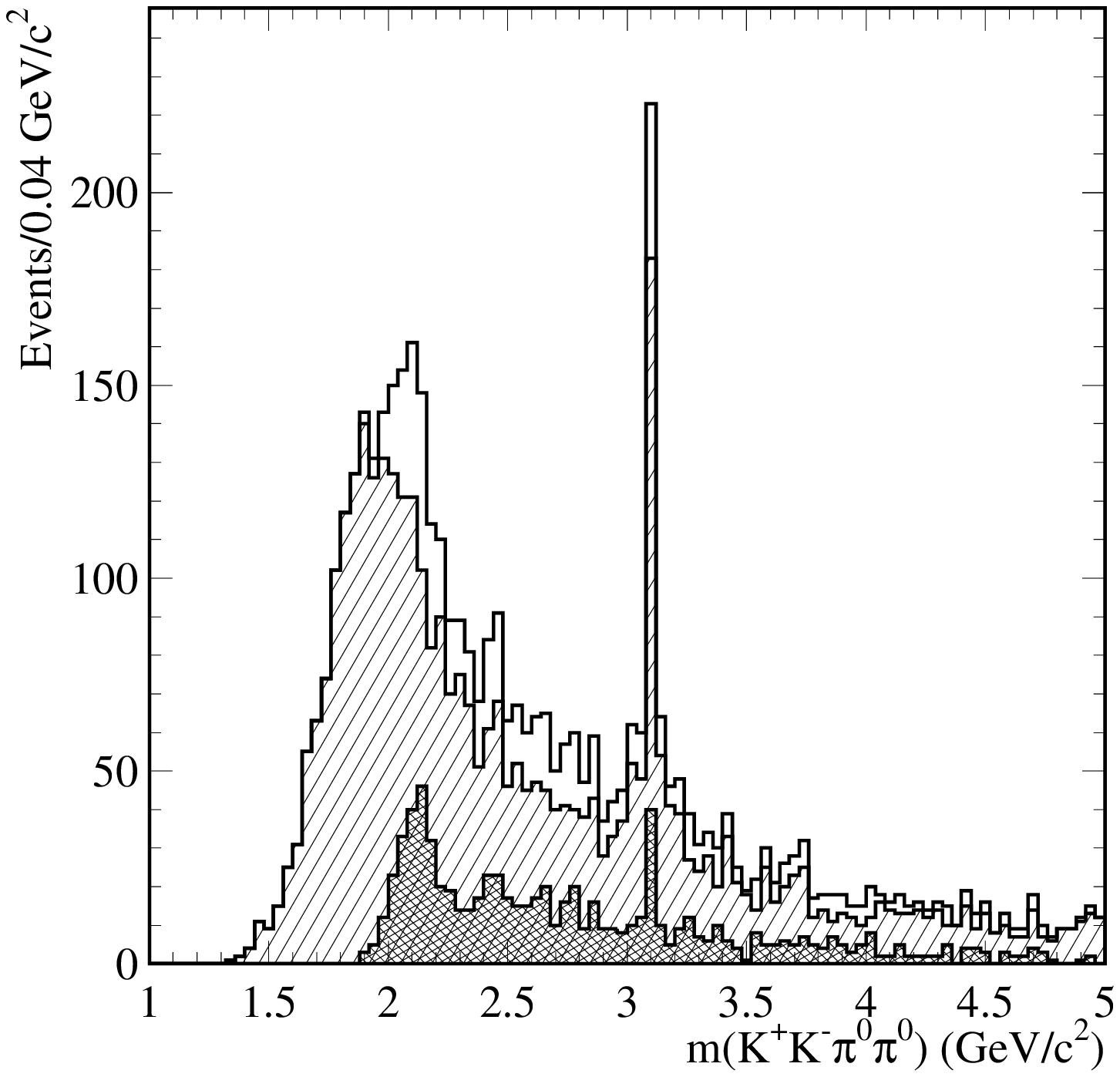}
\vspace{-0.6cm}
\caption{ 
   The $K^+K^-\ppz$ invariant mass distribution for all selected
   events (open histogram), and for those with a \ppz mass
   inside (cross-hatched) or outside (hatched) the $f_0$ band
   as defined in the text.
   }
\label{f0nu}
\end{figure}

Since this $f_0(980)$ band appears to contain a large fraction of 
the events within the structure,
we now consider all selected events with a dipion mass inside or
outside this range.
Figure~\ref{f0nu}
shows the mass distribution for all selected \KKppnt events as the
open histogram,
and the subsets of events with $\ppz$ mass inside and outside the range 
0.85--1.10~\gevcc as the hatched and cross-hatched histograms, respectively.
It is evident that the $K^+ K^- f_0$ channel contains the majority of
the structure in the 2.0--2.6~\gevcc range.

We show the corresponding distributions for the \KKppch events in
Fig.~\ref{f0ch}. 
Due to the presence of the $\rho^0$, 
the relative $f_0$ contribution is much smaller in this final state,
but the events in the $f_0$ band show clear indications of structure in
the 2.0--2.4~\gevcc region.
The remaining events may also have structure in this region, but the
statistical significance is marginal and it could be due to other
sources, such as the $\phi f_2(1270)$ threshold at 2.3~\gevcc.

\begin{figure}[tbh]
\vspace{0.3cm}
\includegraphics[width=0.85\linewidth]{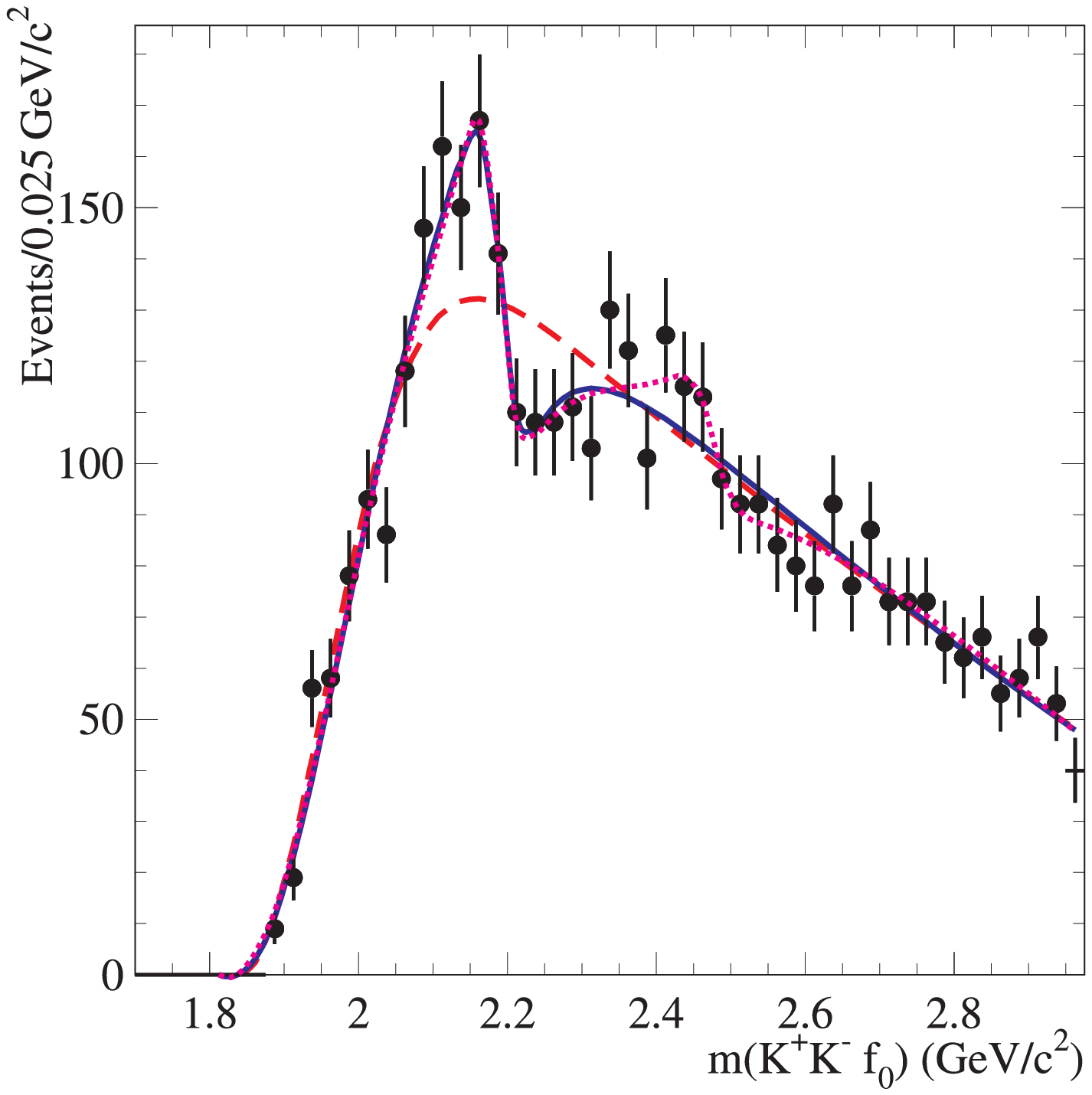}
\vspace{-0.3cm}
\caption{
   The $K^+K^-\pipi$ invariant mass distribution in the $\Kp\Km f_0(980)$
   threshold region for events with a \pipi mass inside the $f_0$ band.
   The lines represent the results of the fits including no (dashed), one
   (solid) and two (dotted) resonances described in the text.
   }
\label{pipifit}
%
\vspace{0.3cm}
\includegraphics[width=0.83\linewidth]{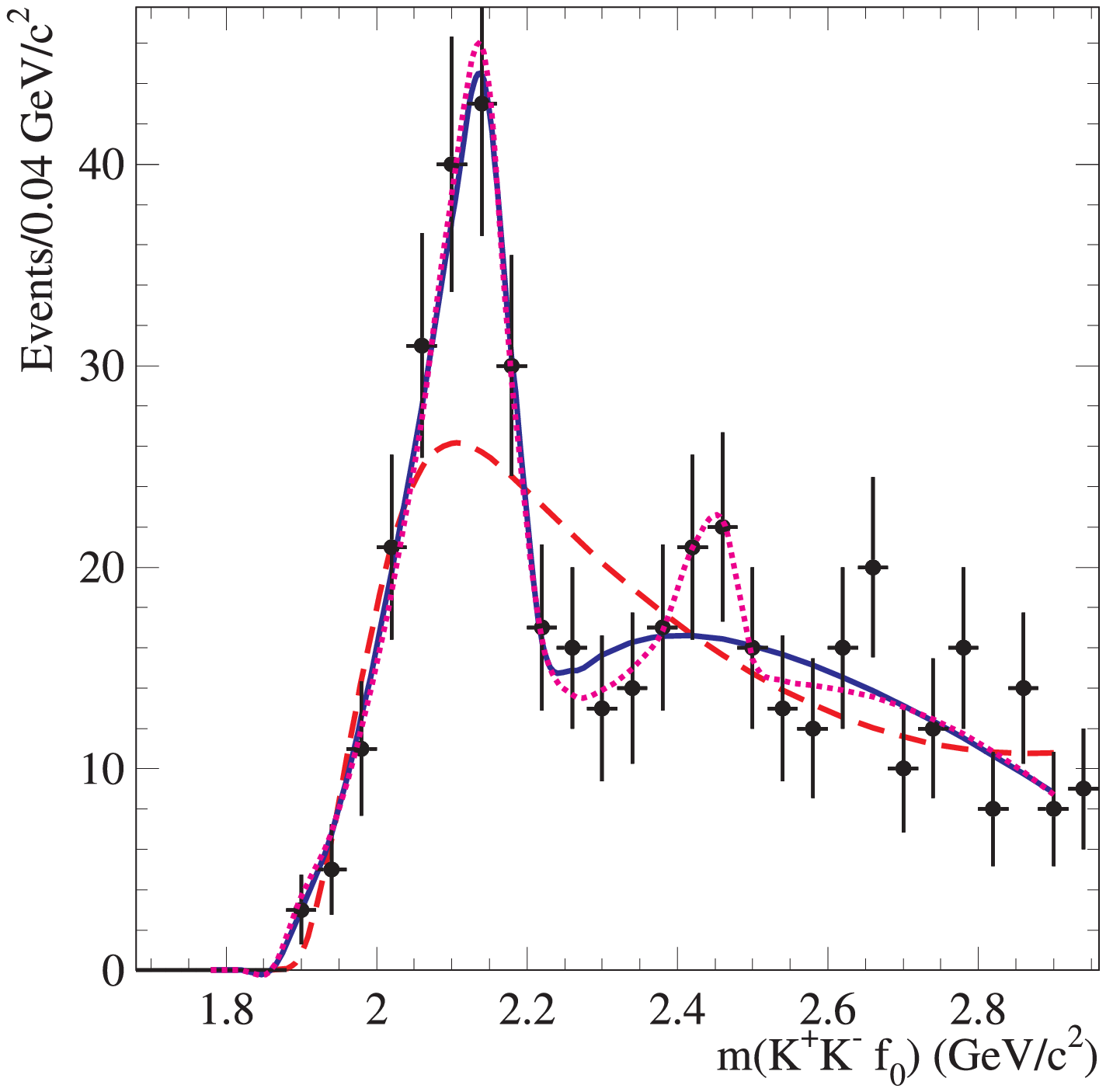}
\vspace{-0.3cm}
\caption{
   The $K^+K^-\ppz$ invariant mass distribution in the $\Kp\Km f_0(980)$
   threshold region for events with a \ppz mass inside the $f_0$ band.
   The lines represent the results of the fits including no (dashed), one
   (solid) and two (dotted) resonances described in the text.
}
\label{ppzfit}
\end{figure}

\begin{table*}[tbh]
\caption{
  Summary of parameters obtained from the fits described in the text 
  to the \KKppch and \KKppnt events with dipion mass in the $f_0(980)$ band.
  An asterisk denotes a value that was fixed in that fit.
  }
\label{fittab2}
\vspace{-0.1cm}
\begin{center}
\begin{ruledtabular}
\begin{tabular}{l c c c c c c } 
    &  \multicolumn{2}{c}{No Resonance} 
    &  \multicolumn{2}{c}{One Resonance} 
    &  \multicolumn{2}{c}{Two Resonances}                        \\
Fit &  \KKppch & \KKppnt & \KKppch & \KKppnt & \KKppch & \KKppnt \\
\hline
$N_{nr}$          &  7204$\pm$775   &  991$\pm$202
                  &  8466$\pm$334   &  722$\pm$112
                  &  6502$\pm$476   &  117$\pm$89     \\
$a_1$             & 0.181$\pm$0.012 & 0.134$\pm$0.017
                  & 0.224$\pm$0.024 & 0.197$\pm$0.048
                  & 0.201$\pm$0.035 & 0.143$\pm$0.053 \\
$a_2$             &$-$0.75$\pm$0.21 &$-$1.47$\pm$0.38
                  &$-$0.89$\pm$0.17 &$-$0.36$\pm$0.10
                  &$-$0.44$\pm$0.15 &  5.80$\pm$2.36  \\
$a_3$             &   0.09$\pm$0.17 &   0.75$\pm$0.35
                  &   0.17$\pm$0.08 &$-$0.28$\pm$0.14
                  &$-$0.15$\pm$0.12 &$-$5.26$\pm$1.75 \\
$a_4$             &   0.75*         &   0.50*
                  &   0.75*         &   0.50*
                  &   0.75*         &   0.50*         \\
$N_1$             & 0*              & 0*
                  &  116$\pm$95     & 149$\pm$36
                  &  163$\pm$70     & 192$\pm$44      \\
$m_1$ (\gevcc)    &        --       &       --
                  & 2.192$\pm$0.014 & 2.169$\pm$0.020
                  & 2.187$\pm$0.013 & 2.154$\pm$0.029 \\
$\Gamma_1$ (\gev) &        --       &       --
                  & 0.071$\pm$0.021 & 0.102$\pm$0.027
                  & 0.066$\pm$0.018 & 0.110$\pm$0.022 \\
$\psi_1$ (rad)    &         --      &        --
                  &$-$0.60$\pm$0.41 &$-$1.02$\pm$0.19
                  &$-$1.10$\pm$0.14 &$-$1.04$\pm$0.23 \\
$N_2$             & 0*              & 0*
                  & 0*              & 0*
                  &  16$\pm$16      & 6$\pm$5         \\
$m_2$ (\gevcc)    &        --       &       --
                  &        --       &       --
                  & 2.47$\pm$0.07   & 2.45$\pm$0.04   \\
$\Gamma_2$ (\gev) &        --       &       --
                  &        --       &       --
                  & 0.077$\pm$0.065 & 0.062$\pm$0.102 \\
$\psi_2$ (rad)    &         --      &        --
                  &         --      &        --
                  & 0.28$\pm$1.06   &   1.41$\pm$1.29 \\
\chisq /n.d.f.    &    62.8/41      &     38.1/21
                  &    35.6/37      &     13.0/17
                  &    31.4/34      &      9.7/13     \\
P(\chisq)         &     0.016       &      0.012
                  &     0.54        &      0.74
                  &     0.60        &      0.72       \\
\end{tabular}
\end{ruledtabular}
\end{center}
\end{table*}

Figures~\ref{pipifit} and~\ref{ppzfit} show enlarged views of the mass
distributions within the $f_0$ bands from
Figs.~\ref{f0ch} and~\ref{f0nu}, respectively.
The two-peak structure is more evident here than in the $\phi f_0$ events.
The $0.85\! <\! m(\pi\pi)\! <\! 1.10$~\gevcc requirement gives enough
phase space for $K^+K^-$ invariant mass to cover the region from
threshold to $\sim$1.3~\gevcc for $m(\Kp\Km\pi\pi)\approx 2.15$~\gevcc.
From the measured kaon form factor we expect to find only about two-thirds 
of $\Kp\Km$ P-wave in our fitted $\phi$ peak.
Since the non-ISR and ISR $\pi\pi\pi\pi$ backgrounds have not been subtracted
and the samples contain an unknown mixture of intermediate states,
we fit them with a modified version of Eq.~\ref{bwsig} that allows up
to two resonances,
\begin{eqnarray}
F(\mu)  &  =  &  (a_4 \cdot A_{nr})^2                     \\
        &  +  &  |(1-a_4)A_{nr} + A_{r1}e^{i\psi_1} + A_{r2}e^{i\psi_2}|^2.
\nonumber
\end{eqnarray}
Here, 
the normalization is in terms of events rather than cross section
($\sigma_i \!\!\to\! N_i$)
and a fraction $a_4$ of the non-resonant component does not interfere 
with the resonances.
We first fit the distribution with no resonances (and $a_4\! =\! 1$). 
The results are shown as the dashed lines in Figs.~\ref{pipifit} and~\ref{ppzfit}
and listed in Table~\ref{fittab2}; 
both are inconsistent with the data.

We next include one resonance in the fit.
The parameter $a_4$ is not well constrained by the data and its value 
has a small influence on all other fit parameters except for the
number of events assigned to the resonance, 
so we present results with $a_4$ fixed to the reasonable values of 
0.75 and 0.50 for the \KKppch and \KKppnt data, respectively.
The results are shown as the solid lines in Figs.~\ref{pipifit} 
and~\ref{ppzfit} and listed in Table~\ref{fittab2}.
The fit quality is good in both cases,
the fitted resonance parameters are consistent with those from the
$\phi f_0$ study, and the calculated significance of the structure for
the \KKppch data is similar, 5.2 standard deviations.
The \KKppnt data show much more pronounced structure than in the 
$\phi f_0$ study, allowing a full fit to this sample with a
significance of 5.0 standard deviations.

We then add a second resonance to the fit, keeping $a_4$ fixed and
floating all other parameters.
The results are shown as the dotted lines in Figs.~\ref{pipifit} 
and~\ref{ppzfit}, and listed in Table~\ref{fittab2}.
These fits are also of good quality, but do not change the \chisq CL
or the parameters of the first resonance significantly.
We also perform fits with no interference between the non-resonant
component and any resonance ($a_4\! =\! 1$),
obtaining good quality fits for both one resonance and two resonances
with relative phase $\pi/2$.
The fitted resonance parameters are consistent in all cases, except
that the mass of the first resonance is lower by about 50~\mevcc,
similar to the 30~\mevcc shift seen in the $\phi f_0$ study.

From these studies we conclude that we have observed a new vector structure 
at a mass of about 2150~\mevcc with a significance of over six standard 
deviations.
It decays into $\Kp\Km f_0(980)$, with the $\Kp\Km$ pair produced 
predominantly via the $\phi(1020)$.
There is an additional structure at about 2400~\mevcc,
and the two structures can be described by either two resonances or a
single resonance that interferes with the non-resonant $\Kp\Km f_0(980)$
process.
More data and searches in other final states are needed to understand
the nature of these structures.

If the main structure is due to a resonance, then it is relatively
narrow and might be interpreted as the strange analog of the recently 
observed charmed Y(4260) state~\cite{y4260}, which decays to
$J/\psi\pipi$.
The value of 
$\BR_{\phi f_0} \cdot \Gamma_{ee} = (2.5\pm 0.8\pm 0.4)~\ev$
measured here is similar to the value of
$\BR_{Y\to J/\psi\pipi} \cdot \Gamma^Y_{ee} = (5.5\pm 1.0\pm 0.8)~\ev$
reported in Ref.~\cite{y4260}.


\begin{figure}[tbh]
\includegraphics[width=0.9\linewidth]{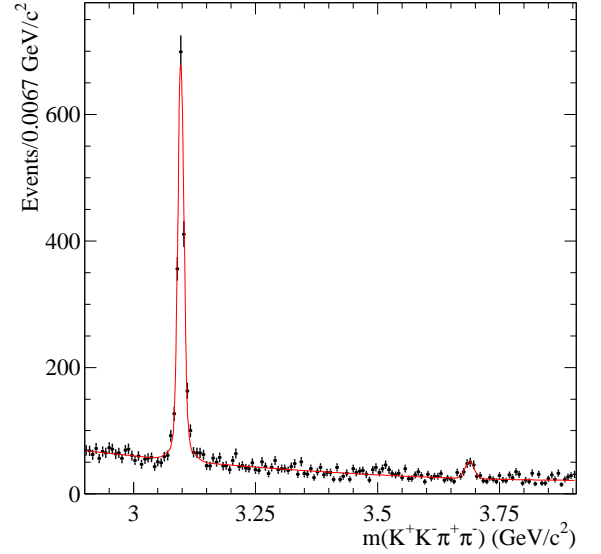}
\vspace{-0.4cm}
\caption{
  Raw invariant mass distribution for all selected 
  $\epem \!\!\to\! \KKppch$ events in the charmonium region.
  The line represents the result of the fit described in the text.
  }
\label{jpsi}
\end{figure}

\section{\boldmath The Charmonium Region}
\label{sec:charmonium}

\begin{figure}[tbh]
\includegraphics[width=0.9\linewidth]{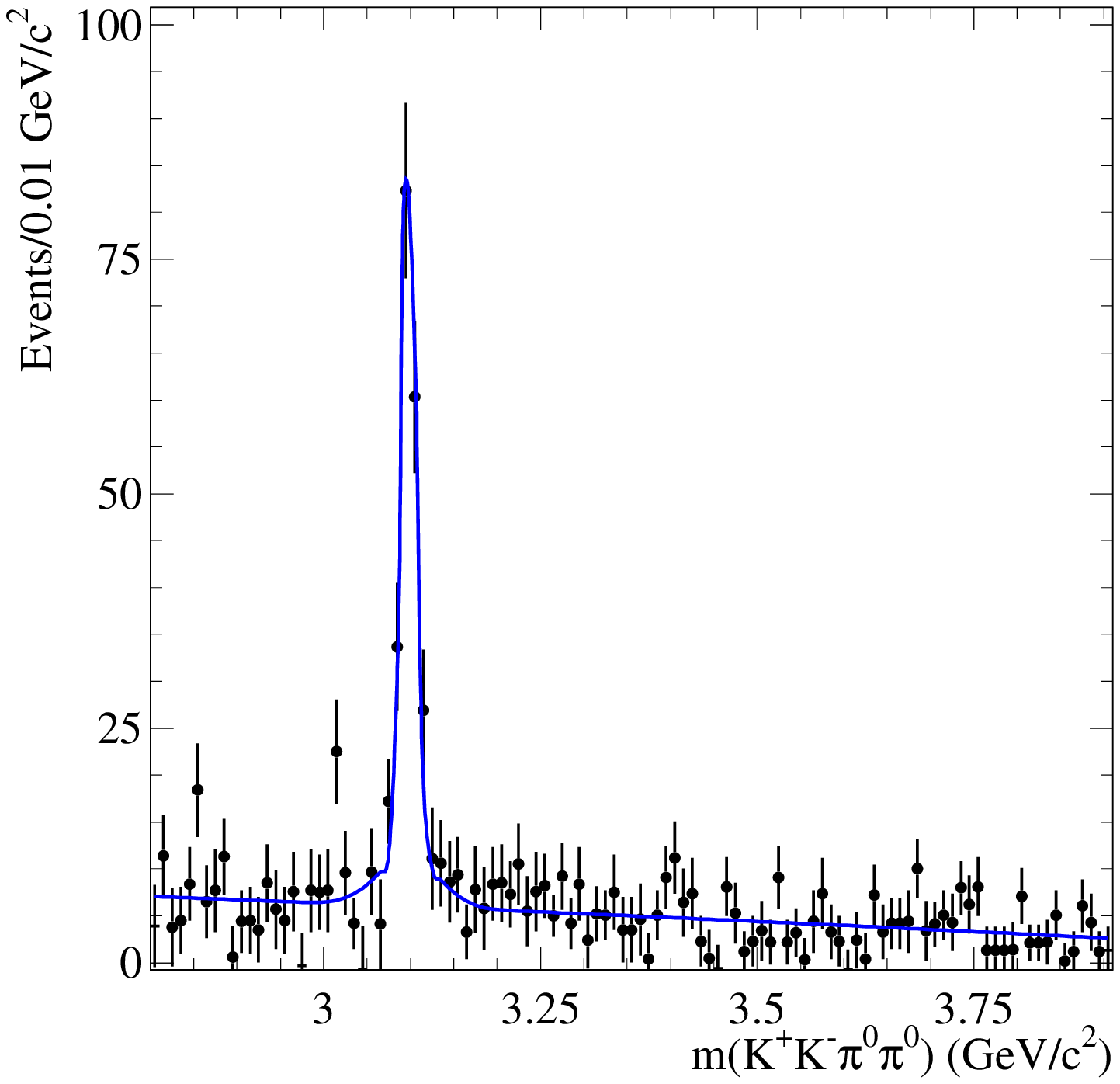}
\vspace{-0.4cm}
\caption{
  Invariant mass distribution for $\epem \!\!\to\! \KKppnt$ events
  in the charmonium region, after partial background subtraction.
  The line represents the result of the fit described in the text.
  }
\label{jpsi2pi0}
%
\includegraphics[width=0.9\linewidth]{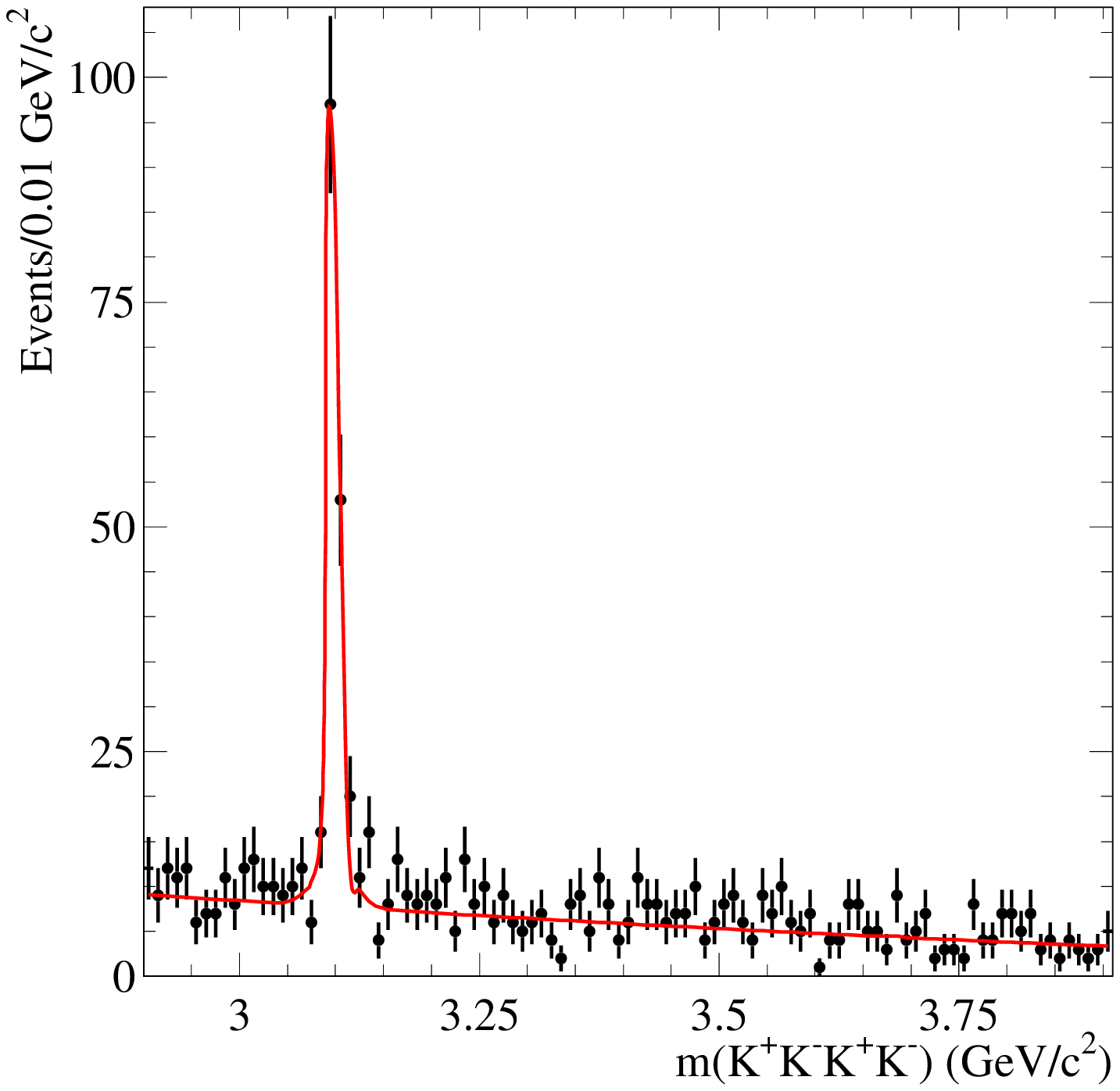}
\vspace{-0.5cm}
\caption{
  Raw invariant mass distribution for all selected 
  $\epem \!\!\to\! \KKKK$ events in the charmonium region.
  The line represents the result of the fit described in the text.
  }
\label{jpsi4k}
\end{figure}

The data at masses above 3~\gevcc can be used to measure or set limits
for the branching fractions of narrow resonances, such as charmonia,
and the narrow $J/\psi$ and $\psi(2S)$ peaks allow measurements of our mass
scale and resolution.
Figures~\ref{jpsi}, \ref{jpsi2pi0} and~\ref{jpsi4k} show the invariant
mass distributions for the selected \KKppch, \KKppnt and \KKKK events,
respectively, in this region, 
with finer binning than in the corresponding Figs.~\ref{2k2pi_babar}, 
~\ref{2k2pi0_babar} and~\ref{4k_babar}.
We do not subtract any background from the \KKppch or \KKKK data,
since it is small and nearly uniformly distributed,
but we use the \chiKKppnt control region to subtract part of the ISR 
background from the \KKppnt data.
Signals from the $J/\psi$ are visible in all three distributions,
and the $\psi(2S)$ is visible in the \KKppch mode.

We fit each of these distributions using a sum of two Gaussian functions 
to describe the $J/\psi$ and $\psi (2S)$ signals plus a polynomial to 
describe the remainder of the distribution.
We take the signal function parameters from the simulation, but let
the overall mean and width float in the fit, along with the amplitude
and the coefficients of the polynomial.
The fits are of good quality and are shown as the curves on
Figs.~\ref{jpsi}, \ref{jpsi2pi0} and~\ref{jpsi4k}.
In all cases, the fitted mean value is within 1~\mevcc of the
PDG~\cite{PDG} $J/\psi$ or $\psi (2S)$ mass,
and the width is consistent within 10\% with the simulated resolution 
discussed in Sec.~\ref{sec:xs2k2pi}, \ref{sec:2k2pi0xs} or~\ref{sec:4kxs}.

The fits yield 
$1586\pm58$ events in the $J/\psi$ peak for the \KKppch final state, 
 $203\pm16$ events for \KKppnt and
 $156\pm15$ events for \KKKK.
From these numbers of observed events in each final state $f$,
$N_{J/\psi \!\to\! f}$, 
we calculate the product of the $J/\psi$ branching fraction to $f$ and
the $J/\psi$ electronic width: 
\begin{equation}
     \BR_{J/\psi \!\to\! f} \cdot \Gamma^{J/\psi}_{ee}  =
 \frac{N_{J/\psi \!\to\! f} \cdot m_{J/\psi}^2}
      {6\pi^2 \cdot d{\cal L}/dE \cdot \epsilon_f(m_{J/\psi}) \cdot C}
 ~~~, \\
\label{jpsicalc}
\end{equation}
where
$d{\cal L}/dE\! =\! 89.8~\invnb/\mev$ and $\epsilon_f(m_{J/\psi})$ 
are the ISR luminosity and corrected selection efficiency, respectively,
at the $J/\psi$ mass
and $C$ is the conversion constant.  
We estimate $\epsilon_{\KKppch} \! =\! 0.202$, 
$\epsilon_{\KKppnt} \! =\! 0.069$ and $\epsilon_{\KKKK} \! =\! 0.176$.

Using $\Gamma^{J/\psi}_{ee}\! =\! 5.40\pm0.18~\kev$~\cite{PDG}, 
we obtain the branching fractions listed in Table~\ref{jpsitab},
along with the measured products and the current PDG values.
The systematic errors include a 3\% uncertainty on $\Gamma^{J/\psi}_{ee}$.
The branching fractions to \KKppch and \KKKK are more precise than the
current PDG values,
which were dominated by our previous results of 
(6.25$\pm$0.80)$\times$10$^{-3}$ and (7.4$\pm$1.8)$\times$10$^{-4}$,
respectively~\cite{isr4pi}.
This is the first measurement of the \KKppnt branching fraction.

These fits also yield 91$\pm$15 \KKppch events in the $\psi(2S)$ peak,
but no other significant signals.
We expect 6.3 events from
$\psi(2S) \!\to\! J/\psi\pipi \!\!\to\! \KKppch$
from the relevant branching fractions~\cite{PDG},
which is less than the statistical error.
Subtracting this contribution and using a calculation analogous 
to Eq.~\ref{jpsicalc}, with $d{\cal L}/dE\! =\! 115.3~\invnb/\mev$,
we obtain the product of the $\psi(2S) \!\!\to\! \KKppch$ branching fraction
and its electronic width.
Dividing by the world average value of $\Gamma^{\psi(2S)}_{ee}$~\cite{PDG},
we obtain the branching fraction listed in Table~\ref{jpsitab};
it is consistent with the current PDG value~\cite{PDG}.

\begin{figure}[tbh]
\includegraphics[width=0.9\linewidth]{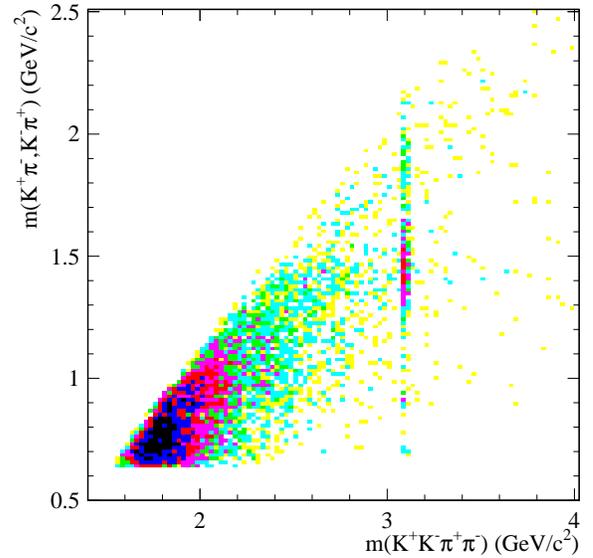}
\vspace{-0.6cm}
\caption{
  The $K^\pm\pi^\mp$ invariant mass versus \KKppch invariant mass for 
  events with the other $K^\mp\pi^\pm$ combination in the
  $K^{*0}(892)$ bands of Fig.~\ref{kkstar}(a).  The overlapped region
  is taken only once.
  }
\label{jpsi_kkstarvs2k2pi}
\end{figure}

\begin{figure}[tbh]
\includegraphics[width=0.9\linewidth]{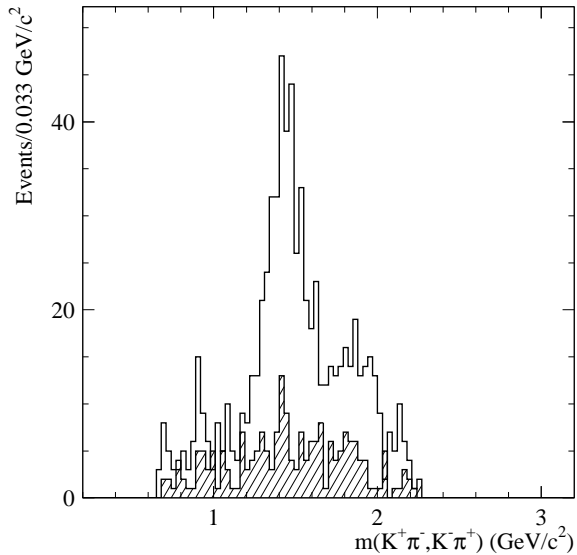}
\vspace{-0.6cm}
\caption{
  The $K^\pm\pi^\mp$ mass projection from Fig.~\ref{jpsi_kkstarvs2k2pi}
  for events with a \KKppch mass within 50~\mevcc of the $J/\psi$ mass
  (open histogram) and 50--100~\mevcc below (hatched).
  }
\label{jpsi_kk2}
\end{figure}

As noted in Sec.~\ref{sec:kaons} and shown in Fig.~\ref{kkstar}, 
the \KKppch final state is dominated by the $K^{*0}(892)K\pi$ channels,
with a small fraction seen in the $K^{*0}(892) \Kbar_2^{*0}(1430) + c.c.$
channels.
Figure~\ref{jpsi_kkstarvs2k2pi} shows a scatter plot of the invariant
mass of a $K^\pm\pi^\mp$ pair versus that of the \KKppch system in events with
the other $K^\mp\pi^\pm$ pair near the $K^{*0}(892)$ mass,
i.e. within the bands in Fig.~\ref{kkstar}(a) with overlapped region
taken only once. 
There is a large concentration of entries in the $J/\psi$ band with 
$K^\pm\pi^\mp$ masses near 1430~\mevcc,
but no solid evidence for a horizontal band corresponding to
$\Kbar_2^{*0}(1430)$ production other than in $J/\psi$ decays.  
We show the $K^\pm\pi^\mp$ mass projection for the subset of events
with a \KKppch mass within 50~\mevcc of the $J/\psi$ mass 
in Fig.~\ref{jpsi_kk2} as the open histogram.  
The hatched histogram is the projection for events with a \KKppch mass
between 50 and 100~\mevcc below the $J/\psi$ mass.


The $J/\psi$ component appears to be dominated by the $K_2^{*0}(1430)$.
Also seen is a small signal from $K^{*0}(892)$ 
indicating the $K^{*0}(892)\bar K^{*0}(892)$ decay of $J/\psi$:
this is also seen
as an enhancement in the vertical $J/\psi$ band in Fig.~\ref{jpsi_kkstarvs2k2pi}. 
The enhancement at
1.8~\gevcc of Fig.~\ref{jpsi_kk2} can be explained by the $J/\psi$ decay
into $K^{*0}(892) K_2(1770) + c.c.$ (or $K^{*0}(892) K_2(1820) +
c.c.$), a mode which has not previously been reported.
Subtracting the number of side-band events from the number in the
$J/\psi$ mass window,  
we obtain 317$\pm$23 events with a $K^\pm\pi^\mp$ mass in the range 
1200--1700~\mevcc,
which we take as a measure of $J/\psi$ decays into
$K^{*0}(892) \Kbar_2^{*0}(1430)$,
$25 \pm 8$ events in the
0.8--1.0~\gevcc window for  the $K^{*0}(892)\bar K^{*0}(892)$ decay and
$110 \pm 14$ events for the $K^{*0}(892) K_2(1770)$ 
or $K^{*0}(892) K_2(1830)$ final state in the 1.7--2.0~\gevcc region.
We convert these to branching fractions using Eq.~\ref{jpsicalc} and
dividing by the known branching fractions of excited kaons~\cite{PDG}.
The results are listed in Table~\ref{jpsitab}: 
they are considerably more precise than the PDG values.
We cannot calculate $B_{J/\psi\to K^{*0} K_2 (1770)}$ because no
branching fractions of $K_2 (1770)$ or $K_2 (1830)$ to $K\pi$ are reported.

\begin{figure}[tbh]
\includegraphics[width=0.9\linewidth]{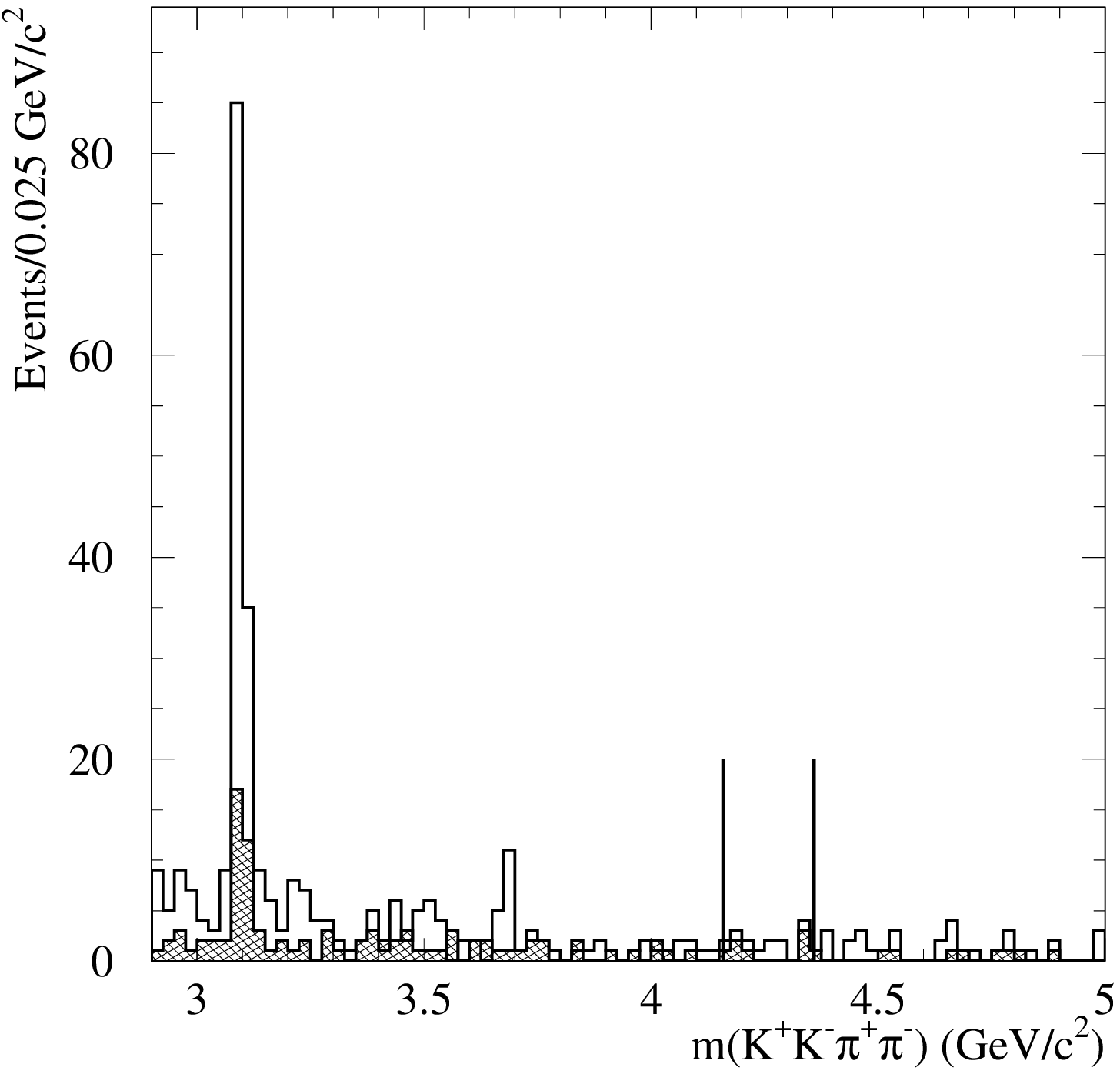}
\vspace{-0.6cm}
\caption{
  Raw invariant mass distributions for
  candidate $\epem \!\!\to\! \phi\pipi$ events (open histogram) and
  events in the $\phi$ side bands of Fig.~\ref{phif0_sel}(c)
  (cross-hatched) in the charmonium region.
  The vertical lines indicate the region used for the $Y(4260)$ search.
  }
\label{jpsi_phi2pi}
%
\includegraphics[width=0.9\linewidth]{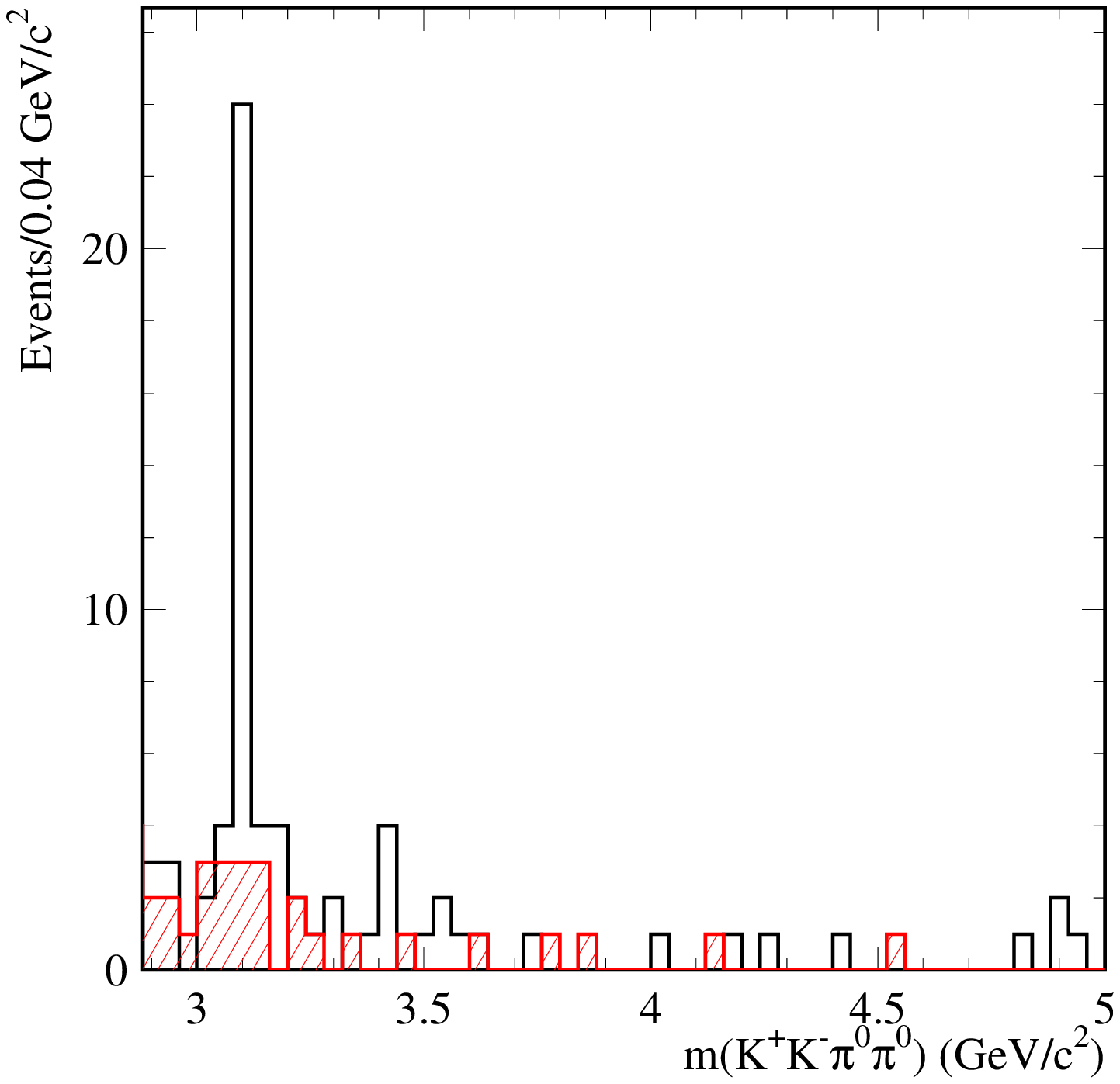}
\vspace{-0.6cm}
\caption{
  Raw invariant mass distributions for
  candidate $\epem \!\!\to\! \phi\ppz$ events (open histogram) and
  events in the \chiKKppnt control region (hatched) in the charmonium region.
  }
\label{jpsi_phi2pi0}
\end{figure}

We study decays into $\phi\pipi$ and  $\phi\ppz$ using the
mass distributions shown in Figs.~\ref{jpsi_phi2pi} and~\ref{jpsi_phi2pi0}, 
respectively.
The open histograms are for the events with a $\Kp\Km$ mass within the
$\phi$ bands of Figs.~\ref{phif0_sel}(c) and~\ref{phif0_sel2}(c).
The cross-hatched histogram in Fig.~\ref{jpsi_phi2pi} is from the
$\phi$ sidebands of Fig.~\ref{phif0_sel}(c) and represents the
dominant background in the $\phi\pipi$ mode.
The hatched histogram in Fig.~\ref{jpsi_phi2pi0} is from the
\chiKKppnt control region and represents the dominant background in
the $\phi\ppz$ mode.
Subtracting these backgrounds,
we find 103$\pm$12 $J/\psi \!\to\! \phi\pipi$ events,
23$\pm$6 $J/\psi \!\to\! \phi\ppz$ events,
and 10$\pm$4 $\psi(2S) \!\to\! \phi\pipi$ events.
We convert these to branching fractions and list them in Table~\ref{jpsitab}.
This is the first measurement of the $J/\psi \!\to\! \phi\ppz$ 
branching fraction, 
and the other two are consistent with current PDG values.

If the $Y(4260)$ has a substantial branching fraction into $\phi\pipi$,
then we would expect to see a signal in Fig.~\ref{jpsi_phi2pi}.
In the mass range $|m(\phi\pipi)-m(Y)|<0.1$~\gevcc,
we find 10 events,
and assuming a uniform distribution we estimate 9.2 background events from
the 3.8--5.0~\gevcc region. 
This corresponds to a signal of $0.8\pm 3.3$ events or a limit of $< 5$
events at the 90\% C.L. 
Using $d{\cal L}/dE\! =\! 147.7~\invnb/\mev$ at the $Y(4260)$ mass,
we calculate
$\BR_{Y\to   \phi\pipi} \cdot \Gamma^{Y}_{ee}\! <\! 0.4~\ev $ 
which is well below the value of 
$\BR_{Y\to J/\psi\pipi} \cdot \Gamma^{Y}_{ee} = 
(5.5\pm 1.0\pm 0.8)~\ev$~\cite{y4260}. 
No $Y(4260)$ signal is seen in any other mode studied here.

\begin{figure}[tbh]
\includegraphics[width=0.9\linewidth]{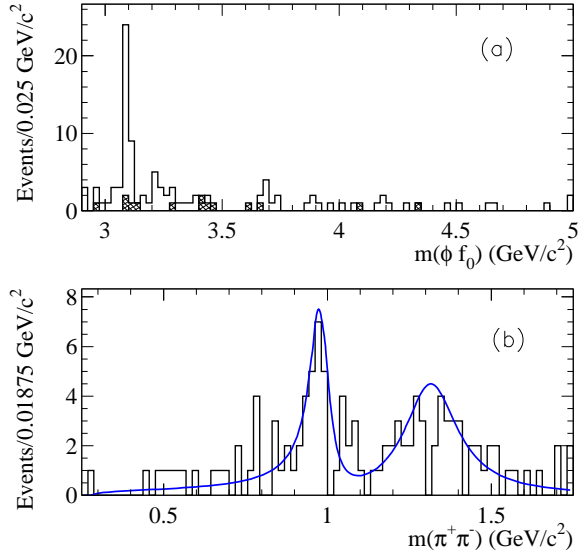}
\vspace{-0.6cm}
\caption{
  (a) Raw invariant mass distribution for candidate $\phi f_0$,
  $f_0 \!\!\to\! \pipi$ events (open histogram) and
  events in the $\phi$ side bands (cross-hatched)
  in the charmonium region;
  (b) the $\pipi$ invariant mass distribution for $\phi \pipi$ events from
  the $J/\psi$ peak of Fig.~\ref{jpsi_phi2pi}.
  The line represents the result of the fit described in the text.
  }
\label{jpsi_phif0}
\end{figure}

\begin{figure}[tbh]
\includegraphics[width=0.9\linewidth]{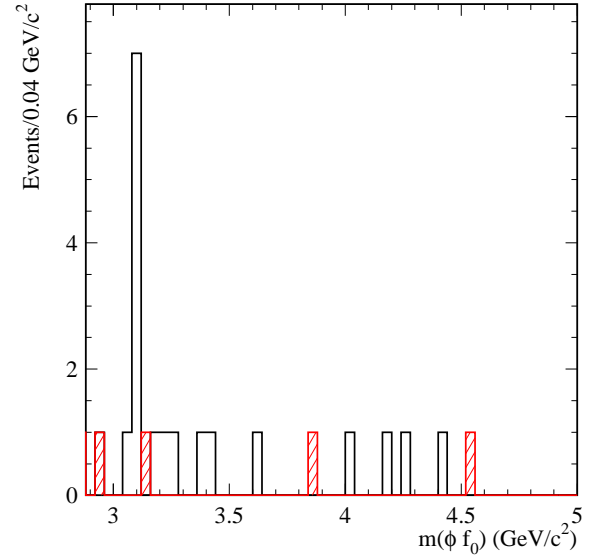}
\vspace{-0.6cm}
\caption{
  Raw invariant mass distribution for candidate $\phi f_0$,
  $f_0 \!\!\to\! \ppz$ events (open histogram) and
  events in the \chiKKppnt control region (hatched)
  in the charmonium region.
  }
\label{jpsi_phif0pi0}
\end{figure}

Figures~\ref{jpsi_phif0}(a) and~\ref{jpsi_phif0pi0} show the
corresponding mass distributions for $\phi f_0$ events,
i.e. the subsets of the events in Figs.~\ref{jpsi_phi2pi} 
and~\ref{jpsi_phi2pi0} with a dipion mass in the range 0.85--1.10~\gevcc.
Signals at the $J/\psi$ mass are visible in both cases, but
$\phi f_0$ is not the dominant mode of the $J/\psi \!\to\! \phi\pipi$ decay.
Figure~\ref{jpsi_phif0}(b) shows the $\pipi$ invariant mass
distribution for events in the $J/\psi$ peak of Fig.~\ref{jpsi_phi2pi},
$3.05\! <\! m(\KKppch)\! <\! 3.15$~\gevcc.
A two-peak structure is visible that can interpreted as due to 
the $f_0(980)$ and $f_2(1270)$ resonances.
Fitting the distribution in Fig.~\ref{jpsi_phif0}(b) with a sum of two 
Breit-Wigner functions
with parameters fixed to PDG values~\cite{PDG},
we find $19.5\pm 4.5$ $J/\psi \!\to\! \phi f_0$ events and
$44\pm 7$ $J/\psi \!\to\! \phi f_2$ events.
From Fig.~\ref{jpsi_phif0pi0} we estimate 
$7.0 \pm 2.8$ $\phi f_0$ events in the $\ppz$ mode. 


\begin{table*}[tbh]
\caption{
  Summary of the $J/\psi$ and $\psi(2S)$ branching fractions measured
  in this article.
  }
\label{jpsitab}
\begin{ruledtabular}
\begin{tabular}{r@{$\cdot$}l  r@{.}l@{$\pm$}l@{$\pm$}l 
                              r@{.}l@{$\pm$}l@{$\pm$}l
                              r@{.}l@{$\pm$}l } 
\multicolumn{2}{c}{Measured} & \multicolumn{4}{c}{Measured}    &  
\multicolumn{7}{c}{$J/\psi$ or $\psi(2S)$ Branching Fraction  (10$^{-3}$)}\\
\multicolumn{2}{c}{Quantity} & \multicolumn{4}{c}{Value (\ev)} &
\multicolumn{4}{c}{Calculated, this work}    & 
\multicolumn{3}{c}{PDG2006} \\
\hline
$\Gamma^{J/\psi}_{ee}$  &  $\BR_{J/\psi  \to \KKppch}$  &
  36&3 & 1.3 & 2.1  &   6&72 & 0.24 & 0.40  &   6&2 & 0.7  \\

$\Gamma^{J/\psi}_{ee}$  &  $\BR_{J/\psi  \to \KKppnt}$  &
  13&6 & 1.1 & 1.3  &   2&52 & 0.20 & 0.25  & \multicolumn{3}{c}{no entry} \\

$\Gamma^{J/\psi}_{ee}$  &  $\BR_{J/\psi  \to \KKKK}  $  &
   4&11& 0.39& 0.30 &   0&76 & 0.07 & 0.06  &   0&78 & 0.14\\[0.4cm]

$\Gamma^{J/\psi}_{ee}$  &  $\BR_{J/\psi  \to K^{*0} \Kbar_2^{*0}}
                      \cdot \BR_{K^{*0}  \to K\pi}
                      \cdot \BR_{K_2^{*0}\to K\pi}   $  \hspace*{0.4cm}&
   7&3 & 0.5 & 0.6  &   2&7  & 0.2  & 0.2   &   6&7  & 2.6\\

$\Gamma^{J/\psi}_{ee} $ & $\BR_{J/\psi\to K^{*0}\bar K{*0}}
                        \cdot\BR_{K^{*0}\to K\pi} 
                         \cdot\BR_{\bar K{*0}\to K\pi}$ &
  0&57 & 0.18 & 0.05 &  0&11 & 0.04 & 0.01 & 
$<$0&\multicolumn{2}{l}{5 at 90\% C.L.} \\

$\Gamma^{J/\psi}_{ee} $ & $B_{J/\psi\to K^{*0} K_2 (1770)}
                          \cdot\BR_{K^{*0}\to K\pi} 
                           \cdot\BR_{K_2\to K\pi}$ &
2&5 & 0.3 & 0.2 & \multicolumn{3}{c}{ -- } &
\multicolumn{3}{c}{\hspace{2.0cm}no entry} \\

$\Gamma^{J/\psi}_{ee}$  &  $\BR_{J/\psi  \to \phi\pipi}
                      \cdot \BR_{\phi    \to \Kp \Km}$  &
   2&61& 0.30& 0.18 &   0&98 & 0.11 & 0.07  &   0&94 & 0.15\\

$\Gamma^{J/\psi}_{ee}$  &  $\BR_{J/\psi  \to \phi\ppz} 
                      \cdot \BR_{\phi    \to \Kp \Km}$  &
   1&54& 0.40& 0.16 &   0&58 & 0.15 & 0.06  & \multicolumn{3}{c}{no entry} \\

$\Gamma^{J/\psi}_{ee}$  &  $\BR_{J/\psi  \to \phi f_0}
                      \cdot \BR_{  \phi  \to \Kp\Km}
                      \cdot \BR_{  f_0   \to \pipi}  $  &
   0&50& 0.11& 0.04 &   0&28 & 0.07 & 0.02  &   0&32 & 0.09 (s=1.9)\\

$\Gamma^{J/\psi}_{ee}$  &  $\BR_{J/\psi  \to \phi f_0}
                      \cdot \BR_{  \phi  \to \Kp\Km}
                      \cdot \BR_{  f_0   \to \ppz}   $  &
   0&47& 0.19& 0.05 &   0&54 & 0.21&  0.05  &   0&32 & 0.09 (s=1.9) \\

$\Gamma^{J/\psi}_{ee}$  &  $\BR_{J/\psi  \to \phi f_2}
                      \cdot \BR_{  \phi  \to K^+K^-}
                      \cdot \BR_{  f_2   \to \pipi}  $  &
   1&12& 0.18& 0.09 &   0&50 & 0.08 & 0.04  &
$<$0&\multicolumn{2}{l}{37 at 90\% C.L.} \\[0.4cm]

$\Gamma^{\psi(2S)}_{ee}$  &  $\BR_{\psi(2S) \to \KKppch} $  &
   2&56& 0.42& 0.16 &   1&2  & 0.2  & 0.08   &   0&72 & 0.05 \\

$\Gamma^{\psi(2S)}_{ee}$  &  $\BR_{\psi(2S) \to \phi\pipi}
                        \cdot \BR_{\phi     \to \Kp \Km} $  &
   0&28& 0.11& 0.02 &   0&27 & 0.11 & 0.02  &   0&113& 0.029\\ 

$\Gamma^{\psi(2S)}_{ee}$  &  $\BR_{\psi(2S) \to \phi f_0}
                        \cdot \BR_{\phi     \to K^+K^-}
                        \cdot \BR_{ f_0     \to \pipi}   $  &
  \hspace*{0.5cm}    0&17  & 0.08  & 0.02 \hspace*{0.6cm} &
  \hspace*{0.5cm}    0&26  & 0.12  & 0.03 \hspace*{0.6cm} &
  \hspace*{0.5cm}    0&090 & 0.033        \hspace*{0.6cm} 

\end{tabular}
\end{ruledtabular}
\end{table*}

Using Eq.~\ref{jpsicalc} and dividing by the appropriate branching
fractions, 
we obtain the $J/\psi$ branching fractions listed in Table~\ref{jpsitab}.
The measurements of $\BR_{J/\psi\to\phi f_0}$ in the \pipi and \ppz 
decay modes of the $f_0$ are consistent with each other and with the
PDG value, and combined they have roughly the same precision as listed
in the PDG~\cite{PDG}.
This is the first measurement of $\BR_{J/\psi\to\phi f_2}$, and the
value is consistent with the previous upper limit~\cite{PDG}.
We also observe $6\pm 3$ $\psi(2S) \!\to\! \phi f_0$, $f_0 \!\!\to\! \pipi$ 
events, which we convert to the branching fraction listed in 
Table~\ref{jpsitab}; 
it is consistent with the PDG value~\cite{PDG}, assuming
$\BR_{f_0\to\pipi} = 2/3$.  

In the $Y(4260)$ region we have 4 events with an estimated background  
of about 1 event.
This corresponds to 3$\pm$2 events, or a 90\% CL upper limit of 5.6 events.
We convert this to the limits
\begin{eqnarray*}
 \BR_{Y \to \phi f_0} \cdot \Gamma^{Y}_{ee} 
                      \cdot \BR_{\phi \to K^+  K^-}
                      \cdot \BR_{ f_0 \to \pipi} 
      &  <  &  0.14  ~\ev\ ,\\
 \BR_{Y \to \phi f_0} \cdot \Gamma^{Y}_{ee}  &  <  &  0.43~\ev, ~~90\% CL,
\end{eqnarray*} 
which is again much lower than the corresponding quantity for the
$Y(4260) \!\!\to\! J/\psi \pipi$ decay.


\section{Summary}
\label{sec:Summary}
\noindent

We use the excellent charged particle tracking and identification, and
photon detection of the \babar\ detector to fully reconstruct events
of the type 
$\epem \!\!\to\! \gamma\epem \!\!\to\! \gamma\KKppch$,  
$\gamma\KKppnt$ and  
$\gamma\KKKK$,  
where the $\gamma$ is radiated from the initial state $e^+$ or $e^-$.
Such events are equivalent to direct \epem annihilation at an
effective c.m.\@ energy corresponding to the mass of the hadronic system,
and we study the annihilation into these three final states at low
\Ecm, from their respective production thresholds up to 5~\gev.
The \KKppch and \KKKK measurements are consistent with, and supersede,
our previous results~\cite{isr4pi}.
This is the first measurement of the \KKppnt final state, although
some of the results were also presented in Ref.~\cite{phif0prd}.

The systematic uncertainties on the normalization of the
$\epem \!\!\to\! \KKppch$, \KKppnt and \KKKK cross sections are 
8\%, 10\% and 9\%, respectively, for $\Ecm \! <\! 3$~\gev, 
and 10\%, 14\% and 13\% in the 3--5~\gev range.
The obtained cross sections are considerably more precise than previous measurements and 
cover this low energy range completely, so they
provide important input to calculations of the hadronic corrections to
the anomalous magnetic moment of the muon and the fine structure
constant at the $Z^0$ mass.

These final states exhibit complex resonant substructure. 
In the \KKppch mode we measure the cross sections for the first time for
the specific channels
$\epem \!\!\to\! K^{*0}(890)\Km\pip$, $\phi\pipi$ and $\phi f_0$.
We also observe signals for the $\rho^0(770)$, $K_1(1270)$, $K_1(1400)$, 
$K_2^{*0}(1430)$ and $f_2^{*0}(1270)$ resonances.
It is difficult to disentangle these contributions to the final state,
and we make no attempt to do so in this paper. 
We note that 
the $\rho^0$ signal is consistent with being due entirely to $K_1$ decays,
and 
the total cross section is dominated by the $K^{*0}(892)\Km\pip+c.c.$ channels,
but there is no significant signal for 
$\epem \!\!\to\! K^{*0}(892)\Kbar^{*0}(892)$.

In the \KKppnt mode we measure cross sections for 
$\epem \!\!\to\! \phi f_0$ 
and observe signals for the $K^{*\pm}(892)$ and $K_2^{*\pm}(1430)$ 
resonances.
Again, the total cross section is dominated by the
$K^{*\pm}(892)K^\mp\piz$ channels,
and there is no signal for 
$\epem \!\!\to\! K^{*+}(892)K^{*-}(892)$.
The \KKppnt final state is not accessible to intermediate states
containing $K_1$ resonances,
and we note that the cross section is roughly a factor of four smaller over
most of the range than the \KKppch cross section,
consistent with $K^*K\pi$ dominance with a factor of two isospin
suppression of the \ppz final state and another factor of two for the
relative branching fractions of the neutral and charged $K^*$ to
charged kaons.

In the \KKKK mode we find 
$\epem \!\!\to\! \phi\Kp\Km$ to be the dominant channel.
With the current data sample we can say little about the other $\Kp\Km$
combination, except that there is an enhancement near threshold,
consistent with the $\phi f_0$ channel, and that in $J/\psi$ decays
there is structure in the 1.5--2.0~\gev region, consistent with that
observed by BES~\cite{bes4k}.

The $\phi f_0$ cross section measured in the \KKppch final state shows 
structure around 2.15~\gev and possibly 2.4~\gev,
and the corresponding measurement in the \KKppnt final state is consistent,
as reported in Ref.~\cite{phif0prd}.
Further investigation here reveals consistent results in the \KKKK
final state
and clear signals in the $\Kp\Km f_0$ channels, with 
$f_0 \!\!\to\! \pipi$ and \ppz.
The signals are predominantly from $\phi f_0$, but the relaxation of
the \KpKm mass requirement reveals a strong signal in the \KKppnt
final state.
This structure can be interpreted as a strange partner 
(with c-quarks replaced by s-quarks)
of the recently observed $Y(4260)$, which has the analogous decay mode
$J/\psi\pipi$,
or as an $\ssbar\ssbar$ state that decays predominantly to $\phi f_0$.

We also study charmonium decays into these final states and their
intermediate channels.
All nine of the $J/\psi$ branching fractions and one of the three 
$\psi(2S)$ branching fractions listed in Table~\ref{jpsitab} are as
precise or more precise than the current world averages.
We do not observe the $Y(4260)$ in any decay mode.
In particular, we find that the branching fraction for the
$Y(4260) \!\!\to\! \phi\pipi$ decay, that a glueball model~\cite{shin}
predicts, is less than one-tenth of that to $J/\psi\pipi$.

\section{Acknowledgments}
\label{sec:Acknowledgments}
We are grateful for the 
extraordinary contributions of our \pep2\ colleagues in
achieving the excellent luminosity and machine conditions
that have made this work possible.
The success of this project also relies critically on the 
expertise and dedication of the computing organizations that 
support \babar.
The collaborating institutions wish to thank 
SLAC for its support and the kind hospitality extended to them. 
This work is supported by the
US Department of Energy
and National Science Foundation, the
Natural Sciences and Engineering Research Council (Canada),
the Commissariat \`a l'Energie Atomique and
Institut National de Physique Nucl\'eaire et de Physique des Particules
(France), the
Bundesministerium f\"ur Bildung und Forschung and
Deutsche Forschungsgemeinschaft
(Germany), the
Istituto Nazionale di Fisica Nucleare (Italy),
the Foundation for Fundamental Research on Matter (The Netherlands),
the Research Council of Norway, the
Ministry of Education and Science of the Russian Federation, 
Ministerio de Educaci\'on y Ciencia (Spain), and the
Science and Technology Facilities Council (United Kingdom).
Individuals have received support from 
the Marie-Curie IEF program (European Union) and
the A. P. Sloan Foundation.

\end{document}